\documentclass{jfm}

\usepackage{multirow}
\usepackage[normalem]{ulem}

\newcommand{\parone}[2] {\df{\partial #1}{\partial#2}}

\newcommand{\partwo}[2] {\df{\partial^2 #1}{\partial#2^2}}

\newcommand{\derivone}[2]{\df{\mathrm{d}#1}{\mathrm{d}#2}}

\newcommand{\df}[2]{\displaystyle{\frac{#1}{#2}}}

\newcommand{\tens}[1]{\mathsfbi{#1}}  

\usepackage{subcaption}
\usepackage[font=footnotesize]{subcaption} 
\usepackage{amssymb}
\usepackage{multirow}
\usepackage{array}
\usepackage{hyperref}
\usepackage{graphicx}
\usepackage{epstopdf, epsfig}
\usepackage{color}
\usepackage{amsmath,mathtools}
\usepackage{placeins}
\usepackage{changes}
\usepackage{mathrsfs}
\usepackage[toc,page]{appendix}
\allowdisplaybreaks
\usepackage{tikz}

\begin{document}
\shorttitle{Bulk, boundaries and microswimmers
}
\shortauthor{S. Maretvadakethope, A. L. Hazel, B. Vasiev  and R. N. Bearon}

\title{ 
The interplay between bulk flow and boundary conditions on the distribution of micro-swimmers in channel flow
}

\author{Smitha Maretvadakethope\aff{1}\corresp{\email{sm6412@liverpool.ac.uk}},
 Andrew L. Hazel\aff{2}, Bakhti Vasiev\aff{1} \and  Rachel N. Bearon\aff{1}}

\affiliation{\aff{1}Department of Mathematical Sciences, University of Liverpool,\\
Liverpool, L69 7ZL, UK
\aff{2}Department of Mathematics, University of Manchester, \\
Manchester, M13 9PL, UK}

\maketitle

\begin{abstract}
{ While previous experimental and numerical studies of dilute micro-swimmer suspensions have focused on the behaviours of swimmers in the bulk flow and near boundaries, models typically do not account for the interplay between bulk flow and the choice of boundary conditions imposed in continuum models. In our work, we highlight the effect of boundary conditions on the bulk flow distributions, such as through the development of boundary layers or secondary peaks of cell accumulation in bulk-flow swimmer dynamics. For the case of a dilute swimmer suspension in Poiseuille flow, we compare the distribution (in physical and orientation space) obtained  from individual based stochastic models with those  from continuum models, and identify mathematically sensible continuum boundary conditions for different physical scenarios (i.e. specular reflection, uniform random reflection and absorbing boundaries). We identify that the spread of preferred cell orientations is dependent on the interplay between rotation driven by the shear flow (Jeffery orbits) and rotational diffusion. We  find that in the absence of hydrodynamic wall-interactions, swimmers preferentially approach the walls perpendicular to the surface in the presence of high rotational diffusion, and that the preferential approach of swimmers to the walls is shape-dependent at low rotational diffusion (when suspensions tend towards a fully deterministic case). In the latter case, the preferred orientations  are nearly parallel to the surface for elongated swimmers and nearly perpendicular to the surface for near-spherical swimmers. Furthermore, we highlight the effects of swimmer geometries and shear throughout the bulk-flow on swimmer trajectories and show how the full history of bulk-flow dynamics affects the orientation distributions of micro-swimmer wall incidence.
}

\end{abstract}

\begin{keywords}
\end{keywords}
\section{Introduction} 

     Microorganisms are ubiquitous and can be found in disparate systems like soils, surfaces, and fluids. While microorganisms are not all harmful, and some are  important for the daily processes of larger lifeforms, like  gut bacteria in humans \citep{rinninella2019healthy} and microalgae in the marine food-chain \citep{arrigo2005marine}, there exist a number of pathogenic or toxic microorganisms \citep{hallegraeff2004manual}. Pathogenic bacteria are sources of infections and infectious diseases, ranging from typhoid fever (\textit{Salmonella} \textit{typhi}), to tuberculosis (\textit{Mycobacterium} \textit{tuberculosis}), pneumonia (Streptococcus, Pseudomonas), and food illnesses (other Salmonella) \citep{rowe1997multidrug,gordon2018microbe,cohen2007group,bodey1983infections,hardy1999food,ohl2001salmonella}. Meanwhile, harmful algal blooms \citep{anderson2021evidence} 
     can produce highly potent neurotoxins (e.g. \textit{Alexandrium} \textit{catenella}), block sunlight for aquatic plants, and lead to hypoxic and anoxic water \citep{mohd2020prolonged}. A neurotoxin build-up can lead to serious injury or death in marine animals, freshwater animals, and humans. The motility of many microorganisms  \citep{jarrell2008surprisingly,kearns2010field} makes them effective pathogens \citep{ottemann1997roles} especially when using medical equipment. For example, biofilms can develop inside medical devices, such as catheters, and subsequent upstream motility of the microorganisms can then lead to infection \citep{figueroa2020coli}. To develop improved insertion devices it is essential to understand the  behaviours of motile microorganism suspensions in sheared flows, especially as the microorganisms approach surfaces. Harmful microorganisms can also contaminate water transport infrastructure, and if not dealt with early on (or prevented from colonising surfaces) can lead to illness, serious injury, or death in local populations which consume the water. The prevention of such contamination is important for population well-being and also the associated industries which seek to meet governmental regulation targets.   
    
Since motile microorganisms are exceedingly small and typically 
on the micron scale \citep{childress1981mechanics}, swimming microorganisms perceive the fluids through which they traverse as highly viscous environments, and adapt their behaviour for motility in a regime with negligible inertia (Stokes flow). 
     For this traversal, some motile microorganisms have developed long, slender appendages, known as flagella,  which can create propulsion through various means \citep{brennen1977fluid}. Bacteria swim by bundling their appendages and rotating them  via specialised motors at flagellar bases; 
     sperm pass waves along their tails \citep{lauga2016bacterial}; and microalgae \citep{goldstein2015green} use  different strokes (recovery and effective strokes) to create asymmetry with various degrees of coordination (e.g. breaststroke motion in Chlamydomonas or metachronal waves in Volvox).

     A field of much recent interest has been the study of microswimmers near walls, whether these be hydrodynamic interactions, the mechanisms of reorientation, or accumulation to form biofilms. 
      Experiments in confined environments have shown swimming cells to be attracted to surfaces with some authors hypothesising that the hydrodynamic interaction of the cells with the walls realigns bacteria  parallel to the walls \citep{berke2008hydrodynamic} whilst puller-type algae 
      {(front actuated swimmers which pull in the fluid from the direction of propulsion)} approach walls at steep angles \citep{buchner2021hopping}. 
     In microfluidic channels, the phenomenon of upstream swimming has  been observed for bacteria \citep{hill2007hydrodynamic,kaya2009characterization} where \emph{E. coli} swimming in a region below a critical flow speed can reorient and swim against the direction of fluid flow. However, in the presence of strong flow,  swimming is dominated by  fluid advection, and cells are transported downstream. 
     In three-dimensions, \emph{E. coli} have also been observed to swim in clockwise circles near rigid surfaces \citep{frymier1995three,vigeant1997interactions,giacche2010hydrodynamic}. 
     Three-dimensional models for monotrichous bacteria near walls \citep{park2019flagellated}, which account for hydrodynamic interactions via regularised Stokeslets and the method of images, have also highlighted the importance of body aspect ratios to the inclination angles near walls and the radii of circular trajectories along walls, while finding that flagellar length affects whether bacteria can leave the wall.
 Meanwhile, numerical models without hydrodynamic interactions propose that the reorientation of swimmers interacting with walls can be explained purely mechanistically, by hitting a wall, maintaining orientation for a finite time scale,  rotating via Brownian rotation, and swimming away 
{\citep{li2009accumulation,li2011accumulation,costanzo2012transport,elgeti2013wall}}.
    In this paper, we will study microswimmer distributions and microswimmer wall interactions for a dilute suspension  via continuum modelling and stochastic individual based simulations. 
    {Here we do not account for inter-cellular nor cell-wall
hydrodynamic interactions, instead focusing on the impact of the bulk flow and swimmer geometry on cell trajectories, and explore a range of simplified boundary interactions. We can neglect the inter-cellular hydrodynamics as the suspensions are dilute.}

    We are interested in the relationship between the bulk flow and attachment dynamics, that occur  through swimmer-wall interactions. 
 To study the bulk behaviours of suspensions of microswimmers, continuum models have been developed to capture collective dynamics. These are developed as an alternative to expensive individual-based simulations. 
     These types of models have been used to study several suspension phenomena such as bioconvection \citep{pedley1992hydrodynamic}, downwelling gyrotactic swimming \citep{fung2020bifurcation} or determining how sheared flow can lead to layer formation below surface levels for gyrotactic swimmers  \citep{maretvadakethope2019instability}.
     Early continuum type models include advection-diffusion equations as introduced by \cite{kessler1986individual} where deterministic, directional dynamics are captured via  advection terms, and diffusion terms act to capture the randomness of microswimmers.  
    For gyrotactic swimmers, \cite{pedley1990new}  developed a model which allowed both the directional swimming and  the diffusion coefficient to be modified by the flow. It also accounted for reorientation of non-spherical particles by incorporating the reorientation of cells as described by Jeffery's equation \citep{jeffery1922motion,hinch1972effect}. This is particularly important due to the assumption that cells in a volume element swim relative to the fluid in the direction of cell orientation. 
 Another continuum model of note is the Smoluchowski equation, which models active suspensions using continuum kinetic theories, as reviewed in detail by \cite{saintillan2013active}. The Smoluchowski equation describes the cell distribution via a probability distribution function dependent on time, physical space and orientational space. For three-dimensional physical space, the problem has seven-dimensional dependence and is rarely solved in full generality due to the computational cost. 
     To reduce the problem the effective transport coefficients for the advection and diffusivity can be estimated by only using the local flow dynamics, and in generalised Taylor dispersion (GTD) the diffusivity is approximated from the probability distribution function of a tracer particle in orientation and physical space \citep{hill2002taylor,manela2003generalized, frankel1993taylor}.
 Although the GTD model is more accurate than the \cite{pedley1990new} model at high shear rates \citep{croze2013dispersion,croze2017gyrotactic,fung2020bifurcation}, it can fail for straining dominated flows.  
     A recent new transport model \citep{fung2021local} combines a transformation of the Smoluchowski equation into a transport equation with drift and dispersion terms approximated as functions of local flow fields, allowing it to be applied for any global flow field.
In our study of boundaries and bulk distributions 
we will consider a two-dimensional physical space Smoluchowski equation which reduces the problem to three-dimensional dependencies. 
{The results from our study will have implications on broadening the validity of models such as the doubly periodic Poiseuille flow models \citep{vennamneni2020shear}, justifying their application in capturing the dynamics and cell distributions for  bounded domains.}

    Given that the geometry of swimmers (particularly their aspect ratios) affect swimmer orientations in the bulk flow, the orientation distributions for swimmers interacting with walls are affected as well, thus prompting our study into determining how bulk flow and cell shape play a role in how microswimmers approach walls.
     Furthermore, there is the problem of determining appropriate boundary conditions to be used in continuum models, such as in \cite{bearon2015trapping} and \cite{ezhilan2015transport}. 
     It is possible to introduce a no-flux condition or Dirichlet conditions, or combination of the two. For the case of a two-dimensional equilibrium solution, the no-flux condition corresponds to the integral of the flux terms over all orientations being zero at the wall. 
This condition by itself does not specify the probability density of orientation distributions 
{ at the wall}, and 
{is not}  a sufficient condition to obtain a unique solution. 
    In \cite{bearon2015trapping} and \cite{ezhilan2015transport}  a point-wise no-flux boundary condition was proposed for a finite element solution, imposing that the flux in every direction must be zero 
    {for all microscale orientations}. However, this is not a sensible boundary condition because the formulation of the two-dimensional equilibrium Smoluchowski equation  leads to unrealistic cell densities in a boundary layer. We note that while some continuum models impose the additional constraint of perfect symmetry in  azimuthal angles and  spatial changes in orientation at boundaries that satisfy no-flux \citep{jiang2020dispersion}, this is not the only additional constraint which can satisfy the no-flux condition. We also note that in individual based dynamics, there exist various boundary interactions for Brownian swimmers 
    \citep{jakuszeit2019diffusion}, such as specular reflection \citep{volpe2014simulation, kumar2021taylor},  
     and different types of surface sliding models \citep{sipos2015hydrodynamic,spagnolie2015geometric,zeitz2017active}. Potential-free methods \citep{kumar2021taylor, peng2020upstream} have also been developed to study suspension dynamics.
In our study we consider suspensions in channels with height $W=426\mu$m (see table \ref{table:1}) and typical bacterial lengths of 1-2$\mu$m. The separation of length scales allows us neglect particle size. 
We  approximate surface interactions as point-like 
\citep{saintillan2013active,ezhilan2015transport} without concern about swimmer exclusion areas at the wall, as required when studying swimmers in microfluidic channels \citep{chen2021shape}. We further ignore hydrodynamic interactions with walls for simplicity, allowing us to study various pinball-like wall interactions. 

In this paper we develop and analyse dynamics captured by  two types of mathematical models (continuum models and stochastic individual based models) to determine sensible continuum model boundary conditions for different types of physical wall-interactions. We also study the underlying bulk-flow behaviours which lead to different distributions of wall interactions. 
    We will introduce a conservation equation (\S \ref{sec:conservation}, \S\ref{sec:2d}) and outline the numerical methods for solving the conservation problem (\S \ref{Sec:Methods}) via an an individual based stochastic method (\S \ref{Sec:IBMmethod}) and continuum model (\S \ref{sec:contModel}). 
   We will consider three types of particle-wall interactions (specular reflection, uniform random reflection and wall absorption) using individual based stochastic models and determine sensible corresponding continuum models. 
    We compare the relationships between  specular reflection at wall boundaries with a continuum {doubly periodic Poiseuille flow model} 
    (\S \ref{sec:DPPoiFlow}); between randomised reflections and continuum model with constant Dirichlet wall conditions  (\S \ref{sec:RandomReflection}); and between perfectly absorbing boundaries and a continuum model with zero Dirichlet constant wall conditions (\S \ref{sec:PerfAbsorption}).
      Finally, we will also analyse the role of shape, shear and diffusion dependent bulk flow dynamics on wall-interaction behaviour (\S \ref{sec:BulkFlowWallApproach}) and develop a novel accumulation index to quantify the importance of underlying deterministic trajectories  on wall interactions (\S \ref{sec:DetermineUnderlying}).

\section{Methods \label{Sec:Methods}}
\subsection{Conservation equation for $\psi$ \label{sec:conservation}}
We begin by considering the conservation equation for the probability distribution of microswimmers $\psi(\boldsymbol{x},\boldsymbol{p},t)$ that is dependent on swimmer position, $\boldsymbol{x}$, swimmer orientation $\boldsymbol{p}$, and time $t$,

\begin{align}
    \parone{\psi}{t}+\boldsymbol{\nabla}_{\boldsymbol{x}}\cdot (\Dot{\boldsymbol{x}}\psi)
    +\boldsymbol{\nabla}_{\boldsymbol{p}}\cdot (\Dot{\boldsymbol{p}}\psi)=0, \label{eq:FullConservationEqn}
\end{align}
 where $\boldsymbol{\nabla}_x$ and $\boldsymbol{\nabla}_p$ are the gradient operators in physical space and orientational space on a unit sphere of orientations $\Omega$, respectively. The translational flux, $\Dot{\boldsymbol{x}}$, and orientational flux, $\Dot{\boldsymbol{p}}$, as given in \cite{saintillan2013active}
 , are
 \begin{align}
    \Dot{\boldsymbol{x}}&=\boldsymbol{u}+V_s\boldsymbol{p}-D_T\boldsymbol{\nabla}_{\boldsymbol{x}}\ln{\psi},
    \\
    \Dot{\boldsymbol{p}}&=\beta\boldsymbol{p}\cdot\tens{E}\cdot(\tens{I}-\boldsymbol{pp})+\frac{1}{2}\boldsymbol{\omega}\times\boldsymbol{p}-d_r\boldsymbol{\nabla}_{\boldsymbol{p}}\ln{\psi}.
\end{align}
 The translational flux is dependent on the fluid velocity $\boldsymbol{u}$, the cell swimming at speed $V_s$ in direction $\boldsymbol{p}$, and translational diffusion $D_T$.  The orientational flux for an asymmetric swimmer with a shape factor (Bretherton constant) $\beta$, consists of the rotation characterised by the rate-of-strain tensor $\tens{E}$, background vorticity $\boldsymbol{\omega}$, and Brownian rotational diffusion $d_r$. The shape factor $\beta$ is restricted to $0\leq\beta<1$ for prolate shapes, where $\beta=0$ corresponds to spherical swimmers.
 
On integrating the conservation equation \ref{eq:FullConservationEqn} over all orientations, we obtain
\begin{align}
    \parone{}{t}\int_\Omega \psi(\boldsymbol{x}, \boldsymbol{p},t)\mathrm{d}\boldsymbol{p}+\boldsymbol{\nabla}_{\boldsymbol{x}}\cdot \boldsymbol{J}=0
\end{align}
with flux term
\begin{align}
\boldsymbol{J}=\int_\Omega((\boldsymbol{u}+V_s\boldsymbol{p})\psi-D_T\boldsymbol{\nabla}_{\boldsymbol{x}}\psi)\mathrm{d}\boldsymbol{p}.
\end{align}

To  satisfy a no-flux condition through the walls in a confined geometry corresponding to reflective boundary conditions, such as the specular or random reflection of individual cells, we can impose
\begin{align}
    \boldsymbol{J}\cdot \hat{\boldsymbol{n}}=0,
\end{align}
where $\hat{\boldsymbol{n}}$ is normal to the wall. Due to the non-penetration of the fluid at the walls, this can be simplified to 
\begin{align}
\left[\int_\Omega(V_s\boldsymbol{p}\psi-D_T\boldsymbol{\nabla}_{\boldsymbol{x}}\psi)\mathrm{d}\boldsymbol{p}\right]\cdot \hat{\boldsymbol{n}}=0.
\end{align}
Note, that no-flux does not hold for absorbing boundaries. 

\begin{figure}
\begin{subfigure}[H!]{0.7\textwidth}
     \centering        \includegraphics[width=.99\textwidth]{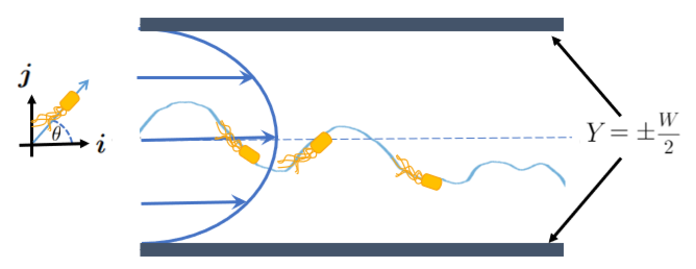}
     \caption{ \label{sketchfigure} }
\end{subfigure}
\begin{subfigure}[H!]{0.25\textwidth}
          \includegraphics[width=.99\textwidth]{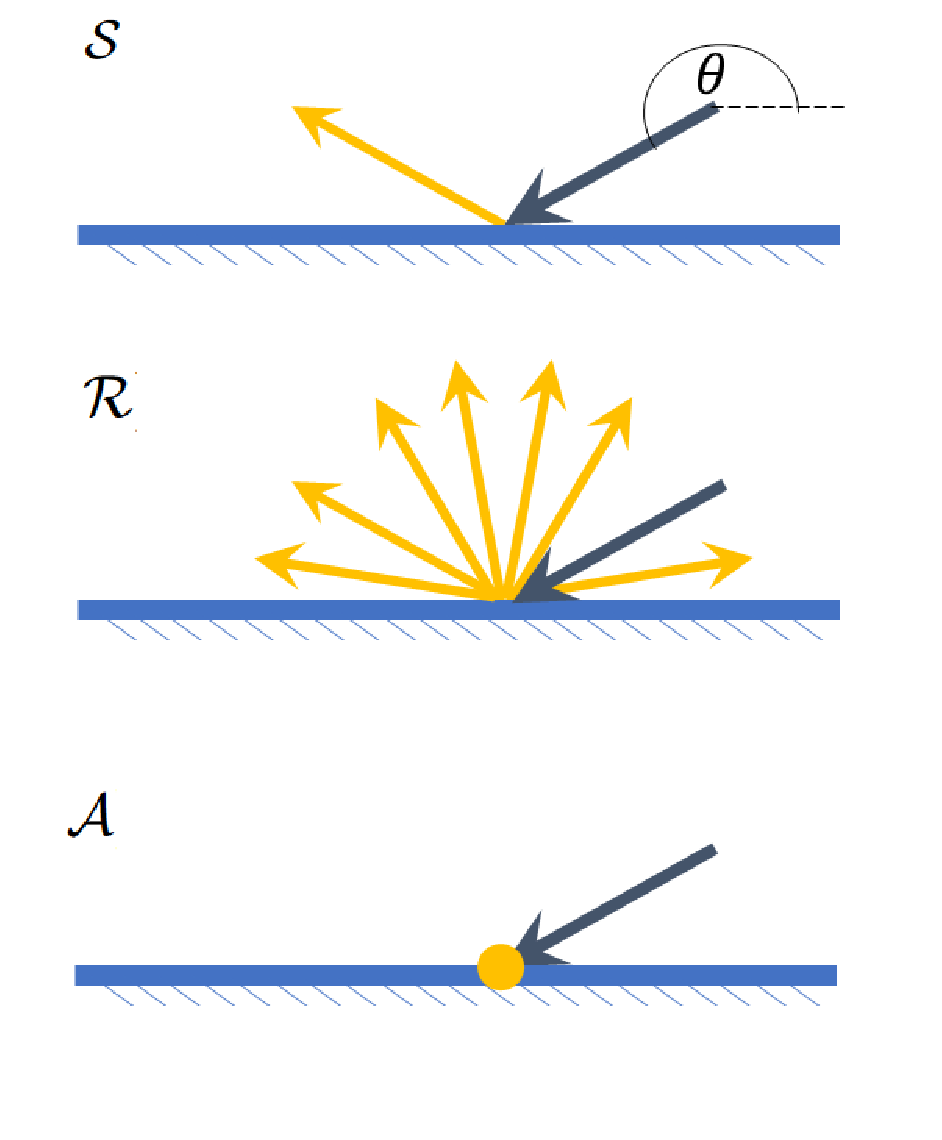}
          \caption{ \label{sketchBC} }
\end{subfigure}
\centering
\begin{subfigure}[H!]{0.4\textwidth}
    {    \includegraphics[width=.99\textwidth]{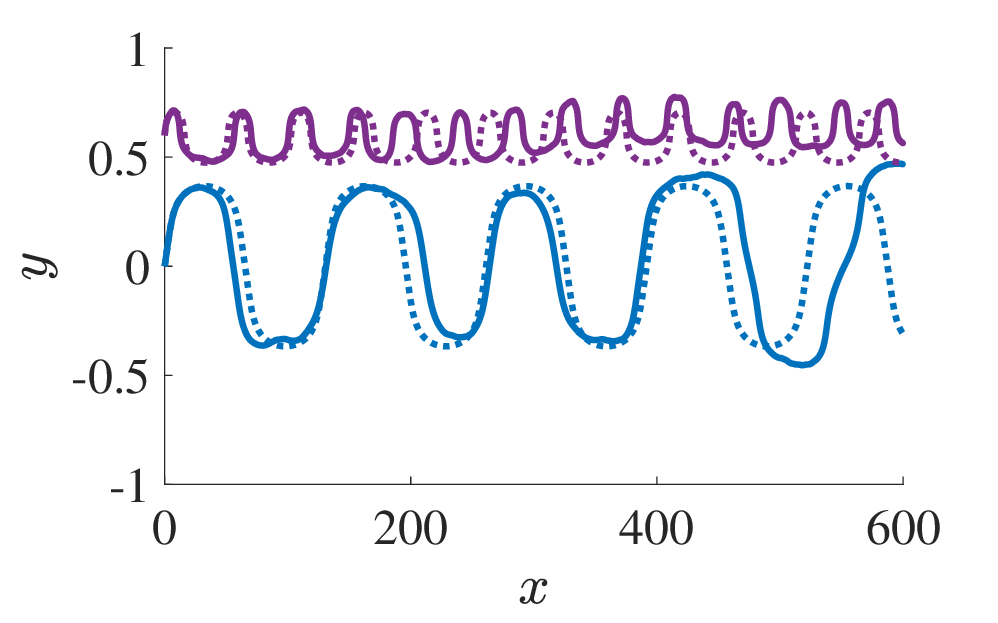}}%
          \caption{\label{ExampleTrajectories}}
\end{subfigure}
\caption{
    (a) Schematic of two-dimensional Poiseuille flow and individual swimmer trajectories. Swimmers are not drawn to scale. (b) Schematic of specular reflection $\mathcal{S}$, uniform random reflection $\mathcal{R}$, and absorbing boundary $\mathcal{A}$ effects. (c) Sample trajectories computed by the IBM model in a dimensionless channel, in the absence of translational diffusion effects, for $\beta=0.99, \nu=0.04$ and initial positions $x_0=0$, $y_0=0,0.6$. Dotted lines correspond to fully deterministic trajectories  and solid lines correspond to trajectories with rotational effects, $Pe=10^4$. }
\end{figure} 
 
 \subsection{Two-dimensional channel flow\label{sec:2d}}
 To expand upon the study of two-dimensional channel flow as motivated by experiments \citep{rusconi2014bacterial} and numerical studies \citep{bearon2015trapping,vennamneni2020shear}, let us consider a horizontal channel of height $W$ (as shown in figure \ref{sketchfigure}), such that for a coordinate system $(X,Y)$ with orthonormal base vectors $\boldsymbol{i}, \boldsymbol{j}$, the channel walls are at positions $Y=\pm W/2$. Suppose there is a parabolic flow through the channel with velocity
 \begin{equation}
     \boldsymbol{u}=U\left( 1-4\left( \frac{Y}{W}  \right)^2
     \right)\boldsymbol{i},
 \end{equation}
where $U$ is the centreline flow speed of the channel.

{We also take the cell orientation to be constrained in the two-dimensional plane, so that}  the direction of orientation $\boldsymbol{p}$ can be defined in terms of the angle $\theta$ measured from the horizontal:
\begin{align}
    \boldsymbol{p}=\cos\theta\boldsymbol{i}+\sin\theta\boldsymbol{j}.
\end{align}

{
\begin{table}
\centering
\begin{tabular}{l  l l} 
& &\\
 Channel width & $W$ & 425$\mu$m\\ 
 Centreline flow velocity&$U$&  1.25mms$^{-1}$\\
Swimming velocity&$V_s$&50--125$\mu$ms$^{-1}$\\
Rotational diffusion& $d_r$& 6$\times10^{-4}$s$^{-1}$--6s$^{-1}$\\
  Brownian translational diffusion& $D_T$&  $2\times10^{-9}\mathrm{cm}^2\mathrm{s}^{-1}$\\
Rotational P\'eclet number& $Pe=2U/Wd_r$ &1--$10^4$ \\
Translational P\'eclet number& $Pe_T=WU/2D_T$ &1--$10^6$ \\
Velocity ratio& $\nu=V_s/U$ & 0.04-0.1 \\
[1ex] 
\end{tabular}
\caption{Scaled parameter variables, based on values reported by \cite{rusconi2014bacterial,berg1993random} and used by \cite{bearon2015trapping}.}
\label{table:1}
 \label{table:2}
\end{table}
We non-dimensionalise the system with length and time scales $L=\frac{W}{2}$ and $T=\frac{W}{2U}$, respectively, such that our coordinate system can be redefined as $(x,y)=\left(\frac{2X}{W},\frac{2Y}{W}\right)$, with boundaries  located at $y=\pm1$. Taking $\psi$ to be independent of $x$, this leads to the two-dimensional conservation equation 

\begin{align}
    \parone{\psi}{t}+\nu\parone{}{y}(\sin{\theta}\,\psi)-\frac{1}{\mathrm{Pe}_T}\partwo{\psi}{y}+\parone{}{\theta}\left(y(1-\beta\cos2\theta)\psi-\frac{1}{\mathrm{Pe}}\parone{\psi}{\theta} \right)=0 \label{eq:Scale2}
\end{align}
with no-flux boundary condition
\begin{align}
  \left.    \int^{2\pi}_0\left(\nu\sin\theta\psi-\frac{1}{\mathrm{Pe}_T}\parone{\psi}{y}\right)\mathrm{d}\theta\right|_{y=\pm1}=0.
\end{align}
Here, $\nu=\frac{V_s}{U}$ is the ratio of the swimming speed to the centreline velocity, $Pe=\frac{2U}{Wd_r}$ is the  rotational P\'eclet number, and $Pe_T=\frac{WU}{2D_T}$ is the translational P\'eclet number. The parameters used are given in table \ref{table:1}. 
{We also introduce the cell concentration distribution
\begin{align}
 n(y,t)=\int_0^{2\pi}\psi(y,\theta,t)d\theta.  
\end{align}
For the steady state problem, with $\psi$ independent of time,  the time-independent cell concentration distribution is
\begin{align}
 n(y)=\int_0^{2\pi}\psi(y,\theta)d\theta.  
\end{align}
}

\subsection{Numerical methods}
\subsubsection{Stochastic differential equations\label{Sec:IBMmethod}}

The conservation equations in section \ref{sec:2d} can be transformed to an individual-based stochastic model, 
{as there exists an established complete equivalence between forward} Fokker-Planck equations and 
{diffusion processes }
{with a drift coefficient $\boldsymbol{\mu}(\boldsymbol{X}_t,t)$  and diffusion coefficient $\boldsymbol{D}(\boldsymbol{X}_t,t)$} \citep{gardiner2009handbook}. 
{Hence,} Fokker-Planck equations of the form
    \begin{equation}
        \parone{\psi}{t}(\boldsymbol{x},t)=-\sum^n_{i=1}\parone{}{x_i}\bigg[\mu_i(\boldsymbol{x},t)\psi(\boldsymbol{x},t)\bigg]+\sum^n_{i,j=1}\df{\partial^2}{\partial x_i\partial x_j}\bigg[D_{ij}(\boldsymbol{x},t)\psi(\boldsymbol{x},t)\bigg] \label{eq:FP}
    \end{equation}
    have an equivalency to It\^o SDEs of the form
\begin{equation}
    \mathrm{d}\boldsymbol{X}_t=\boldsymbol{\mu}(\boldsymbol{X}_t,t)\mathrm{d}t+\boldsymbol{\sigma}(\boldsymbol{X}_t,t)\mathrm{d}\boldsymbol{W}_t,
\end{equation}
where $\boldsymbol{X}_t=(y(t), \theta(t))$ is the position and orientation vector, $\mathrm{d}t$ is the time step, $\mathrm{d}\boldsymbol{W}_t$ is 
{the} Wiener process,  $\boldsymbol{\mu}(\boldsymbol{X}_t,t)$ is a drift term, and the diffusion effects are captured in $\boldsymbol{\sigma}(\boldsymbol{X}_t,t)$ via the  relation
\begin{equation*}
    \boldsymbol{D}(\boldsymbol{X}_t,t)=\frac{\boldsymbol{\sigma}(\boldsymbol{X}_t,t)\boldsymbol{\sigma}(\boldsymbol{X}_t,t)^T}{2}.
\end{equation*}
    As the two-dimensional channel flow equations are of the form of equation \ref{eq:FP}, this allows for transformation to It\^o SDEs 
    with drift and diffusion terms 
        
\begin{subequations}
        \begin{align}
        \boldsymbol{\mu}(y,\theta,t)&=
        \begin{pmatrix}
        \nu\sin\theta\\ y(1-\beta\cos2\theta)
\end{pmatrix}
        \label{driftTerm},\\
        \boldsymbol{\sigma}(y,\theta,t)&=\begin{pmatrix}
          \sqrt{\frac{2}{\mathrm{Pe}_T}}& 0\\
          0& \sqrt{\frac{2}{\mathrm{Pe}}}
        \end{pmatrix}.\label{diffusionTerm}
        \end{align}
        
\end{subequations}
\noindent 
Taking the limits of  $Pe_T, Pe \rightarrow \infty $ we can extract the case of a purely deterministic system without diffusion. Computationally, this is implemented by replacing the diagonal entries of the matrix by $0.$  The effect of rotational diffusion in $x$-$y$ space is illustrated in figure \ref{ExampleTrajectories}, where  the IBM is augmented with an $x$--direction advection  term (details given in appendix \ref{Appendix:cellTrajectories}).

For the SDE, we consider boundary conditions for three types of physical boundary interactions at walls $y=\pm1$: specular reflection, uniform random reflection and an absorbing boundary (see figure \ref{sketchBC}). 
{In the case of specular reflection (boundary condition $\mathcal{S}$), swimmers with angles of incidence $\theta_i$  instantaneously reorient to $\theta_r=2\pi-\theta_i$
such that $\theta_i,\theta_r\in[0,2\pi)$.
For uniform random reflection (boundary condition $\mathcal{R}$)}
\begin{equation}
     \theta_r =
  \begin{cases}
    \pi+\pi\cdot\mathit{U}(0,1)       & \quad \text{if } \theta_i\in[0,\pi] \text{ at } y=1,\\
    \pi\cdot\mathit{U}(0,1)  & \quad \text{if } \theta_i\in[\pi,2\pi] \text{ at } y=-1,
  \end{cases}
  \label{eq:randomReflection}
\end{equation}
where $\mathit{U}(0,1)$ is a uniformly distributed random number in the interval (0,1). 
{Meanwhile, for a perfectly absorbing boundary (boundary condition $\mathcal{A}$) trajectories terminate upon impact with a wall. }
To calculate the probability distribution $\psi$ from the stochastic IBM  
in bounded domains we run simulations for $10^6$ stochastic swimmers which are  uniformly initialized  over the domain $(\theta,y)\in[0,2\pi)\times[-1,1]$  with sampling step size $\mathrm{d}t=0.1$ with 20 sub-intervals each (which are then calculated to approximate the continuous process better), and normalization condition $\int_0^{2\pi}\int_{-1}^1\psi(\theta,y)\mathrm{d}y \mathrm{d}\theta=4\pi$ for ease of comparison of $\theta$ distributions at the wall (\S \ref{Sec:Specular}). Time step convergence is checked by comparing the concentration distributions obtained with sampling step size $\mathrm{d}t=0.025$.
 For any simulation where we want  distributions at runtime $T_\mathrm{sim}$ the probability distribution is calculated from trajectory end-states. The runtime for endstate convergence (i.e. when doubling the run time does not change the macroscopic properties of the probability distribution such as local cell densities) is shape dependent, and adjusted accordingly. 



For comparison to the specular reflection problem and verification against continuum models we also develop a  double Poiseuille flow model. For this, we  extend the flow domain to allow for a doubly periodic flow profile as shown in figure \ref{DoublePoiseuille}c.
The background fluid flow, $\boldsymbol{u}$,
for domain $y\in[-1,3]$, becomes 
\begin{align}
\boldsymbol{u}=
\begin{cases}
-(1-(y-2)^2){\boldsymbol{i}}&\textrm{ for }y>1\\
    (1-y^2){\boldsymbol{i}}&\textrm{ for }y<1\\
\end{cases} \label{eqn:DP}
\end{align}
 such that the background flow for $y\in[-1,1]$ is identical to the background flow for a  simple Poiseuille flow in the channel. 
 
We implement periodic boundary conditions $\mathcal{DP}$
 \begin{subequations}
\begin{align}
   \psi(\theta,-3)&=\psi(\theta, 1),
   \\
    \psi(0,y)&= \psi(2\pi,y),
\end{align} \label{DP_BC}
\end{subequations}
 and normalisation condition $\int_0^{2\pi}\int_{-2}^2\psi(\theta,y)\mathrm{d}y\mathrm{d}\theta=8\pi$.

\subsubsection{Continuum model\label{sec:contModel}}
To solve the two-dimensional  continuum model for the probability distribution $\psi$, we use a Galerkin finite element method, as in \cite{bearon2015trapping} implemented in the C++ library \texttt{oomph-lib}  \citep{heil2006oomph}. 
We solve for the continuum solution by
 multiplying 
  equation \ref{eq:Scale2} by a $y$ and $\theta$ dependent test function $N(\theta,y)$, integrating over the domain, and integrating by parts, to obtain the weak form
\begin{align}
    &\int_0^{2\pi}\int_{-1}^{1}\parone{\psi}{t}N- \left[ \nu\sin\theta\psi-\frac{1}{Pe_T}\parone{\psi}{y}\right] \parone{N}{y}-\left[y(1-\beta\cos2\theta)\psi-\frac{1}{Pe}\parone{\psi}{\theta} \right]\parone{N}{\theta} \mathrm{d}y \mathrm{d}\theta \label{ScaleIIWeak}
    \\&+\int_0^{2\pi}\left[\left(\nu \sin\theta\psi-\frac{1}{Pe_T} \parone{\psi}{y}\right)N\right]_{-1}^1  \mathrm{d}\theta
   +\int_{-1}^1\left[\left( y(1-\beta\cos2\theta)\psi-\frac{1}{Pe}\parone{\psi}{\theta} \right)N
    \right]_0^{2\pi}\mathrm{d}y
    =0. \nonumber
\end{align}


The equations are discretized using finite elements on a grid $n_\theta \times n_y$, 
with $n_\theta$ and $n_y$ varying dependent on the boundary condition type and P\'eclet number of interest. Across all models, simple periodic boundary conditions are applied in the $\theta$--direction to ensure the angles of orientation wrap around, i.e. $\psi(0,y)=\psi(2\pi,y)=0$ for all $y$.
Three different boundary conditions will be applied for the continuum problem: a doubly periodic Poiseuille flow model $\mathcal{DP}$, a pinned non-zero Dirichlet constraint $\mathcal{D}_C$, and a pinned zero Dirichlet constraint $\mathcal{D}_0$.  

A Dirichlet constraint ($\mathcal{D}_C$) will be imposed for a wall-bounded domain ($\theta \times y)\in[0,2\pi)\times [-1, 1]$ such that $\psi(\theta,\pm1)=C_0$  for all $\theta$. The 
 values of the constant emerges upon the enforcement of the normalisation condition $\int_0^{2\pi}\int_{-1}^1\psi(\theta,y)\mathrm{d}y\mathrm{d}\theta=4\pi$. 
When solving the steady-state equilibrium problem (i.e. the first term in equation \ref{ScaleIIWeak} is set to zero), the elements in the $\theta$--direction are uniformly distributed and the elements in the $y$--direction are non-uniform to allow for higher resolutions near the wall. 
A piece-wise linear scaling is implemented to restrict half the elements to $|y|\geq 0.99$.

For the double Poiseuille flow, we adapt the finite element model by extending the flow domain to allow for a doubly periodic flow profile as shown in figure \ref{DoublePoiseuille}c and shown in equations \ref{eqn:DP}.
  The double Poiseuille flow profile is $\mathcal{C}^0$  continuous in shear and $\mathcal{C}^1$ continuous in velocity at $y=\pm1$. 
 While the extended flow velocity profile introduces a discontinuity in the second derivative of the flow velocity in $y$ about $y=1$, this does not lead to any difficulties with 
 the finite element discretisation, as the implementation only requires continuity of the first derivative.   
  For the steady-state, doubly periodic Poiseuille flow model  we implement periodic boundary conditions as given in equations \ref{DP_BC}, with  normalisation condition $\int_0^{2\pi}\int_{-2}^2\psi(\theta,y)\mathrm{d}y\mathrm{d}\theta=8\pi$.
The elements in the $y$ and $\theta$--directions are uniformly distributed. 
 
 Finally, we have a time evolving double Poiseuille flow (see equation  \ref{eqn:DP}) with  zero Dirichlet constraint ($\mathcal{D}_0$), which will be imposed on the boundary  such that $\psi(\theta,-1)=\psi(\theta,3)=0$  for all $\theta$.} The unsteady system  is evolved using a second-order backward difference (BDF2) time stepper  \citep{ascher1998computer} with time step $\mathrm{{d}t=10^{-4}}$. 
 The elements in the $\theta$--directions are uniformly distributed, and the elements in the $y$--direction are non-uniformly distributed via a tanh function over $y\in[-1,3]$, such that half the elements are restricted to $y<-0.93$ and $y>2.93$, allowing for higher resolutions near the zero boundaries. The initial condition $\psi_0$ at $t=0$ is uniform with a small $\tanh$ correction to match the boundary conditions,
 and satisfies $\int_0^{2\pi}\int_{-1}^3\psi_0(\theta,y)\mathrm{d}y\mathrm{d}\theta=8\pi$.
\section{Results}

First, we study different physical boundary-interaction scenarios and determine  appropriate continuum boundary conditions for capturing their dynamics. 
 We consider three types of wall interactions: specular reflection $\mathcal{S}$ (\S \ref{Sec:Specular}), uniform random reflection $\mathcal{R}$ (\S \ref{sec:RandomReflection}), and absorbing boundary $\mathcal{A}$ (\S \ref{sec:PerfAbsorption}). For each case we find a corresponding continuum model. 
 Then, we consider different bulk flow and particle properties and analyse how they affect  cells approaching boundaries.

\subsection{Specular Reflection \label{Sec:Specular} \label{sec:DPPoiFlow}}


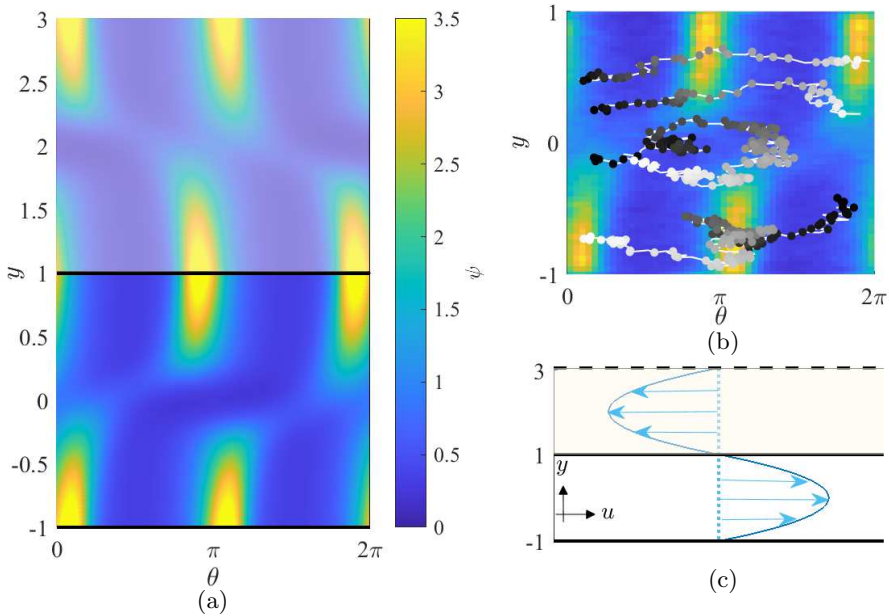
\begin{figure}
    \centering
\begin{tikzpicture}
\node[above right] (img) at (0,0) {\includegraphics[width=.99\textwidth]{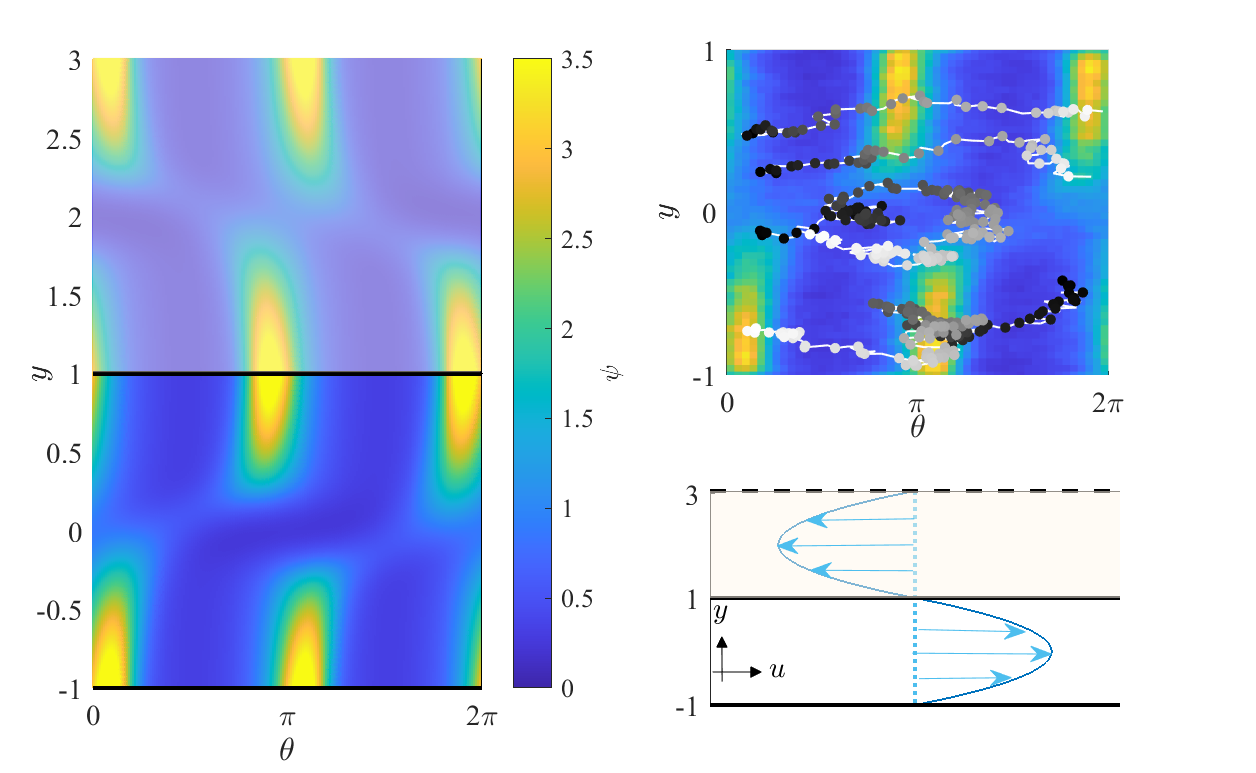}
};
\node at (90pt,1pt) {(a)};
\node at (282pt,98pt) {(b)};
\node at (282pt,9pt) {(c)};
\end{tikzpicture}    
\caption{ \label{DoublePoiseuille} \label{FullIBM}
A comparison of the bulk dynamics in a continuum double Poiseuille model,  to a stochastic  simulation 
with wall-bounded specular reflection 
for $Pe=10$, $\beta=0.99$, $\nu=0.04,$ and $Pe_T=10^6$.  (a) Finite element continuum simulation for ($n_\theta= 100, n_y=500$) double Poiseuille bivariate $\psi$ distribution for flow with periodic boundaries $\mathcal{DP}$; (b) IBM stochastic bivariate $\psi$ distribution for single Poiseuille flow with specular reflection $\mathcal{S}$ at $y=\pm1$.   Example cell trajectories of cells swimming in sheared flow are overlaid in $\theta$-$y$ phase space (white lines), with snapshots in time given by dots along each trajectory (black to white in time); (c) Flow profile for double Poiseuille flow in (a). In (a)\& (b) the colourmap (blue to yellow) indicates the probability distribution of cells in the phase space. 
}
\end{figure} 
{In this section we investigate the IBM with specular reflection $\mathcal{S}$ 
and consider how it compares to a continuum doubly periodic Poiseuille flow $\mathcal{DP}$.
In the literature, double Poiseuille flows have been used for studying low and high shear trapping in the bulk flow of bacterial suspensions  to circumvent the problem of explicitly implementing a boundary \citep{vennamneni2020shear}, but there has been no comparison between the wall dynamics of specular reflection IBMs and continuum doubly periodic Poiseuille models. 

Consider the case of a bounded, stochastic IBM 
 {with boundary condition $\mathcal{S}$} (figure \ref{DoublePoiseuille}b) with $Pe=10$, $Pe_T=10^6$, $\nu=0.04$ and $\beta=0.99$, quantifying rotational diffusion, translational diffusion, velocity ratios, and cell shape, respectively.
The lowest trajectory in figure \ref{FullIBM} highlights a cell trajectory which starts oriented downstream near the bottom wall ($\theta=2\pi)$, reorients rapidly until it is aligned upstream ($\theta=\pi)$, and  enters a region of accumulation (the yellow region) closer to the bottom wall. 
Once there, the cell moves up and down in the channel due to  translational diffusion effects with orientation approximately parallel to the flow direction. If no longer pointed parallel to the flow direction due to rotational diffusion and shear effects, it reorients rapidly again. The extended alignment with the flow direction is a result of the shape-dependent Jeffery orbits, in which local sheared flows cause particle rotation and straining. 
In addition to reorientation due to Jeffery orbits,  there exists some  variation in orientation with the clumping and widening of cell 
    {positions} measured at equal time separations. This results from rotational diffusion counteracting and enhancing reorientations due to Jeffery orbits, respectively. 

The macroscopic areas of accumulation in phase space 
are dependent on both the shape and the motility of the swimmers. The thickness and position of the areas of accumulation are dependent on the balance between deterministic shape dependent effects and diffusion effects. In figures \ref{compareSpecBivariate}a-c we consider probability distribution functions of cell distributions for $\beta=0.99$, $\nu=0.04$ and $Pe_T=10^6.$  For high diffusion (see figure \ref{compareSpecBivariate}a), there exist two regions of accumulation and two regions of depletion of equal widths at each wall, resulting from strong, continuous mixing of cells. With increasing $Pe$ (see \ref{compareSpecBivariate}b-c), the relative diffusive effects decrease and deterministic effects begin to dominate. Due to the nature of Jeffery orbits (which have been described earlier), more cells will be parallel to the flow direction. The reduced diffusion ensures decreased spreading in orientation space, thereby leading to thinner areas of accumulation. For very weak rotational diffusion $Pe=10^4$ (see figure \ref{compareSpecBivariate}c) the areas of accumulation become very thin and cells swim away from the walls leading to peaks of accumulation at $(\theta,y)=(\pi, \pm0.5)$, in agreement with observations by \cite{rusconi2014bacterial} and \cite{zottl2013periodic}.


Next we consider the continuum doubly periodic Poiseuille flow, that serves as a potential alternative for capturing the bulk flow in the bounded domain. Consider the finite element model with a doubly periodic flow profile as shown in figure \ref{DoublePoiseuille}c. 
 Comparing the lower subdomain for the finite element double Poiseuille model $\theta\in[0,2\pi], y\in[-1,1]$ in figure \ref{DoublePoiseuille}a to the bivariate stochastic IBM distribution (figure \ref{DoublePoiseuille}b)
 we find similar bulk-flow dynamics with regions of cell accumulation above $y=-1$, at angles slightly greater than $\theta=0,\pi$. Meanwhile, just below $y=1$, near the `upper wall', there also exist two areas of accumulation of equal intensity, but of flipped geometry, for angles just below $\theta=2\pi$ and $\theta=\pi$. In both cases, these areas of accumulation correspond to swimmers oriented close to the horizontal, but pointing into the wall and out of the wall, respectively. When comparing the probability density distributions of the doubly periodic Poiseuille continuum model  (figures \ref{compareSpecBivariate}d--f) to the IBM with specular reflection (figures \ref{compareSpecBivariate}a-c) for increasing $Pe$, we observe the same sharpening of accumulation areas, with the occurrence of localised peaks in both figures \ref{compareSpecBivariate}c and \ref{compareSpecBivariate}f. The thickness of the areas of accumulation agree as well as the intensity of cell accumulations. 
 
\begin{figure}
    \centering
    
    \begin{tikzpicture}
\node[above right] (img) at (0,0) {
\begin{subfigure}[H!]{0.32\textwidth}
        \centering
        \includegraphics[width=.95\textwidth]{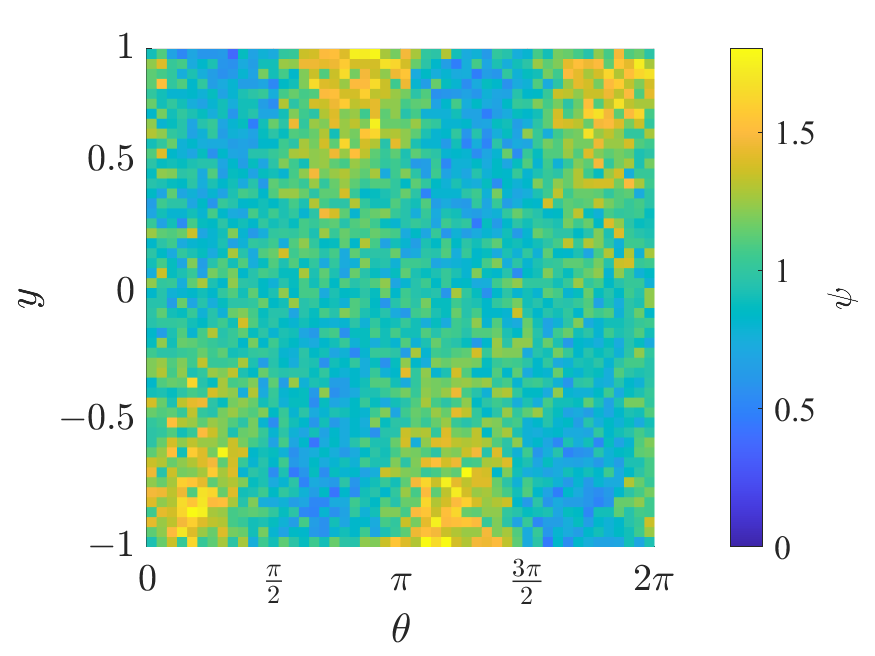}
        \caption{
        \label{SpecPe1}}
    \end{subfigure}
        \begin{subfigure}[H!]{0.32\textwidth}
         \centering        \includegraphics[width=.95\textwidth]{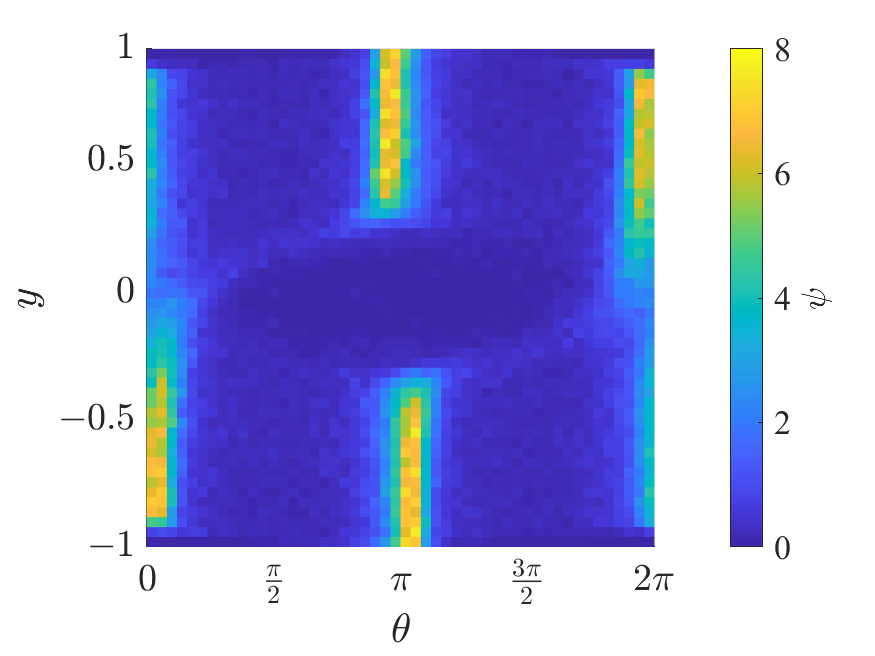}
                \caption{
              \label{SpecPe100}
}
    \end{subfigure}
            \begin{subfigure}[H!]{0.32\textwidth}
         \centering        \includegraphics[width=.95\textwidth]{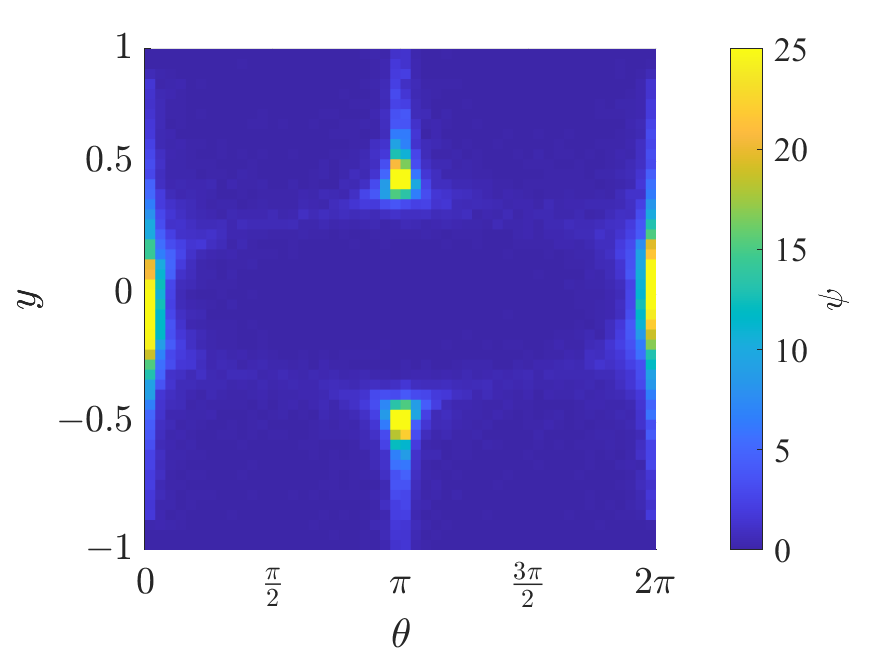}
                \caption{
              \label{SpecPe10000}
}
    \end{subfigure}
};
\node[rotate=90] at (1pt,70pt) {IBM};
\end{tikzpicture} 
 \begin{tikzpicture}
\node[above right] (img) at (0,0) {
\begin{subfigure}[H!]{0.32\textwidth}
        \centering
        \includegraphics[width=.95\textwidth]{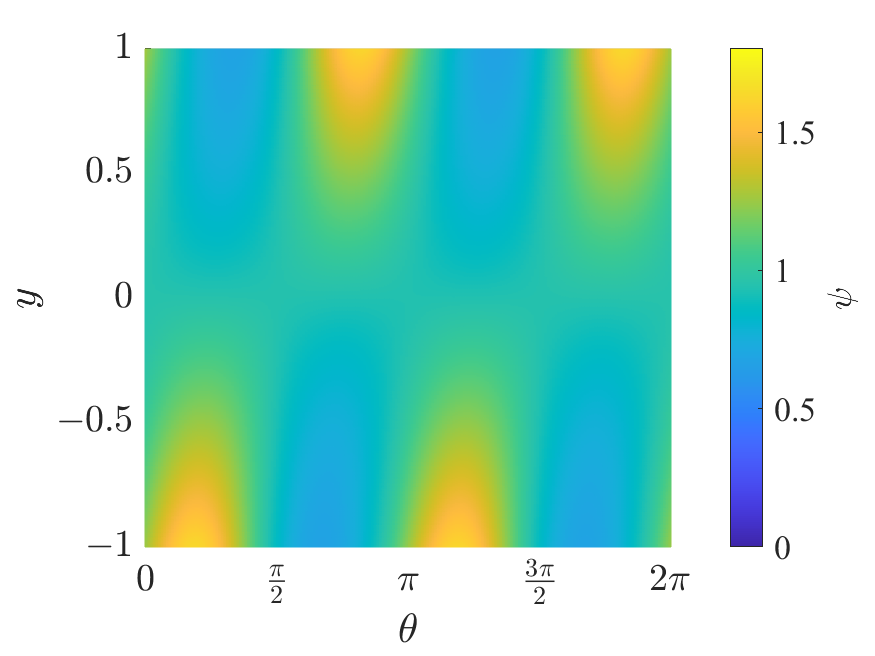}
        \caption{
        \label{SingleDPPe1}}
    \end{subfigure}
        \begin{subfigure}[H!]{0.32\textwidth}
         \centering        \includegraphics[width=.95\textwidth]{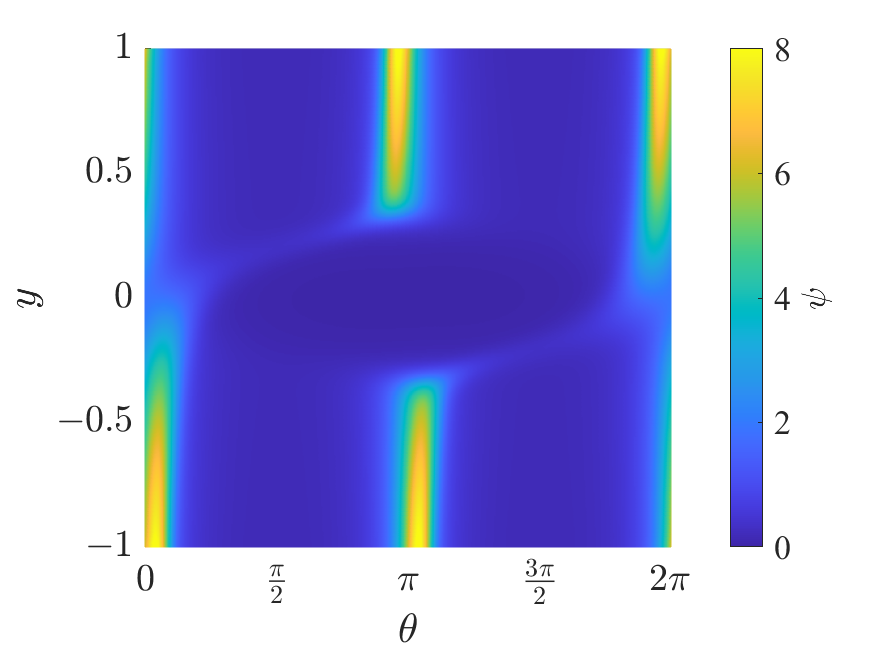}
                \caption{
              \label{SingleDPPe100}
}
    \end{subfigure}
            \begin{subfigure}[H!]{0.32\textwidth}
         \centering        \includegraphics[width=.95\textwidth]{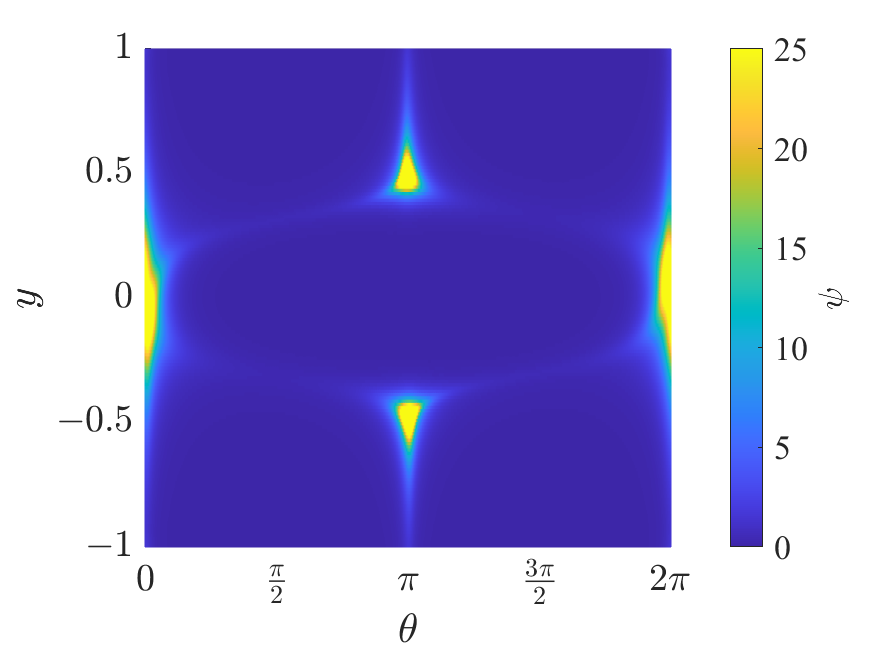}
                \caption{
              \label{SingleDPPe10000}
}
    \end{subfigure}
};
\node[rotate=90] at (1pt,70pt) {Continuum};
\end{tikzpicture} 
\caption{ \label{compareSpecBivariate} 
Comparison of snapshots of bivariate probability density distributions $\psi$, as obtained for converged IBMs with specular wall reflections $\mathcal{S}$ (a)--(c), to equilibrium probability density distributions for doubly periodic continuum models $\mathcal{DP}$ (d)--(f). $\beta=0.99$, $\nu=0.04$ and $Pe_T=10^6$.
 (a)\&(d): $Pe=1$, (b)\&(e): $Pe=100$; and (c)\&(f): $Pe=10^4$.
}
\end{figure} 





\begin{figure}
    \centering
                    \begin{subfigure}[H!]{0.32\textwidth}
         \centering        \includegraphics[width=.99\textwidth]{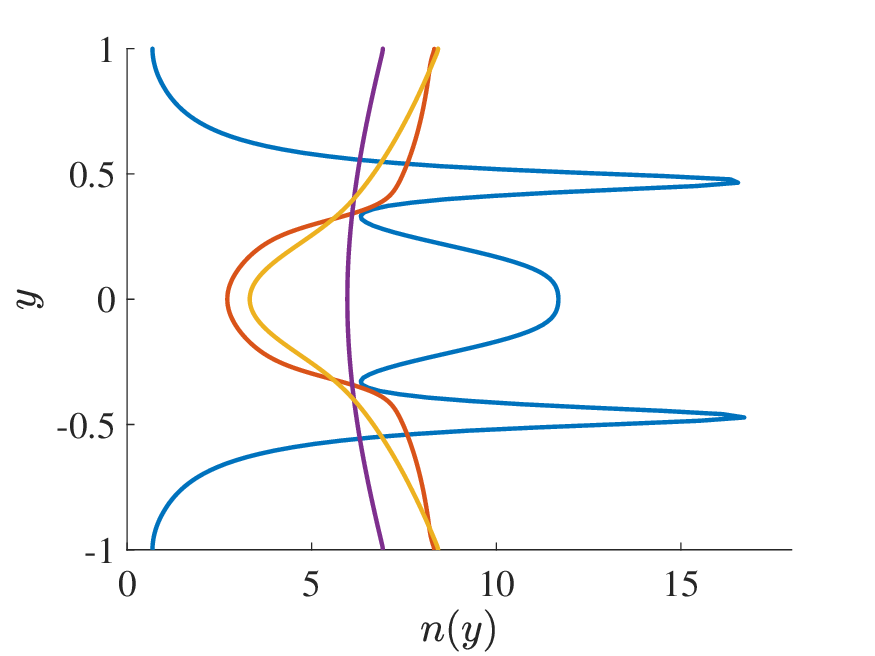}
                \caption{
              \label{FE_DPbeta0.99}
}
    \end{subfigure}
                \begin{subfigure}[H!]{0.32\textwidth}
         \centering        \includegraphics[width=.99\textwidth]{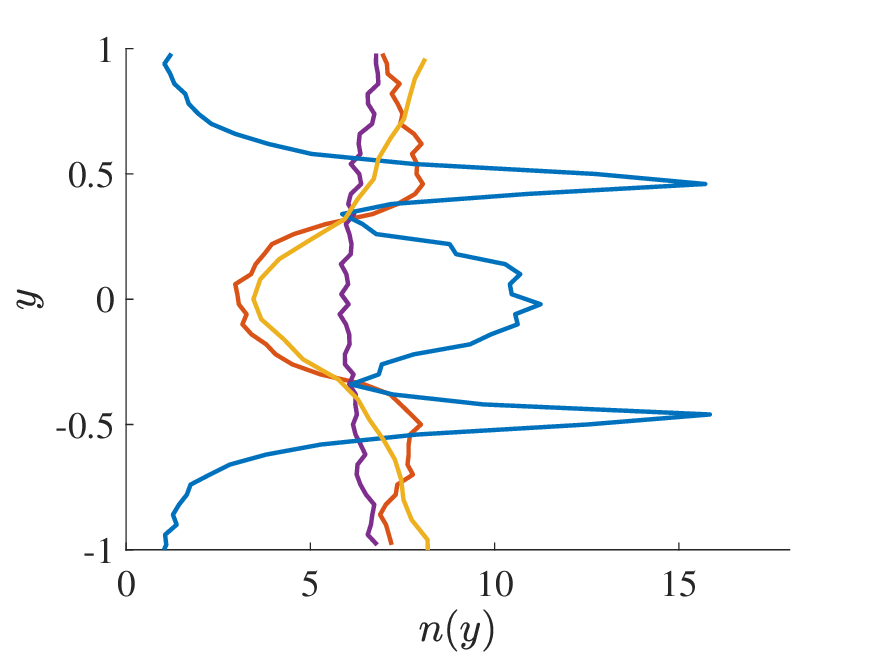}
                \caption{
              \label{IBM_DPbeta0.99}
}
    \end{subfigure}
\begin{subfigure}[H!]{0.32\textwidth}
         \centering        \includegraphics[width=.99\textwidth]{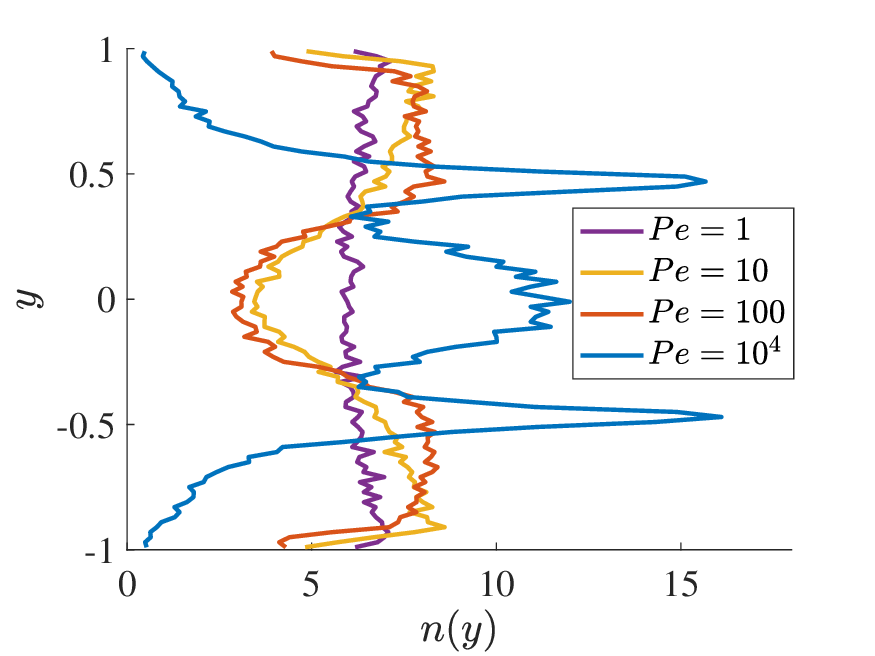}
                \caption{
              \label{IBM_Wallbeta0.99}
}
    \end{subfigure}

\caption{
Cell number density distributions for  (a): the continuum model distribution with doubly periodic Poiseuille; (b): IBM with doubly periodic Poiseuille flow; and (c): the IBM distribution with specular reflection boundary condition. For shape parameters $\beta=0.99$ 
with $Pe=10^4$ (blue),  $Pe=100$ (red), $Pe=10$ (yellow) and $Pe=1$ (purple). 
\label{CellDensityWallBounded}
}

\end{figure} 
\begin{figure}
    \centering
\begin{subfigure}[H!]{0.32\textwidth}
        \centering
        \includegraphics[width=.99\textwidth]{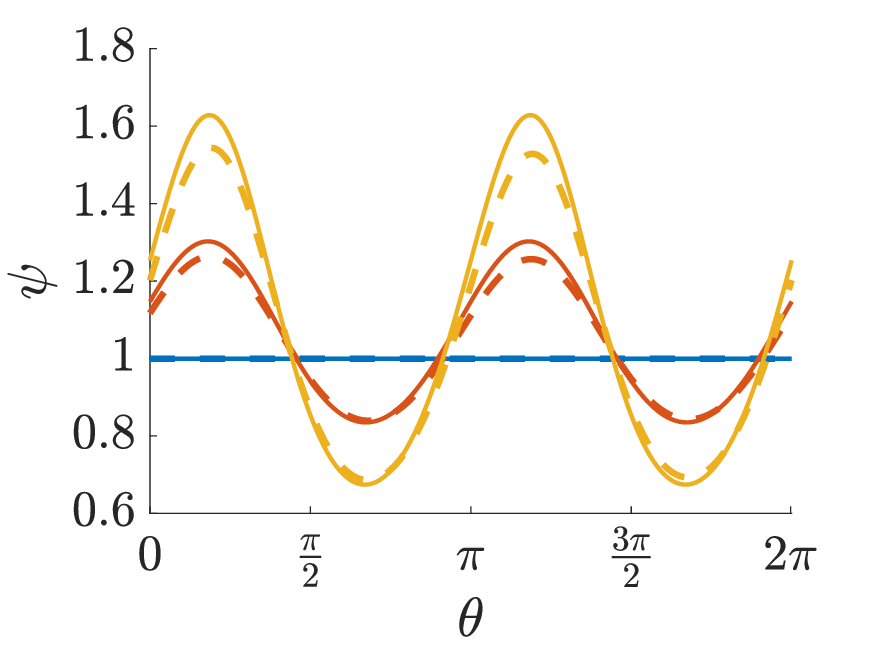}
        \caption{
              \label{DoublePWallPe1}
}
    \end{subfigure}
    \begin{subfigure}[H!]{0.32\textwidth}
        \centering
        \includegraphics[width=.99\textwidth]{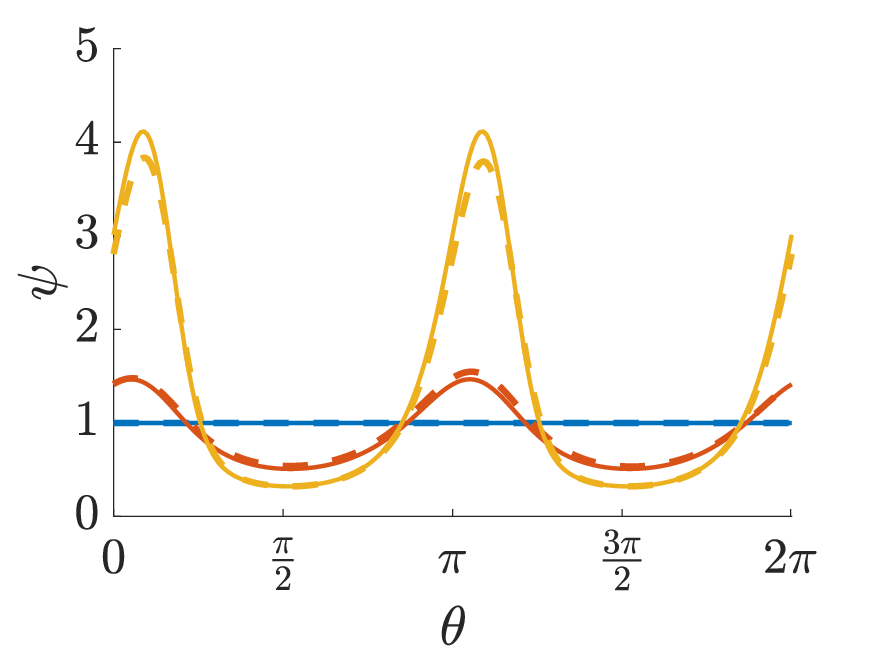}
                        \caption{    
              \label{DoublePWallPe10}
}
    \end{subfigure}
        \begin{subfigure}[H!]{0.32\textwidth}
         \centering        \includegraphics[width=.99\textwidth]{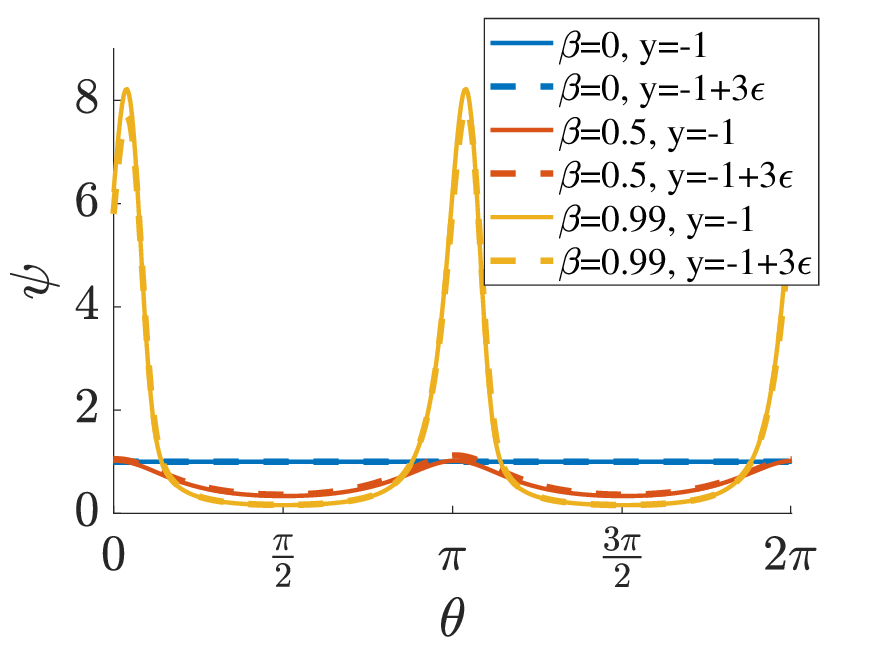}
                \caption{
              \label{DoublePWallPe100}
}
    \end{subfigure}
                \begin{subfigure}[H!]{0.32\textwidth}
         \centering        \includegraphics[width=.99\textwidth]{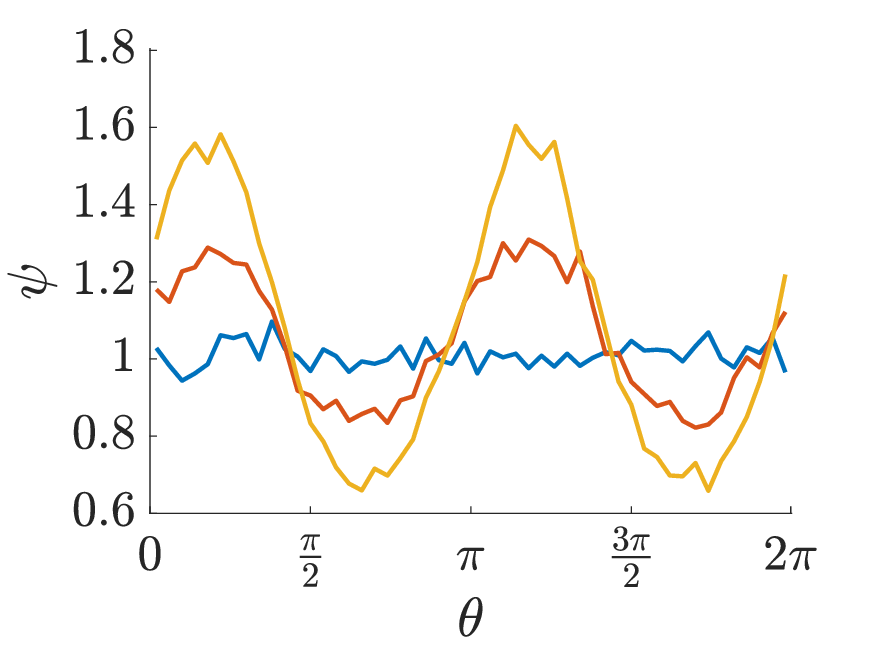}
                \caption{
              \label{IBM_DPWallPe1}
}
    \end{subfigure}
      \begin{subfigure}[H!]{0.32\textwidth}
         \centering        \includegraphics[width=.99\textwidth]{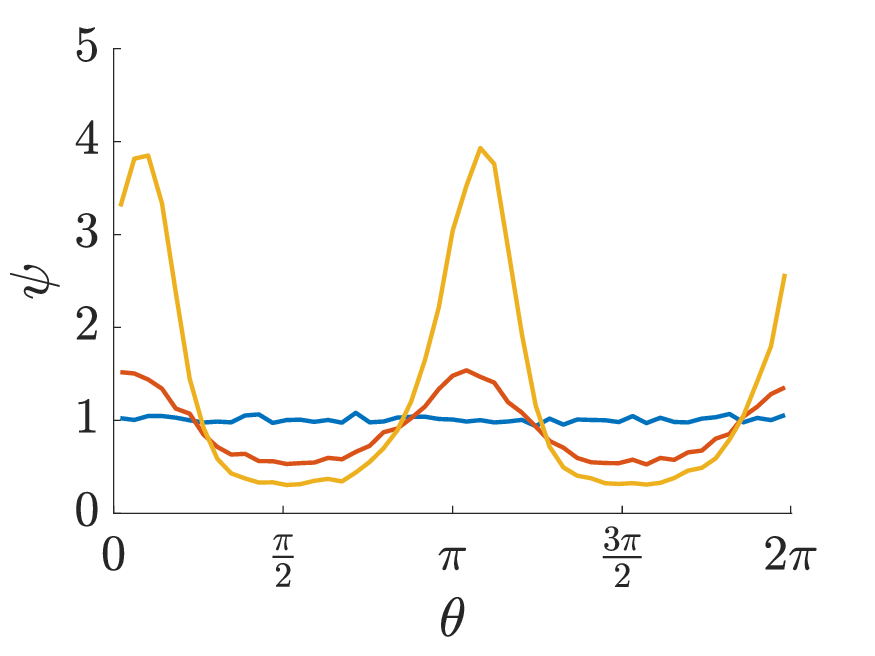}
                \caption{
              \label{IBM_DPWallPe10}
}
    \end{subfigure}            
                \begin{subfigure}[H!]{0.32\textwidth}
         \centering        \includegraphics[width=.99\textwidth]{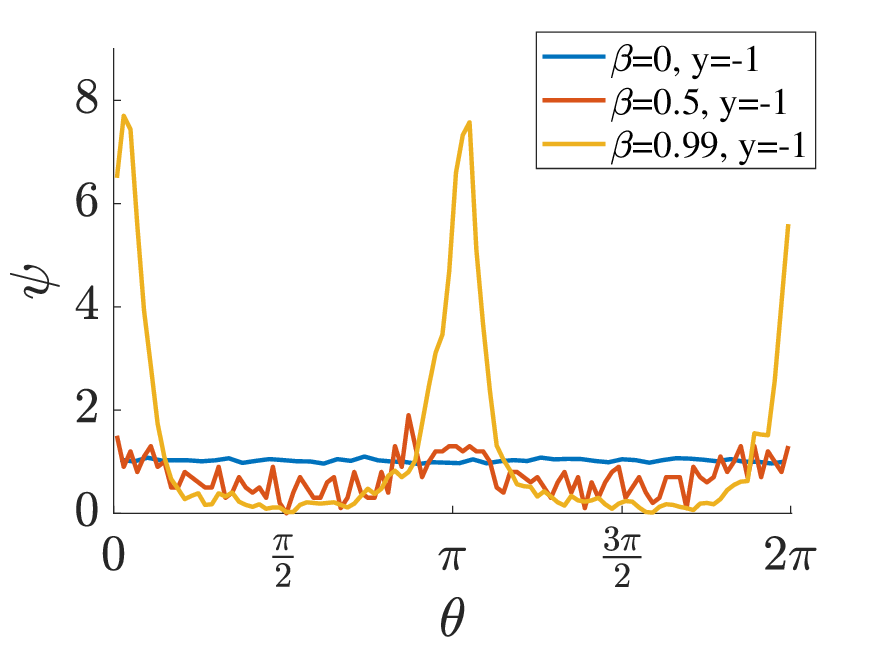}
                \caption{
              \label{IBM_DPWallPe100}
}
    \end{subfigure}
                    \begin{subfigure}[H!]{0.32\textwidth}
         \centering        \includegraphics[width=.99\textwidth]{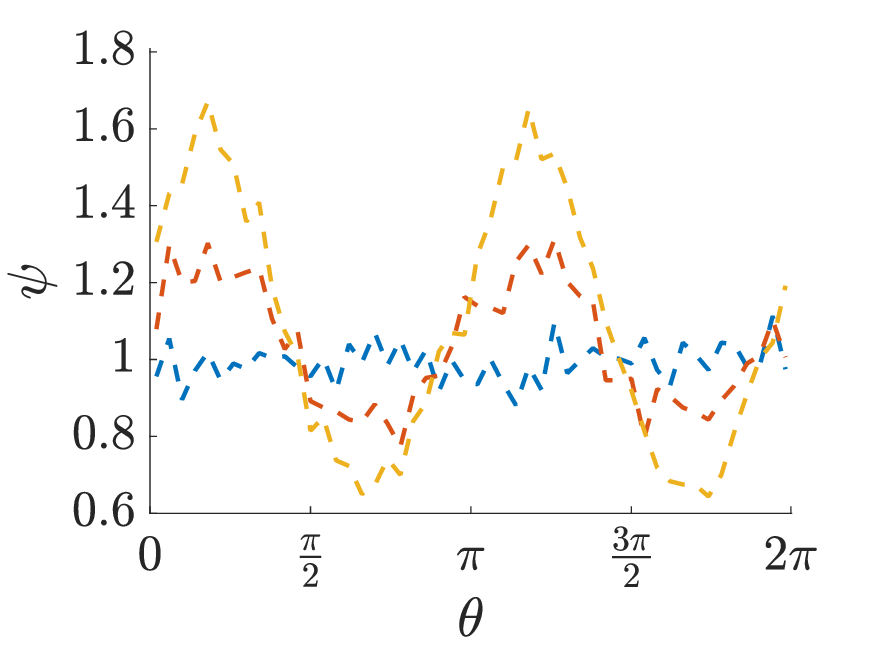}
                \caption{
              \label{IBMWallPe1}
}
    \end{subfigure}
                \begin{subfigure}[H!]{0.32\textwidth}
         \centering        \includegraphics[width=.99\textwidth]{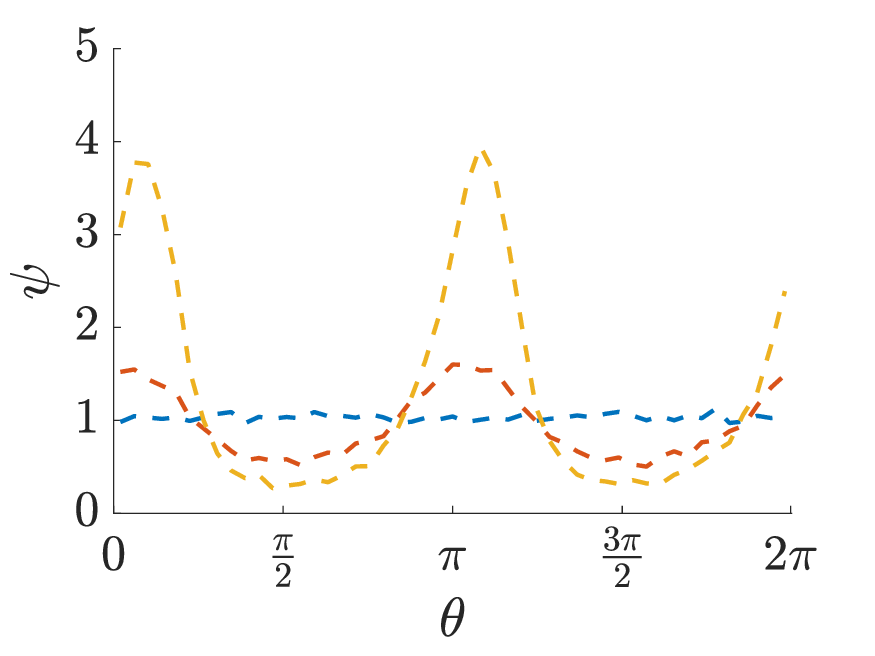}
                \caption{
              \label{IBMWallPe10}
}
    \end{subfigure}
                \begin{subfigure}[H!]{0.32\textwidth}
         \centering        \includegraphics[width=.99\textwidth]{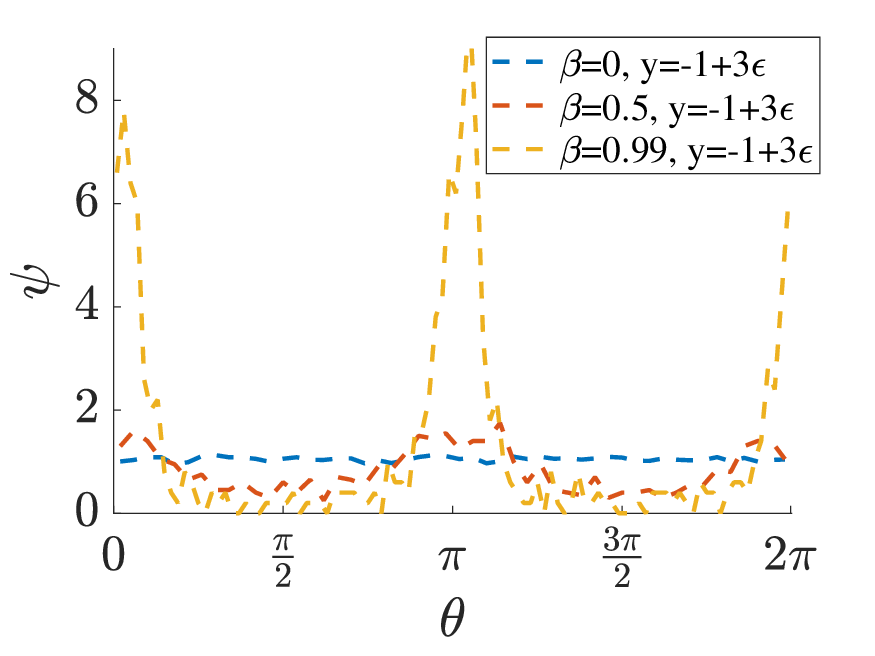}
                \caption{
              \label{IBMWallPe100}
}
    \end{subfigure}
\caption{ \label{WallDistr} 
Comparing the distributions at the wall for varying $Pe$ and $\beta$, between the doubly periodic continuum model and the wall-bounded specular reflection IBM for $\nu=0.04$ and $\beta=0, 0.5, 0.99$. The probability distributions $\psi$ at  $y=-1$ (solid lines) and $y=-1+3\epsilon$ (dashed lines) for the double Poiseuille continuum model, for (a) $Pe=1$ ($n_\theta=100$, $n_y=500$); (b) $Pe=10$ ($n_\theta=200$, $n_y=500$); and (c) $Pe=100$ ($n_\theta=400$, $n_y=200$).
The probability distributions $\psi$  at $y=-1$ for the doubly periodic Poiseuille IBM, for (d) $Pe=1$; (e) $Pe=10$; and (f) $Pe=100$.
The probability distributions $\psi$ near the bottom wall $y=-1+3\epsilon$ for the wall-bounded IBM with specular reflection, for (g) $Pe=1$; (h) $Pe=10$; and (i) $Pe=100$.
}
\end{figure}

Next, we compare the cell concentration distributions of swimmers, $n(y)$, across the channel height $y\in[-1,1]$ for $\beta=0.99$,
for different values of rotational diffusion ($Pe=1, 10, 100,10^4$). Direct comparisons between the doubly periodic continuum model (figure \ref{FE_DPbeta0.99}), a doubly periodic IBM (figure \ref{IBM_DPbeta0.99}), and the wall-bounded IBM with specular reflection $\mathcal{S}$ (figure \ref{IBM_Wallbeta0.99})  show clear agreement in the cell concentrations and accumulations. Deviations between the models are only notable very close to the walls for medium $Pe$ where the IBM with specular reflection  exhibits some cell depletion.  The observed depletion is observed consistently for medium $Pe$, and is a result of specular reflection in the IBM. 
Comparing to the bivariate distribution in figures figure \ref{FullIBM}b and \ref{SpecPe100}, we note that there is a cell depletion around $\theta=0$ for $y=-1$ and $\theta=\pi$ at $y=1$.
A contributing factor to this cell depletion is finite time stepping. Due to the finite nature of time steps in the SDE problem, cell trajectories can drift, leading to reduced cells around $\theta=0$ for $y=-1$ and $\theta=\pi$ at $y=1$. This produces less accurate local distributions as time evolves
(see Appendix \ref{Appendix:Depletion} for details of cell trajectory drift in deterministic problems and at medium P\'eclet numbers).
{ For high $Pe$ the migration of cells towards the channel centre due to low shear trapping \citep{vennamneni2020shear} results in low cell concentrations at the wall. The effects of drifting due to finite time step sizes are small because there are low cell concentrations at the wall.
Meanwhile, for low $Pe$, the high rotational diffusion results in model-dependent depletion being largely counteracted. 
Nevertheless, the observed structures and positions of cell distributions obtained across all three models are in good agreements across the studied range of rotational diffusions, capturing centreline cell depletion measured for medium to high rotational effects (low to medium P\'eclet numbers) as observed in experiments (\cite{rusconi2014bacterial}) as well as numerical and analytical studies (\cite{bearon2015trapping,vennamneni2020shear}). 
The strong agreement in the bulk flow   suggests that the doubly periodic Poiseuille continuum model might be a sensible modification for capturing the cell distributions of elongated swimmers undergoing specular reflection at the walls especially at high and low diffusion effects.}

While the cell concentration distribution $n(y)$ tells us about the agreement in the relationship between the three models in terms of cell accumulation for $\beta=0.99$, it does not allow for any insight into the orientations of the swimmers at or near the walls. In figure \ref{WallDistr} we compare the probability density distributions $\psi(\theta,y)$ for the doubly periodic Poiseuille flow continuum model (figures \ref{DoublePWallPe1}--\ref{DoublePWallPe100}), the IBM double periodic Poiseuille flow case
(figures \ref{IBM_DPWallPe1}--\ref{IBM_DPWallPe100}), and the IBM specular reflection model
(figures \ref{IBMWallPe1}--\ref{IBMWallPe100}), for P\'eclet numbers $Pe=1$ (figures \ref{WallDistr}a, \ref{WallDistr}d, \ref{WallDistr}g), $Pe=10$ (figures \ref{WallDistr}b, \ref{WallDistr}e, \ref{WallDistr}h), and $Pe=100$ (figures \ref{WallDistr}c, \ref{WallDistr}f, \ref{WallDistr}i). For direct comparison between the doubly Poiseuille models we plot the distributions at $y=-1$, given by solid lines, for shape parameter $\beta=0, 0.5, 0.99$. To provide a  comparison between the continuum model and the IBM we need to account for the numerical cell depletion. To capture near-wall cell distributions,  we plot the probability distributions near the walls just beyond the model-induced depletion area at $y_{\mathrm{near}}=-1+\epsilon$ for $\epsilon=0.12$, as dashed lines. 

We note that across the continuum models
 for spherical swimmers ($\beta=0$ given by the blue lines), the orientation distribution is constant, indicating that surface interactions in the absence of hydrodynamic wall interactions, show no preferential orientation. This uniformity is due to spherical swimmers undergoing a constant rate of reorientation in sheared flows as spheres have no preferred direction. This is confirmed further by both IBM models, which do not display any preferential wall interactions orientations for all considered orientational P\'eclet numbers. 
 
 For the case of high rotational diffusion, $Pe=1$, we note that all distributions for non-spherical swimmers peak 
 {at approximately} $\theta=\pi/4$ and $\theta=5\pi/4$ (with troughs 
 {at approximately} $\theta=3\pi/4$ and $\theta=7\pi/4$) across all models, with 
 peak concentrations
 increasing with cell elongation. 
 {As the rotational diffusion decreases, corresponding to an increase in the rotational P\'eclet number, the peaks
shift towards $\theta =0$ and $\theta= \pi$ for all $\beta$, with peaks clearly sharpening for the case of $\beta=0.99$.}
For  $\beta=0.5$ there is a non-monotonic change in $\psi_{peak}$, shifting from $\psi_{peak}\approx1.3$ to $\psi_{peak}\approx1.5$ to $\psi_{peak}\approx1$ for $Pe=1, 10, 100$, respectively. 
 {The shift in peaks has a two-fold origin: the relative roles of advection and swimming (deterministic effects) versus diffusion effects, and the shift in the bulk cell distributions due to high- and low-shear trapping. In the aforementioned case, as rotational diffusion effects decrease (increase from $Pe=1$ to $Pe=10$) the decrease in randomness leads to decreased orientational spreading and sharper peaks. The slight elongation of cells ($\beta=0.5$)  results in cells spending more time aligned parallel to the flow direction. Meanwhile high- and low-shear trapping are phenomena observed by \citet{vennamneni2020shear}, where high-shear trapping refers to the shape and rotational diffusion dependent migration of cells towards channel walls and similarly low-shear trapping refers to the migration of swimmers towards the centreline. In our studies, both high-shear trapping and low-shear trapping are captured for $\beta=0.99$, as evidenced by the high-shear trapping leading the peak of the wall distribution  increasing from $\psi_{peak}\approx 1.6$ to $\psi_{peak}\approx 4$ to $\psi_{peak}\approx 8$, for $Pe=1,10,100$, respectively, before a transition to low-shear trapping for $Pe=10^4$ in figure \ref{CellDensityWallBounded} as the cells move away from the walls. 
}

{We further compare the profiles across the different models.}
For a small P\'eclet number $Pe=1$ (see figures \ref{DoublePWallPe1} and \ref{IBMWallPe1}) the profiles 
at $y_{\mathrm{near}}$ (the dashed lines)  are in good agreement, with similar peak magnitudes and spreads. 
The IBM distributions for $\beta=0$  are in agreement with the doubly Poiseuille cases in figures \ref{DoublePWallPe1}--\ref{IBMWallPe100}. For $Pe=100$ and $\beta=0.99$ (figures \ref{DoublePWallPe100} and \ref{IBMWallPe100}) the central peaks about $\theta=\pi$ are of similar height. 
{
However, the 
central peak 
about $\theta=0$ is slightly 
{larger }
in the individual-based model, as the time-step dependent drifting of particles contributes to depleting the peak profile about $\theta=0$. 
The depletion of the peak is, however, minor as the rotational diffusion is sufficiently large to feed more cells into the depletion areas. It is further worth noting that at $Pe=100$ 
{drifting in long time distributions} is only significant at $y_{\mathrm{near}}$ for  elongated swimmers as the deterministic trajectories of elongated swimmers point more sharply away from the wall about $\theta=0$, leading to an increased radius of {depletion} compared to more spherical swimmers.  }
{Running IBM  double Poiseuille simulations, as shown in figures \ref{WallDistr}d--\ref{WallDistr}f corresponding to $Pe=1,10,100$, we find that the probability distributions at $y=-1$ (solid lines) match with those obtained from the continuum model \ref{WallDistr}a--\ref{WallDistr}c after sufficiently long runtimes, }

{Comparing across all models, we find a remarkably good fit in the probability distributions 
within the studied range of rotational diffusion strengths and elongations. We see that all peak height and width distributions are in agreement across the three models, with only a slight discrepancy between the 
{peak heights} at $Pe=100$ for $\beta=0.99$ in the specular reflection IBM, indicating the limitation of the stochastic specular reflection IBM  at high $Pe$ for strong elongation due to larger spatial drifting}.

{
The good fit between the specular reflection IBM and the doubly periodic Poiseuille continuum model raises the question of how the doubly periodic model's symmetry constraints on $\psi$ and $\parone{\psi}{y}$ at $y=\pm1$ relate to the literature  \citep{jiang2019dispersion,jiang2020dispersion}. From figure \ref{WallDistr}, we note that near the walls, the equilibrium cell probability density distributions satisfy  $\psi(\theta,\pm1)=\psi(\theta+\pi,\pm1)$. 
With the periodic boundary cases (figures \ref{WallDistr}a--f),  the derivative satisfies $\parone{\psi}{y}(\theta,\pm1)=-\parone{\psi}{y}(\theta+\pi,\pm1)$. The combination of these symmetries ensure that the flux condition $J_\theta(\pm1)=-J_{\theta+\pi}(\pm1)$ and the integral no-flux boundary condition and impermeability are satisfied for the equilibrium problem.
We note that this observed boundary flux relationship differs from \citet{jiang2019dispersion,jiang2021transient}, in which for their time evolving continuum models with non-uniform initial condition, the flux condition itself was prescribed to be specular, by imposing the equivalent of $J_\theta(\pm1)=-J_{2\pi-\theta}(\pm1)$, through the constraints $\psi(\theta,\pm1)=\psi(2\pi-\theta,\pm1)$ and $\parone{\psi}{y}(\theta,\pm1)=-\parone{\psi}{y}(2\pi-\theta,\pm1)$, in our coordinate system. This was then compared to a double Poiseuille with half channel flows being in the same direction which has a jump in shear at $y=\pm1$. We note that the imposed  conditions in \cite{jiang2019dispersion} satisfy the no-flux condition, and though the bulk results are consistent across \cite{jiang2019dispersion} and our model, the imposed wall behaviours in \cite{jiang2019dispersion} differ from those that emerge in our equilibrium studies. 
We find from our IBMs that imposing specular reflection in instantaneous wall interactions 
does not result in a long-term 
distribution in the equilibrium solution which satisfies $\psi(\theta,\pm1)=\psi(2\pi-\theta,\pm1)$ as no enhanced cell accumulation is seen near $y=-1$ at values just below $\theta=\pi,2\pi$ (figures \ref{compareSpecBivariate}a--c and figures \ref{WallDistr}g--i). Therefore, the boundary conditions used by \citet{jiang2019dispersion,jiang2021transient} are not consistent with the distribution that naturally emerges from our IBM with specular reflection and that also emerges as the `boundary solution' in the doubly Poiseuille periodic continuum model. As both systems satisfy the no-flux and the expectated bulk flow dynamics, this reinforces the importance and sensitivity in the choice of constraints when developing continuum models for active suspension dynamics. 
}

While we have shown that the dynamics from specular reflection are well captured by a continuum approximation with doubly periodic boundary conditions, we know that microswimmers' surface interactions are not perfect specular reflections with a variety of factors affecting surface scattering \citep{kantsler2013ciliary,thery2021rebound}. A swimmer near the wall may remain oriented upstream for a significant period of time, it may attach to the surface, or it may leave the surface at varying outgoing angles which may be independent of the incident angles. Keeping this in mind, we consider the effects of two further wall-interaction models: perfectly random reflections (\S \ref{sec:RandomReflection}) and perfect absorption (\S \ref{sec:PerfAbsorption})
\subsection{Random reflections 
{(boundary condition $\mathcal{R}$)}
\label{sec:RandomReflection}\label{sec:FurtherBC}}

\begin{figure}
    \centering
    \begin{tikzpicture}
\node[above right] (img) at (0,0) {
                \begin{subfigure}[H!]{0.3\textwidth}
         \centering        \includegraphics[width=.99\textwidth]{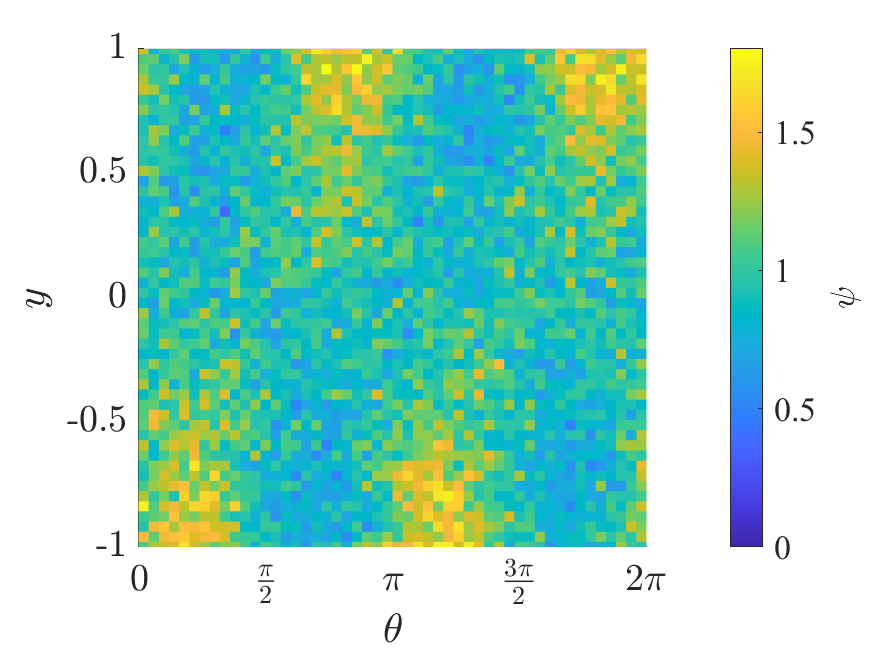}
                \caption{              \label{IBMRandombeta0.99Pe1Bivariate}
}
    \end{subfigure}
                \begin{subfigure}[H!]{0.3\textwidth}
         \centering        \includegraphics[width=.99\textwidth]{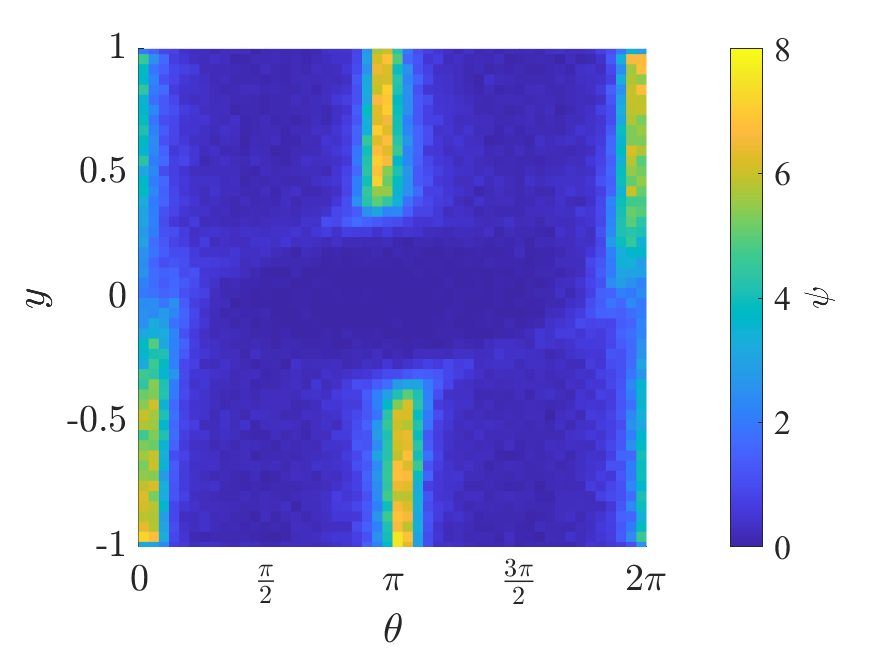}
                \caption{
           \label{IBMRandombeta0.99Pe100Bivariate}
}
    \end{subfigure}
                \begin{subfigure}[H!]{0.3\textwidth}
         \centering        \includegraphics[width=.99\textwidth]{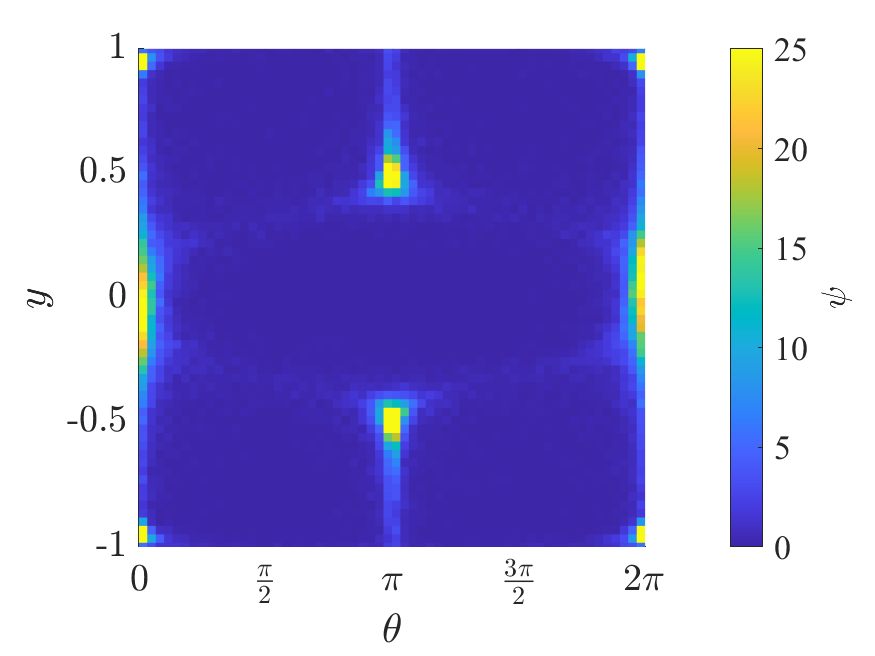}
                \caption{
                \label{IBMRandombeta0.99Pe10000Bivariate}
}
    \end{subfigure}
};
\node[rotate=90] at (1pt,70pt) {IBM};
\end{tikzpicture}
\begin{tikzpicture}
\node[above right] (img) at (0,0) {
                 \begin{subfigure}[H!]{0.3\textwidth}
         \centering        \includegraphics[width=.99\textwidth]{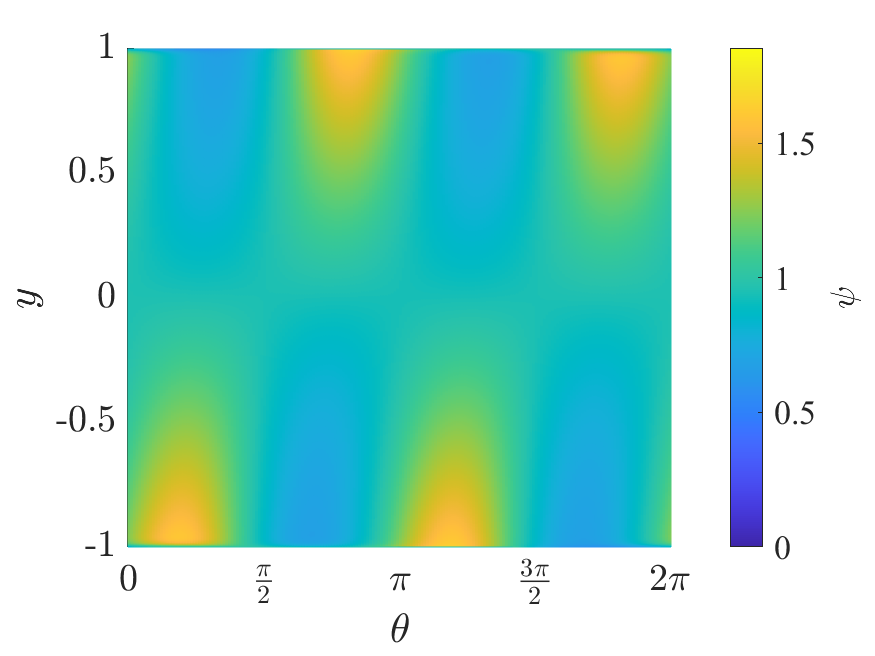}
                \caption{
              \label{FEConstbeta0.99Pe1Bivariate}
}
    \end{subfigure}   
                 \begin{subfigure}[H!]{0.3\textwidth}
         \centering        \includegraphics[width=.99\textwidth]{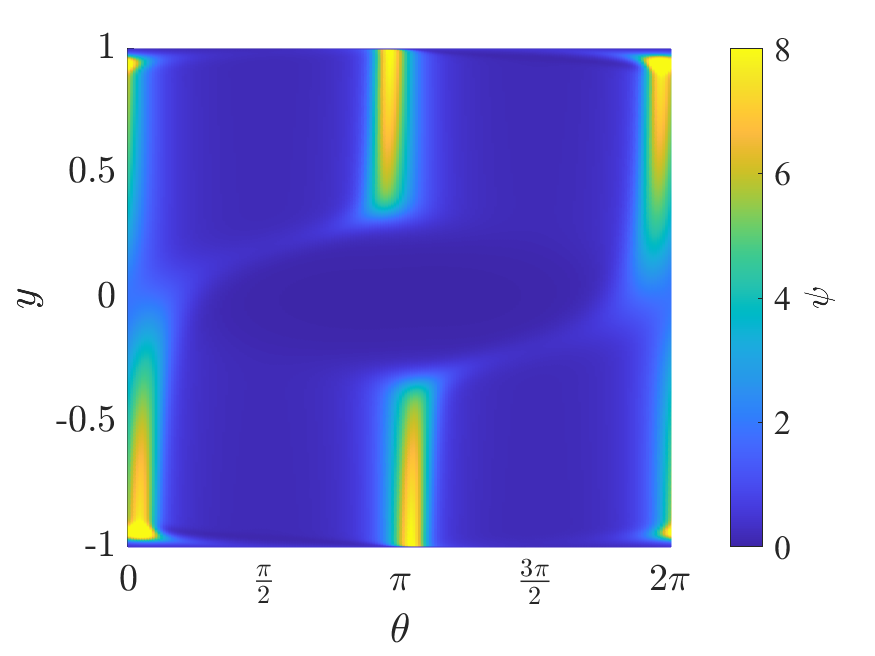}
                \caption{              \label{FEConstbeta0.99Pe100Bivariate}
}
    \end{subfigure}   
                 \begin{subfigure}[H!]{0.3\textwidth}
         \centering        \includegraphics[width=.99\textwidth]{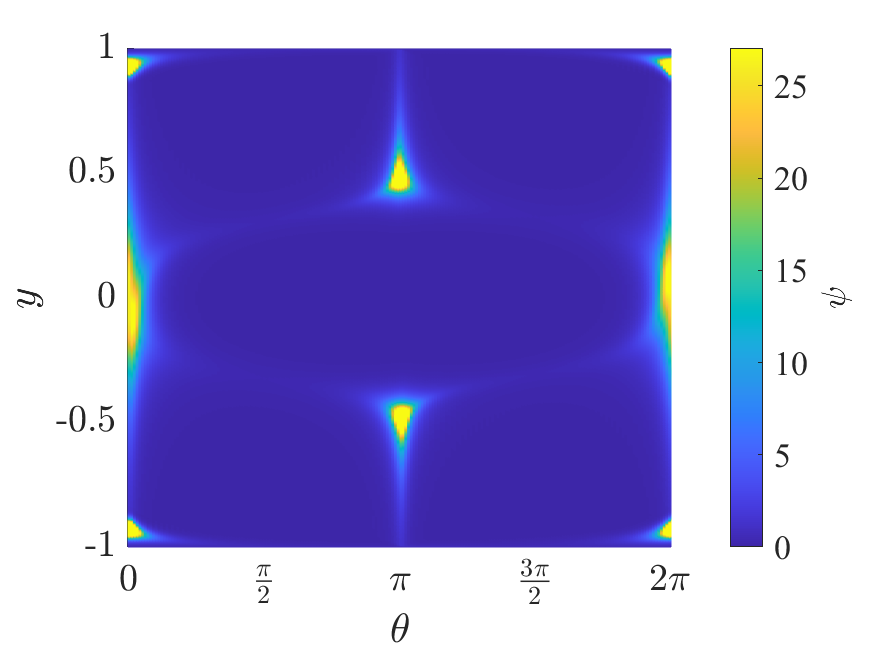}
                \caption{
              \label{FEConstbeta0.99Pe10000Bivariate}
}
    \end{subfigure}   
    };
\node[rotate=90] at (1pt,70pt) {Continuum};
\end{tikzpicture}
\caption{Comparison of snapshots of bivariate probability density distributions $\psi$, as obtained converged IBMs  with random wall reflections (a)--(c), to equilibrium probability density distributions for continuum models with constant boundary condition.
 (a)\&(d): $Pe=1$, (b)\&(e): $Pe=100$; and (c)\&(f): $Pe=10^4$.
\label{FigRandomRefBivariate}
}
\end{figure} 
In this section we consider the effects of random uniform reflections at the boundaries on the equilibrium dynamics of microswimmer suspensions. Suppose we have a random uniform reflection out of the wall for each incident swimmer, independent of the angle of incidence, as described by equation   \ref{eq:randomReflection}.  
 In figures 
\ref{IBMRandombeta0.99Pe1Bivariate}--\ref{IBMRandombeta0.99Pe10000Bivariate}, we see the bivariate cell probability density distribution $\psi$ for $\beta=0.99$ for varying rotational P\'eclet numbers. By inspection, the bulk flow distributions are similar to those found via the doubly periodic continuum model and the specular reflection IBM, with areas of cell accumulations which sharpen with increased $Pe$. However, from figure \ref{IBMRandombeta0.99Pe10000Bivariate} we clearly note the appearance of secondary peaks at $( \theta,y)=(0,\pm 0.93)$ for $Pe=10^4$, and also note  smaller peaks occurring at $(\theta, y)=(0.13, -0.94)$ and $ (\theta,y)=(2\pi-0.13, 0.94)$ for $Pe=100$ (figure \ref{IBMRandombeta0.99Pe100Bivariate}). No clear peak is visible for $Pe=1,$ as rotational effects dominate deterministic secondary structures. We note further, that the upper bound of the aforementioned secondary peaks are bounded at $\theta=0$ by the cusp of the deterministic trajectory originating  from $y=-1$ and $\theta=\pi$ (the yellow separatrix from figure  \ref{PerfectAbsorbBivar_Wallpdfbeta0pt99Pe10000}).

Noting that the secondary peaks are wholly introduced by the uniform reflective conditions, we seek to determine the appropriate continuum model boundary condition to obtain the corresponding bulk dynamics. To capture the uniformity of reflection, and lack of orientation preference upon reflection, we consider %
{
a continuum model with a constant Dirichlet wall-boundary condition (constraint $\mathcal{D}_C$ introduced in \S \ref{sec:contModel}) such that $\psi$ is the same constant on both walls. 
}
From figures \ref{FEConstbeta0.99Pe1Bivariate}--\ref{FEConstbeta0.99Pe10000Bivariate} we find  secondary peaks occurring at the same points in phase-space, indicating that slightly away from the wall there is an enhanced number of cells swimming downstream. 

To further confirm the suitability of comparing the IBM with random uniform reflection to the continuum model with a constant Dirichlet boundary condition, we consider the cell number density distributions in figure \ref{FigRandomRef}.
Comparing figures 
   \ref{IBMRandombeta0.99} and \ref{FEConstbeta0.99} for $\beta=0.99$, we find that both profiles for $Pe=10^4$ (the blue lines) have the same primary peaks about $y=\pm0.5$ as observed for the IBM with specular reflection, but we also get a significant peak in density about 
   {$y=\pm 0.93$} in both figures. 
   {While there are discrepancies in the size of the secondary peaks found  near the wall  via the IBM, we note that their size is limited by the finite time steps. Too large time steps result in cells drifting away from the secondary peaks. We find that there is a computational trade off in the total runtime required to capture the macroscopic effects (such as the peaks and troughs) and the smallness of timesteps required to capture the slim secondary peaks. We further note that in both the continuum and IBM models the minimum cell densities occur at $y\approx \pm0.85$. } 
   We similarly find the distributions for $Pe=100$ to be a good match with the previous models (figure \ref{CellDensityWallBounded}) except at the locations of the secondary peaks. Finally, there is good agreement between the IBM and continuum models for $Pe=1$.

\begin{figure}
    \centering
                \begin{subfigure}[H!]{0.30\textwidth}
         \centering        \includegraphics[width=.99\textwidth]{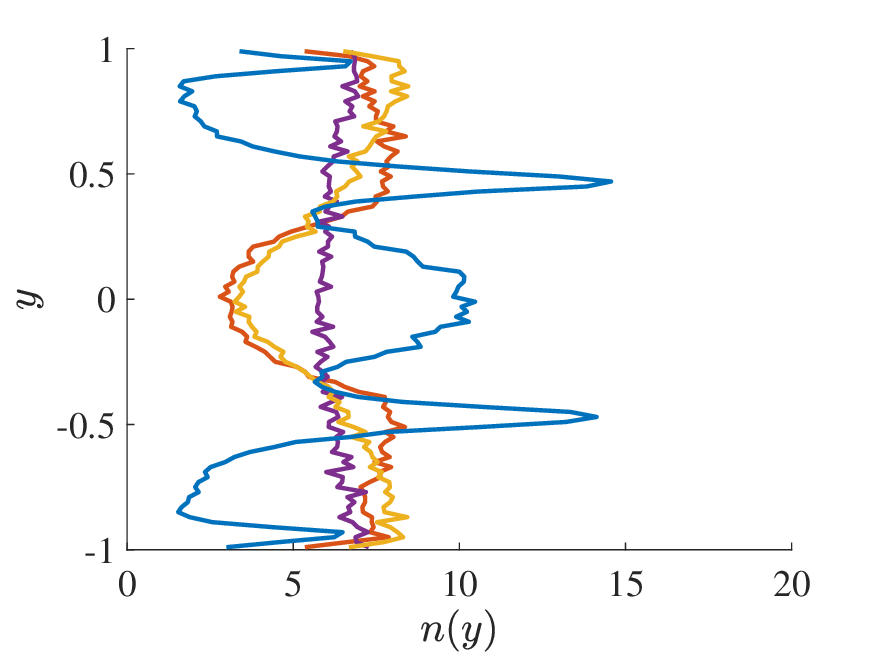}
                \caption{
              \label{IBMRandombeta0.99}
}
    \end{subfigure}
                \begin{subfigure}[H!]{0.30\textwidth}
         \centering        \includegraphics[width=.99\textwidth]{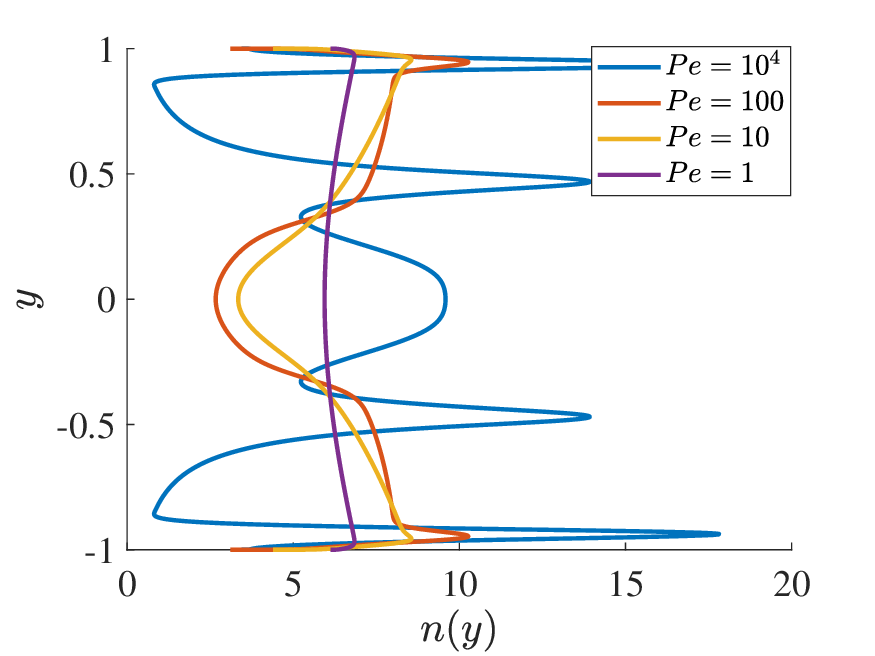}
                \caption{
              \label{FEConstbeta0.99}
}
    \end{subfigure}
    
\caption{
Cell  density distributions for (a): IBM with uniform random wall reflection; and (b): the distribution of continuum model with constant wall distribution. For shape parameters $\beta=0.99$ 
with $Pe=10^4$ (blue),  $Pe=100$ (red),  $Pe=10$ (yellow) and $Pe=1$ (purple). 
\label{FigRandomRef}
}

\end{figure} 

\subsection{Perfectly absorbing boundary 
{(boundary condition $\mathcal{A}$) \label{sec:PerfAbsorption}}}

Our final boundary condition of interest is the case of perfect wall attachment $\mathcal{A}$, i.e. any swimmer that encounters the wall will adhere to it. For ease of comparing the effect on the bulk dynamics we allow for specular reflection at the top wall ($y=1$) while enforcing a perfectly absorbing bottom wall, such that cell trajectories are terminated upon contact with the bottom wall ($y=-1$). It is worth noting that given the presence of diffusive effects, given sufficient time, all cells will attach to the bottom wall. Let us begin by considering a snapshot of the bivariate probability density distributions $\psi$ at time $T=600$ for the stochastic IBM.  
In figure \ref{fig:PerfAbsorbingbeta0pt99}a 
($\beta=0.99$ and $Pe=1$), the distributions show clear depletion {near} the bottom wall, as all cells that have been able to encounter the bottom have attached. In figure \ref{fig:PerfAbsorbingbeta0pt99}b (for $Pe=100$) fewer cells are captured by the bottom wall by time $T=600$, however, there is a clear depletion, and 
{the accumulation band in the bottom half contains approximately 75\% the number of cells compared to their upper-half channel} specular-reflecting counterparts. Finally, we consider figure \ref{fig:PerfAbsorbingbeta0pt99}c (for $Pe=10^4$). The yellow line is a separatrix for fully deterministic cell trajectories, where all cells below the line must interact with the wall, and  all cells above will not in the absence of diffusion.  We find that figure \ref{fig:PerfAbsorbingbeta0pt99}c  mainly has depletion of cells originating below the separatrix, as  cell diffusion is very small. In figure \ref{TimeEvolvePerfectAbsorb_WallPercentage} we see that by $T_{sim}=600$ approximately 30\%, 12\% and 5\% of total cells were absorbed at the bottom wall for $Pe=1, 100,10^4$, respectively, with absorption plateauing earliest for $Pe=10^4$ (yellow line). Therefore, higher diffusion effects yield higher rates of wall absorption for extended times.

In figures \ref{fig:PerfAbsorbingbeta0pt99}d-\ref{fig:PerfAbsorbingbeta0pt99}f, we consider the normalised 
orientation distributions of the cells which have been absorbed at the bottom wall. 
In figure \ref{fig:PerfAbsorbingbeta0pt99}d we  find that in the presence of high diffusion, the wall encounter probability distributions are wide, centred about $\theta_{peak}=3\pi/2$, and the distributions remain unchanged for $T=50,100, 600$. For $Pe=100$ in figure \ref{fig:PerfAbsorbingbeta0pt99}e, the peak near $\theta=\pi$ continuously increases in time. 
This is due to a localised increase in the number of swimmers crossing the separatrix (from figure \ref{fig:PerfAbsorbingbeta0pt99}c) with time. The rotational diffusion is sufficiently weak that deterministic effects dominate and cells are quickly  captured by the absorbing condition just above $\theta=\pi$. Finally, for figure \ref{fig:PerfAbsorbingbeta0pt99}f, we note a similar increase in the peak near $\theta=\pi$. 
In fact, across figures \ref{fig:PerfAbsorbingbeta0pt99}d-f, we find that the peak orientations at which absorptions occurs shifts from $\theta_{peak}=3\pi/2$ towards $\theta_{peak}=\pi$ with decreasing rotational diffusion. The difference in distribution for increasing P\'eclet number is a result of swimming and fluid advection  dominating diffusion effects. The role of diffusion effects on cell interactions with walls will be studied in more detail in \S \ref{sec:BulkFlowWallApproach}.

\begin{figure}
    \centering
\begin{tikzpicture}
\node[above right] (img) at (0,0) {    
    \begin{subfigure}[H!]{0.32\textwidth}
        \centering
        \includegraphics[width=\textwidth]{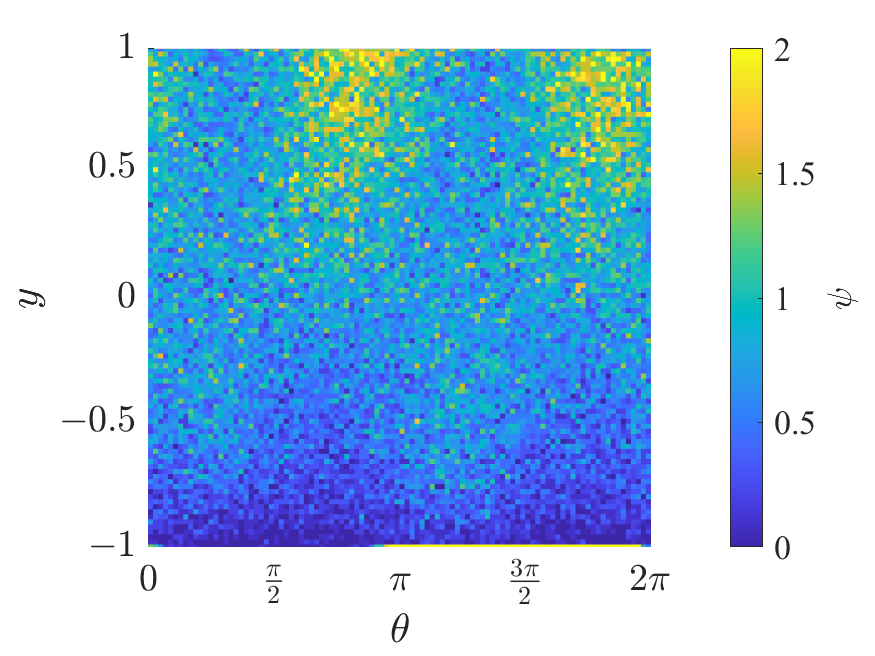}
                        \caption{    
              \label{PerfectAbsorbBivar_Wallpdfbeta0pt99Pe1}
}
    \end{subfigure}
        \begin{subfigure}[H!]{0.32\textwidth}
        \centering
        \includegraphics[width=\textwidth]{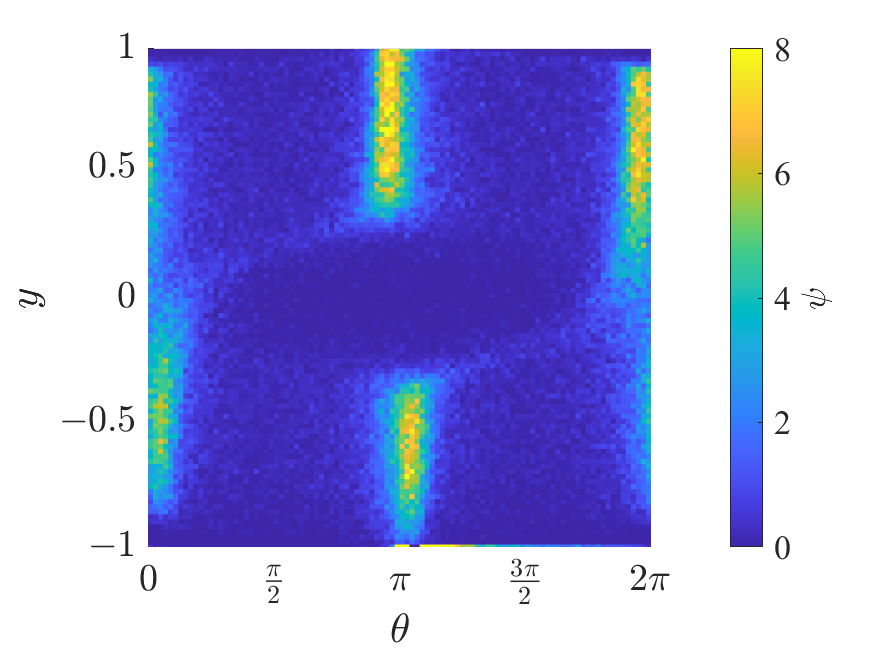}
                        \caption{     
              \label{PerfectAbsorbBivar_Wallpdfbeta0pt99Pe100}
}
    \end{subfigure}
        \begin{subfigure}[H!]{0.32\textwidth}
        \centering
        \includegraphics[width=\textwidth]{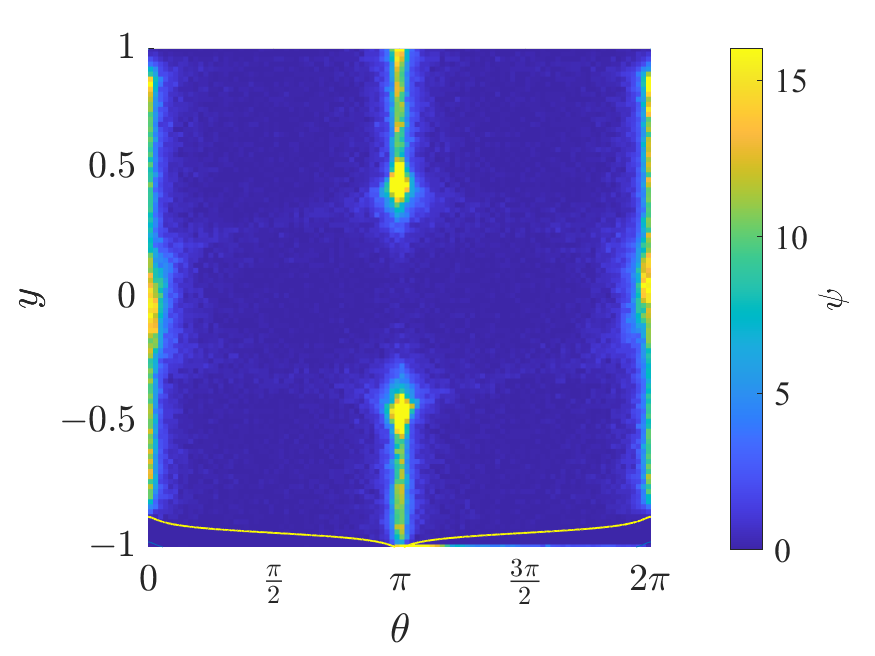}
                        \caption{  \label{PerfectAbsorbBivar_Wallpdfbeta0pt99Pe10000}
}
    \end{subfigure}
    };
\node (A) at (298pt,90pt) {\footnotesize\textcolor{white}{ Region 2}};
\node (B) at (320pt,42pt) {\textcolor{white}{Region 1}};
\node (C) at (344pt,32pt) {.};
\draw[white, ->, to path={-| (\tikztotarget)}]
  (B) edge (C);
\end{tikzpicture}
\begin{subfigure}[H!]{0.32\textwidth}
        \centering
        \includegraphics[width=\textwidth]{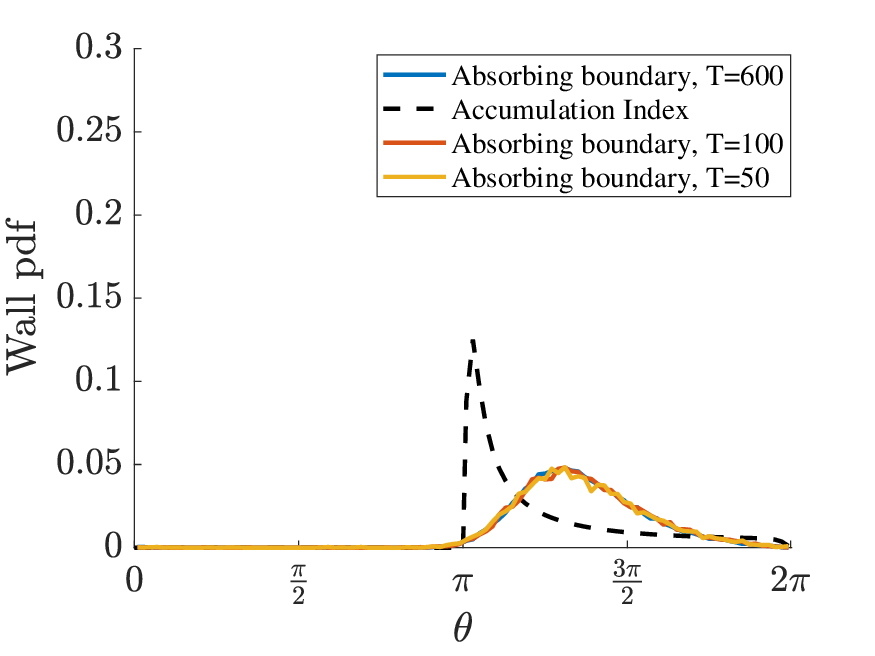}
                        \caption{ \label{PerfectAbsorb_Wallpdfbeta0pt99Pe1}
}
    \end{subfigure}
\begin{subfigure}[H!]{0.32\textwidth}
        \centering
        \includegraphics[width=\textwidth]{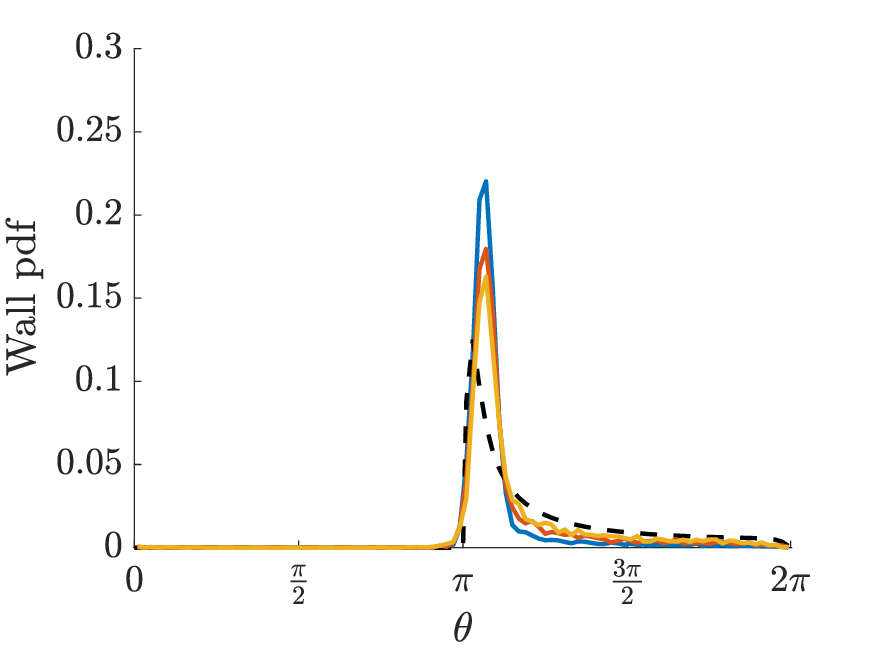}
                        \caption{ \label{PerfectAbsorb_Wallpdfbeta0pt99Pe100}
}
    \end{subfigure}
    \begin{subfigure}[H!]{0.32\textwidth}
        \centering
        \includegraphics[width=\textwidth]{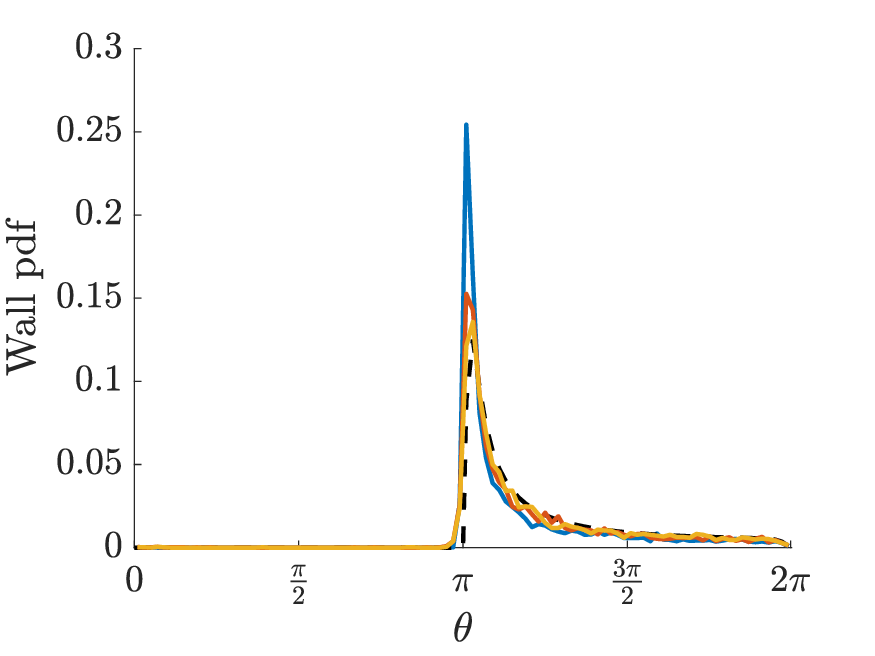}
                        \caption{ \label{PerfectAbsorb_Wallpdfbeta0pt99Pe10000}
}
    \end{subfigure}
        \begin{subfigure}[H!]{0.32\textwidth}
        \centering
        \includegraphics[width=\textwidth]{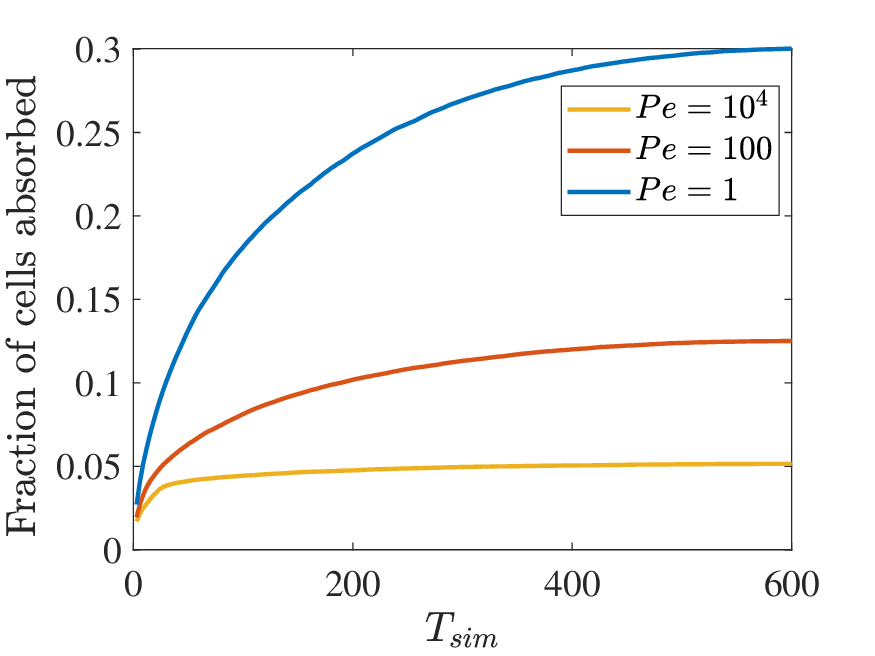}
                        \caption{ \label{TimeEvolvePerfectAbsorb_WallPercentage}
}
    \end{subfigure}
\caption{Snapshots of the effects of a perfectly absorbing wall condition for different $Pe$ at the bottom wall for an IBM (with dynamics at the top wall prescribed by specular reflection) on the bulk dynamics ((a)--(c)) at $T_\mathrm{sim}=600$ and on normalised wall orientation probability distributions for $\beta=0.99$ ((d)--(f))for runtimes $T_\mathrm{sim}=50, 100,600$. The yellow line in (c) is a separatrix between fully deterministic trajectories which interact with the bottom wall and those that don't. In figures (d)--(f), The black dashed lines correspond to wall distributions for $\beta=0.99$ as calculated by the accumulation index (see \S \ref{sec:DetermineUnderlying}).  (a)\&(d):$Pe=1$, (b)\&(e):$Pe=100$ and (c)\&(f):$Pe=10^4$. (g): Time evolution of the fraction of cells absorbed by the bottom wall.
\label{fig:PerfAbsorbingbeta0pt99} 
}
\end{figure} 

\begin{figure}
    \centering
    \begin{tikzpicture}
\node[above right] (img) at (0,0) {
    \begin{subfigure}[H!]{0.45\textwidth}
        \centering
        \includegraphics[width=\textwidth]{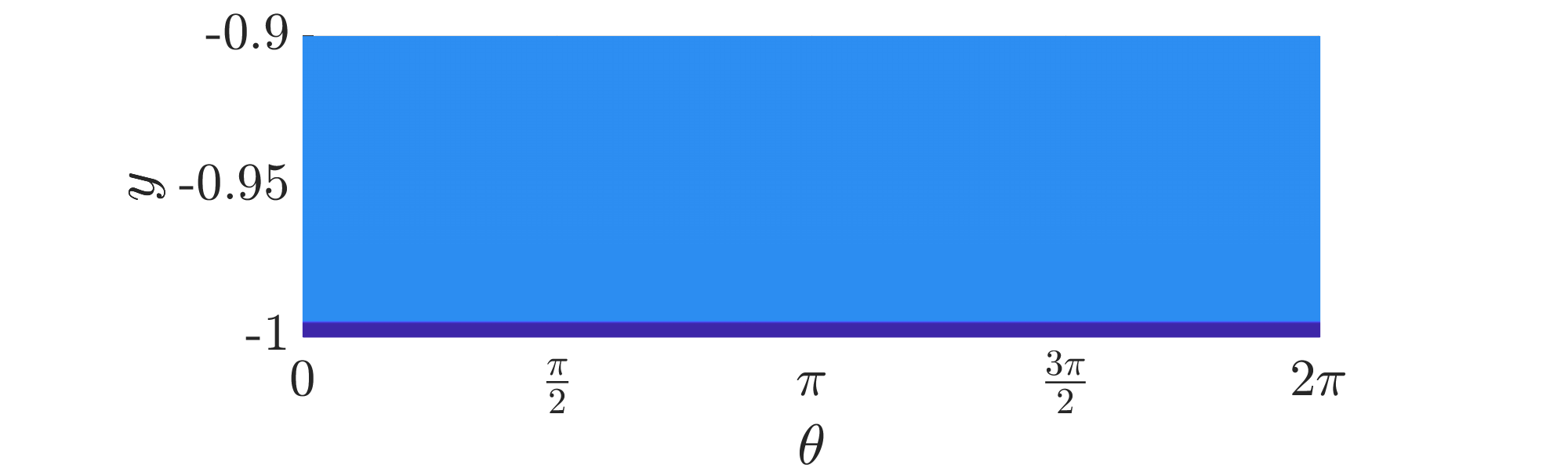}
                        \caption{    
              \label{FEAbsorbingBL0}
}
    \end{subfigure}
    };
\node at (90pt,90pt) {Continuum};
\end{tikzpicture}
\begin{tikzpicture}
\node[above right] (img) at (0,0) {
    \begin{subfigure}[H!]{0.45\textwidth}
        \centering
        \includegraphics[width=\textwidth]{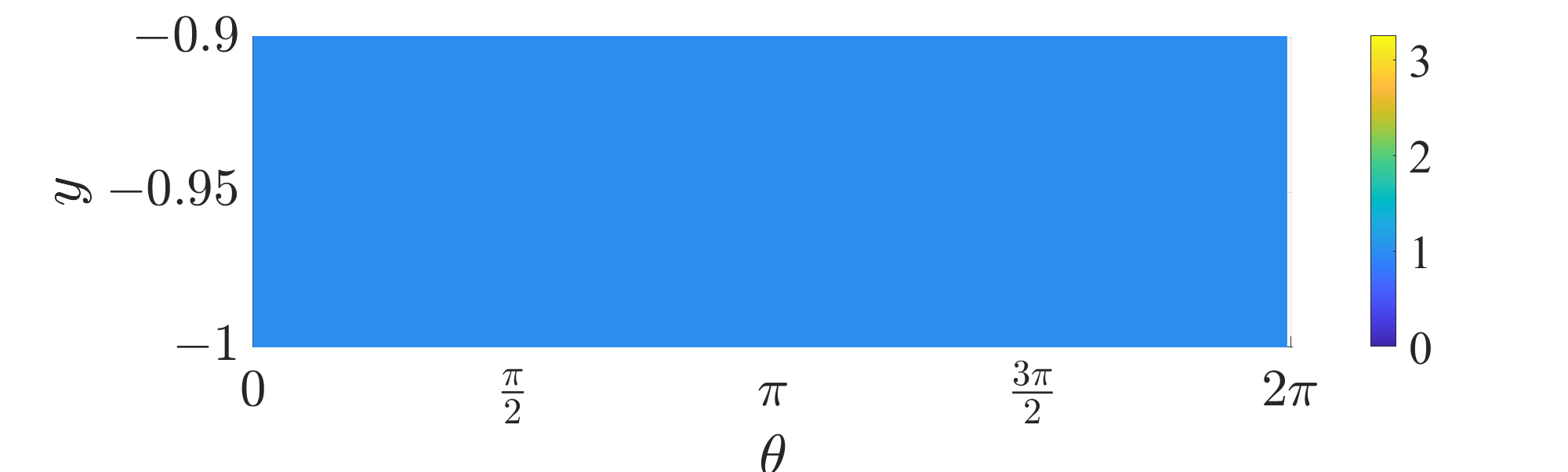}
                        \caption{  
              \label{StochAbsorbingBL0}
}
    \end{subfigure}
    };
\node at (90pt,90pt) {IBM};
\end{tikzpicture}
        \begin{subfigure}[H!]{0.45\textwidth}
        \centering
        \includegraphics[width=\textwidth]{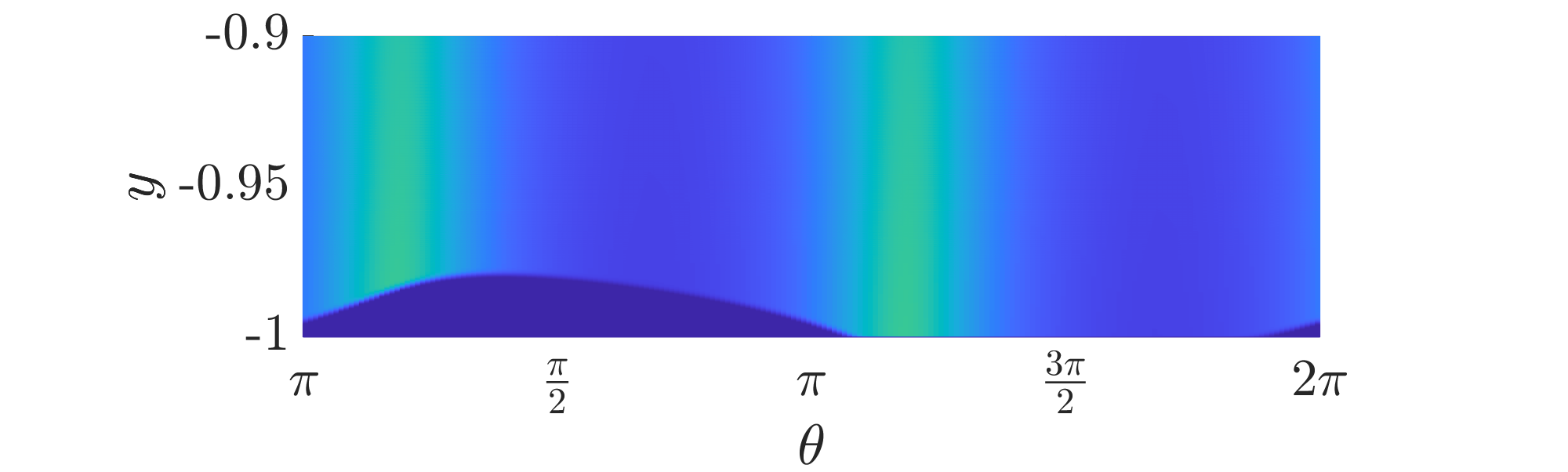}
                        \caption{  \label{FEAbsorbingBL410}
}
    \end{subfigure}
     \begin{subfigure}[H!]{0.45\textwidth}
        \centering
        \includegraphics[width=\textwidth]{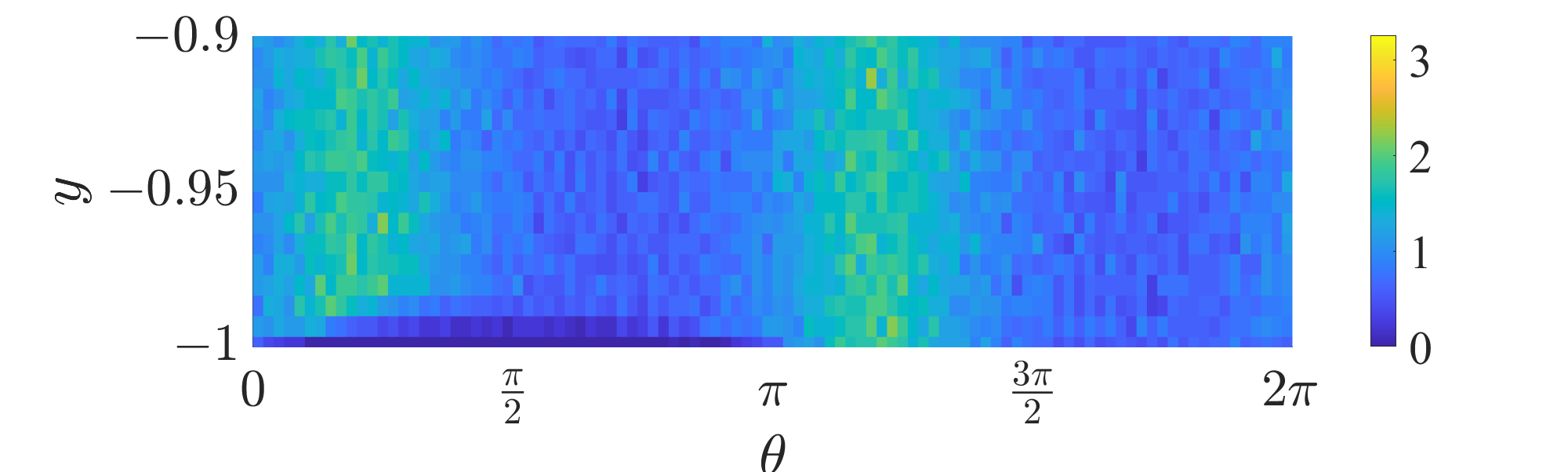}
                        \caption{     
              \label{StochAbsorbingBL400}
}
    \end{subfigure}    
\begin{subfigure}[H!]{0.45\textwidth}
        \centering
        \includegraphics[width=\textwidth]{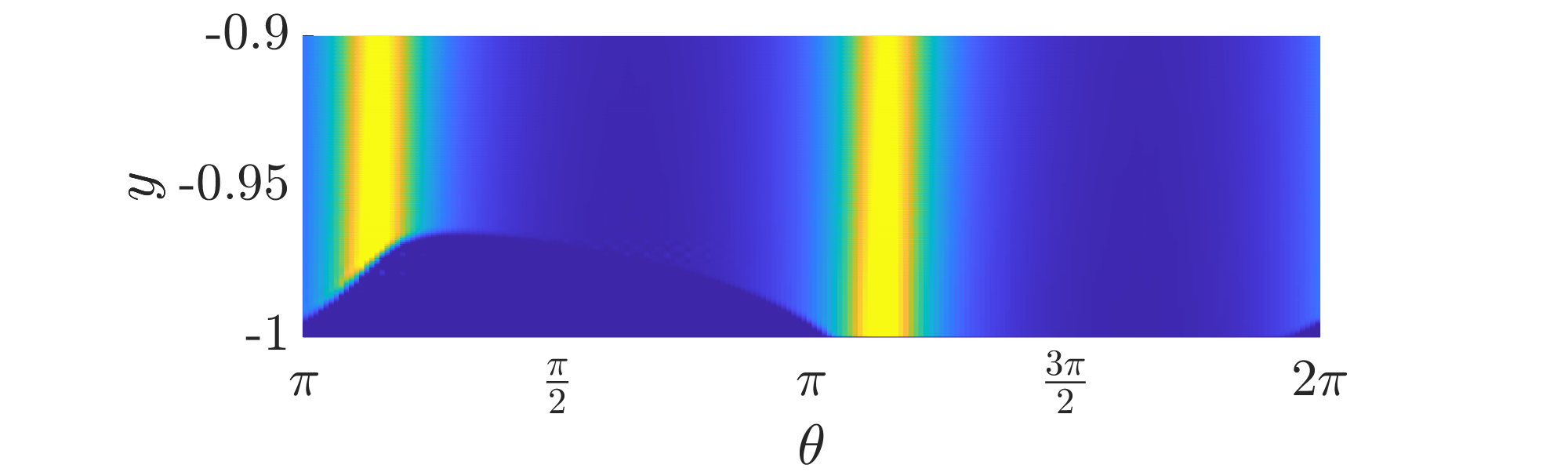}
                        \caption{ \label{FEAbsorbingBL810}
}
    \end{subfigure}
         \begin{subfigure}[H!]{0.45\textwidth}
        \centering
        \includegraphics[width=\textwidth]{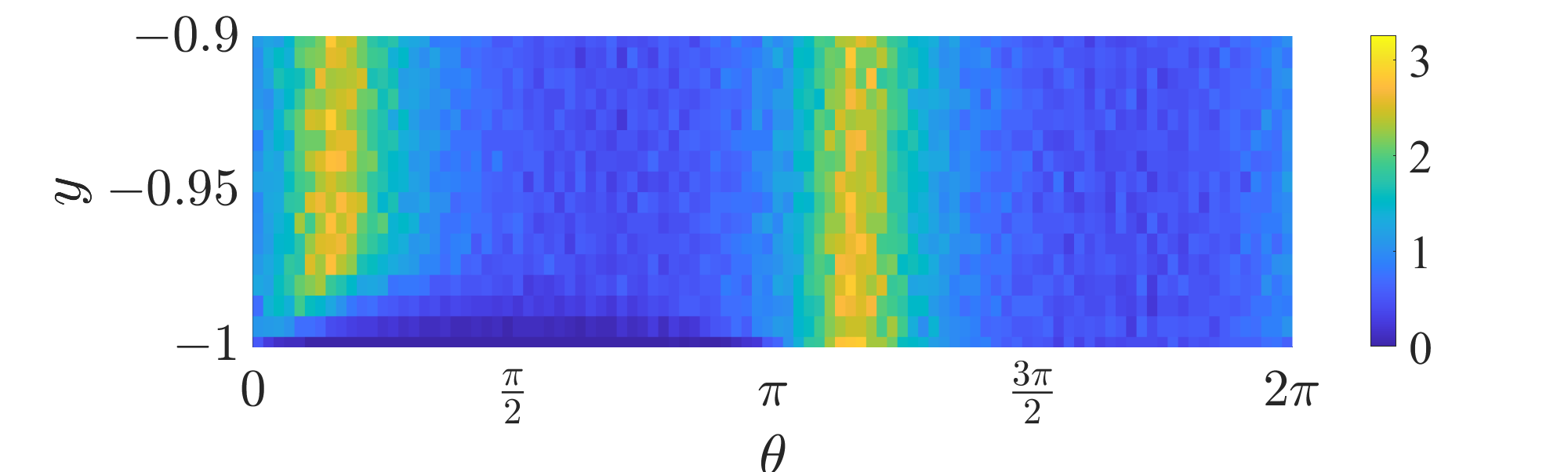}
                        \caption{     
              \label{StochAbsorbingBL800}
}
    \end{subfigure}
\begin{subfigure}[H!]{0.45\textwidth}
        \centering
        \includegraphics[width=\textwidth]{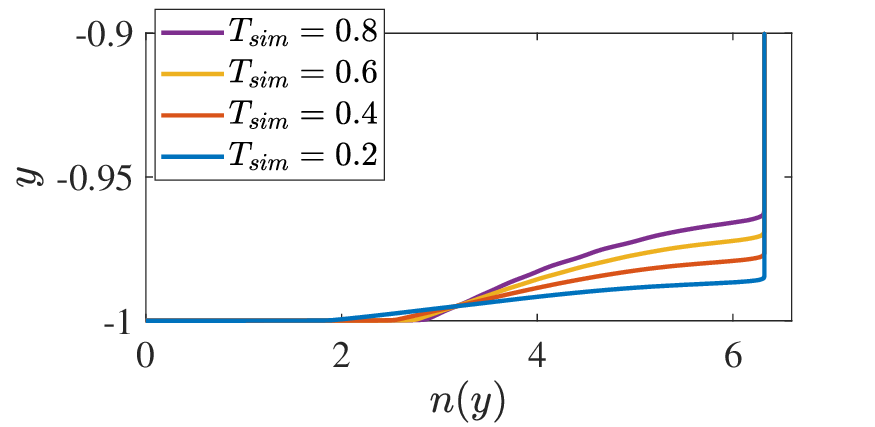}
                        \caption{ \label{FETimeEvolve}
}
    \end{subfigure}
         \begin{subfigure}[H!]{0.45\textwidth}
        \centering
        \includegraphics[width=\textwidth]{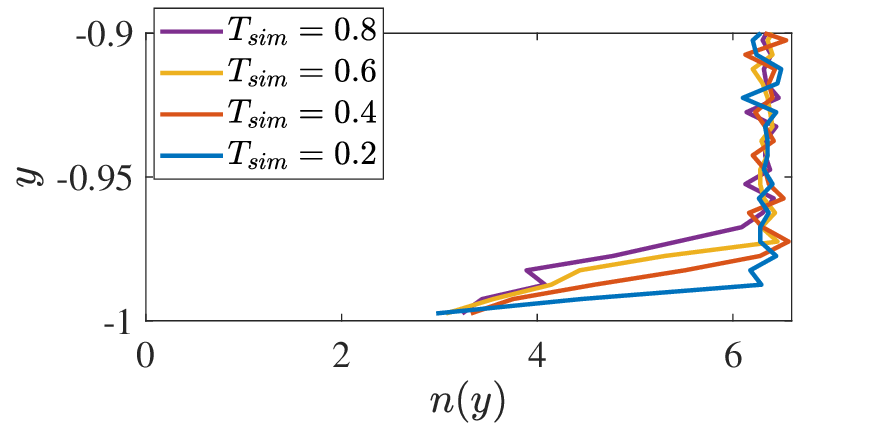}
                        \caption{     
              \label{StochTimeEvolve}
}
    \end{subfigure}    
   \caption{Early time evolution near the bottom boundary of a double Poiseuille absorbing boundary continuum model  ((a), (c), (e), \& (g) ), and the stochastic IBM with specular reflection upper boundary and perfectly absorbing lower boundary ((b), (d), (f),  \&(h)), for $\nu=0.04, \beta=0.99, Pe=10^4, Pe_T=10^6.$ (a)\&(b): $T_\mathrm{sim}=0$; (c)\&(d): $T_\mathrm{sim}=0.4$; and (e)\&(f): $T_\mathrm{sim}=0.8$. 
\label{fig:PerfAbsorbingbeta0pt99TEvolve} Cell concentration $n(y)$ evolution near the lower wall for (g): the continuum model with boundary condition $\mathcal{D}_0$ and (h): the IBM with absorbing boundary condition $\mathcal{A}$.
}
\end{figure} 

We compare these results with the time evolving continuum model where we model the perfectly absorbing wall at $y=-1$ with Dirichlet boundary condition $\mathcal{D}_0$ (see \S\ref{sec:contModel} for details), and use the double Poiseuille profile to capture the reflective boundary condition at $y=1$.
In figure \ref{fig:PerfAbsorbingbeta0pt99TEvolve} we compare the early time evolution of cells near the absorbing boundary using the continuum model with boundary condition $\mathcal{D}_0$ and stochastic boundary condition $\mathcal{A}$. At $t=0$, the stochastic system is initially uniformly distributed (see figure \ref{StochAbsorbingBL0}), and the continuum model (figure \ref{FEAbsorbingBL0}) has a uniform distribution everywhere except at the very thin  boundary layer as defined in \S \ref{sec:contModel}. By $t=0.4$, it is clear from figure \ref{FEAbsorbingBL410} and \ref{StochAbsorbingBL400} that cells near the wall oriented into the wall with $\pi<\theta<2\pi$ move to the bottom wall, and those with orientations $0<\theta<\pi$ swim away from the bottom wall, leaving an area of depletion. Across both models, with increasing time (figures \ref{FEAbsorbingBL810}-\ref{StochAbsorbingBL800}), the depletion area grows, emanating into the bulk domain from $0\leq \theta<\pi$ as cells continue to be absorbed at $y=-1$ and $\pi\leq\theta<2\pi$. This persists despite the emergence of the macroscopic areas of accumulation sharpening in time.
In figures \ref{FETimeEvolve} and \ref{StochTimeEvolve}, we see the time evolution of the cell concentration distributions for absorbing boundaries at different instants in time. The concentration profiles capture the cell depletion away from the wall across both models. The depletion front (the location in $y$ at which we see sharp dip in cell concentration $n(y)$) moves into the bulk at comparable rates across the IBM and continuum models in figures \ref{FETimeEvolve} and \ref{StochTimeEvolve}. While the macroscopic areas of accumulation are already visibly beginning to form (see figures \ref{FEAbsorbingBL0} to \ref{StochAbsorbingBL800}) it is worth noting that in the time-evolving cell concentration profiles the cell distributions are still uniform near the wall aside from the region of cell depletion due to the absorption boundary condition. This stands in contrast to the cell equilibrium profiles for other boundary conditions (see figures \ref{CellDensityWallBounded} and \ref{FigRandomRef}) where there are greater spatial variations in cell  concentration across $y$. This highlights the existence of a faster time scale of interest, which will be discussed further in \S \ref{sec:DetermineUnderlying}.

\begin{figure}
    \centering
    \begin{subfigure}[H!]{0.32\textwidth}
         \centering        \includegraphics[width=.99\textwidth]{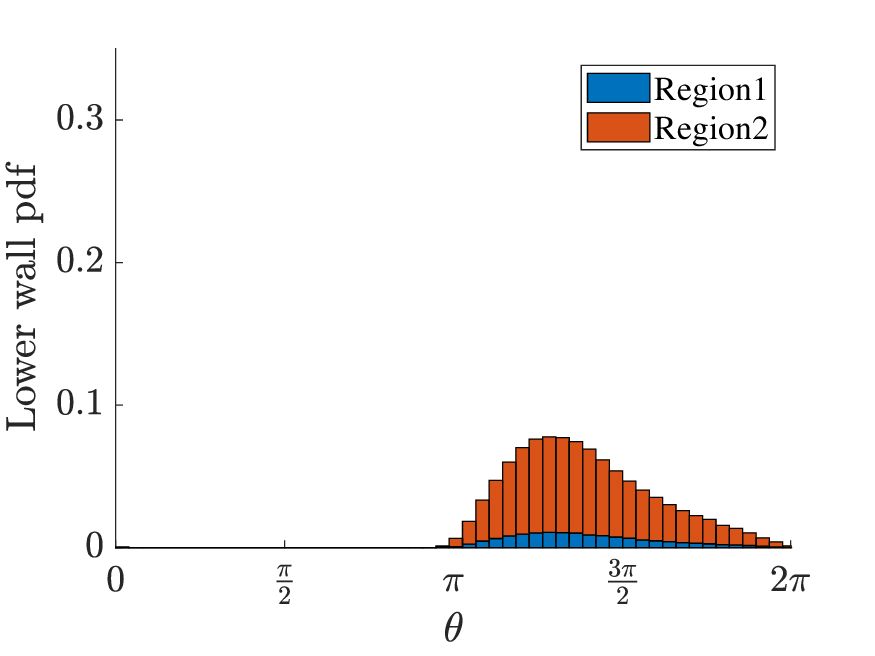}
                \caption{
              \label{IBMStackPe1beta0.99}
}
    \end{subfigure}
                \begin{subfigure}[H!]{0.32\textwidth}
         \centering        \includegraphics[width=.99\textwidth]{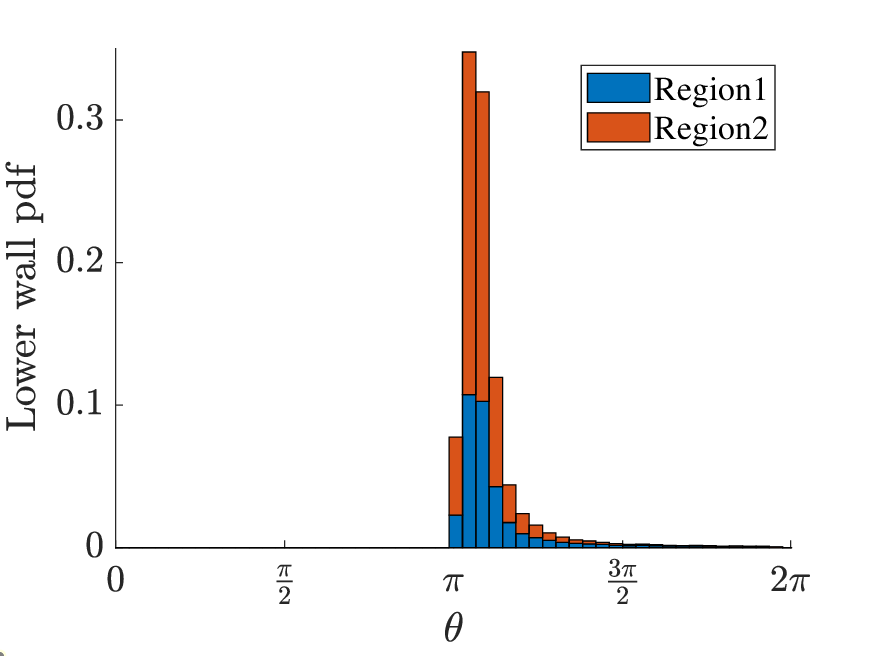}
                \caption{
              \label{IBMStackPe100beta0.99}
}
    \end{subfigure}
 \begin{subfigure}[H!]{0.32\textwidth}
         \centering        
         \includegraphics[width=.99\textwidth]{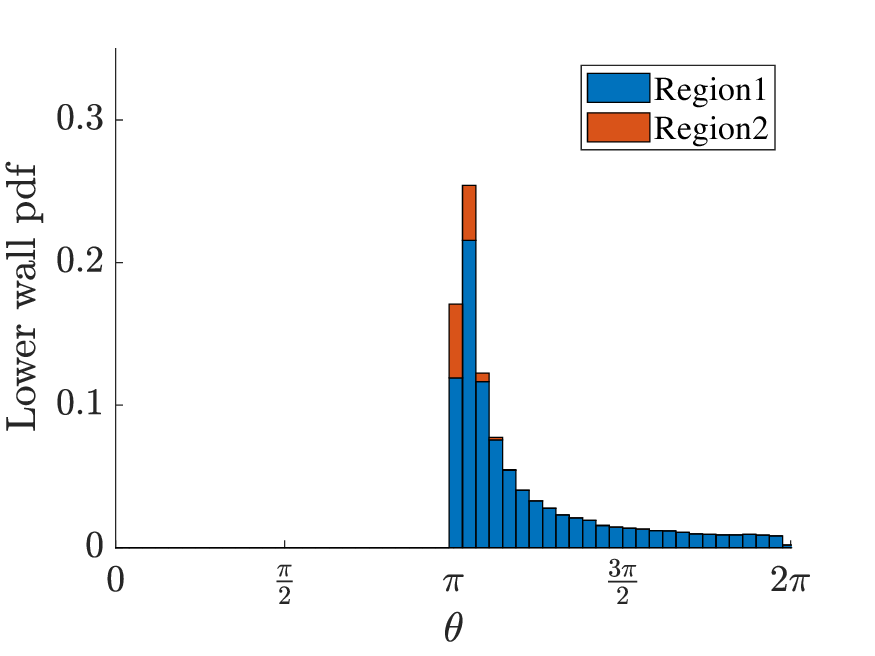}
                \caption{
              \label{IBMStackPe10000beta0.99}
}
    \end{subfigure}                
                \begin{subfigure}[H!]{0.4\textwidth}
         \centering        \includegraphics[width=.99\textwidth]{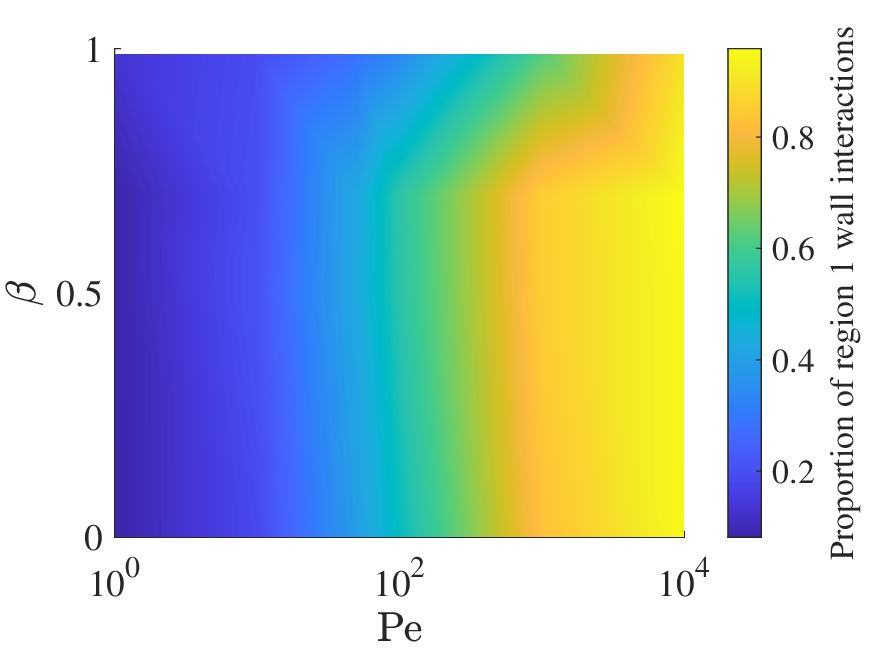}
         \caption{
              \label{ProportionRegion1}
}
    \end{subfigure}  
\caption{
Stacked probability distribution of angle of incidence for particles striking the lower wall ($y=-1$), for $\nu=0.04$ and $Pe_T=10^{6}$. The blue distribution corresponds to particles which are expected to strike the wall in the absence of diffusive effects (originating in region 1), and the red, correspond to particles that would not strike the bottom wall in the absence of diffusive effects (originating in region 2). 
The overall envelope characterises the distribution of cells striking the bottom wall and integrates to 1.
For 
$\beta=0.99$, (a)$Pe=1$, (b)$Pe=100$, and (c)$Pe=10^4$. (d)
 {Ratio of cell--wall interactions with cells originating in region 1 to total cell--wall interactions, 
 for varying $\beta$ and P\'eclet numbers. }
\label{Fig5}
}
\end{figure} 

\subsection{Bulk flow dynamics effects on wall approach \label{sec:BulkFlowWallApproach}}

  {   In this section we analyse how the underlying bulk flow dynamics of swimmers of different shapes in sheared fluids impact their orientations at wall-approach in the $\theta$-$y$ space. 
     This is of particular interest as  individual swimmer dynamics inform how suspensions interact with the walls, and sheds insights into why swimmers of different geometries are more likely to interact with  walls at different preferred orientations, thereby affecting their likelihood of wall attachment and biofilm formation.}
     
 \subsubsection{Diffusive wall approach}
From the perfectly absorbing IBM (\S \ref{sec:PerfAbsorption}) we know that as time evolves the orientation of the cells as they interact with boundaries evolves (see figure \ref{fig:PerfAbsorbingbeta0pt99}d--f). The time evolution and spread of orientations at the point of wall interaction is shown to be diffusion dependent.
We can analyse the extent of diffusion dependence by considering two regions of cell origin. If cells are initially uniformly distributed in the phase space, in the deterministic case we can clearly divide the cells in the phase space into two regions with respect to  the yellow deterministic streamline in figure \ref{fig:PerfAbsorbingbeta0pt99}c. 
We call the area of interactions (below the yellow line) `Region 1', and the area above `Region 2'. In the absence of diffusion,  only Region 1 cells  interact with the walls. However, with increasing diffusion, larger quantities of microswimmers cross the deterministic streamlines, and more cells from Region 2 interact with the walls. The shift in interactions is captured in figure \ref{Fig5} for fixed IBM runtime $T_\mathrm{sim}=600$ through stacked probability distributions, where the total number of wall interactions across 51 bins are normalised to 1. For the case of $Pe=10^4$ with $\beta=0.99$, the orientation distribution peaks tend toward $\theta\rightarrow \pi$ as $\beta\rightarrow1$ (see figure \ref{IBMStackPe10000beta0.99}). This means that there is increased upstream alignment with elongation. 
{From figure   \ref{Fig5}d we find that in this low rotational diffusion case, over} 80\% of all wall interaction originate from region 1, and 
{this percentage decreases} monotonically with $Pe$, irrespective of swimmer shape. An increase in rotational diffusivity, corresponding to $Pe=1$ (figure \ref{Fig5}a) shifts the peak of the distribution  $\theta_{peak}\rightarrow3\pi/2$. 
\subsubsection{Deterministic wall approach and underlying dynamics \label{sec:DetermineUnderlying}}
To further understand what happens when there is an absorbing wall we develop a novel accumulation index to capture the orientation for the fully deterministic case. 
\begin{figure}
    \centering
\begin{subfigure}[H!]{0.32\textwidth}
        \centering
        \includegraphics[width=\textwidth]{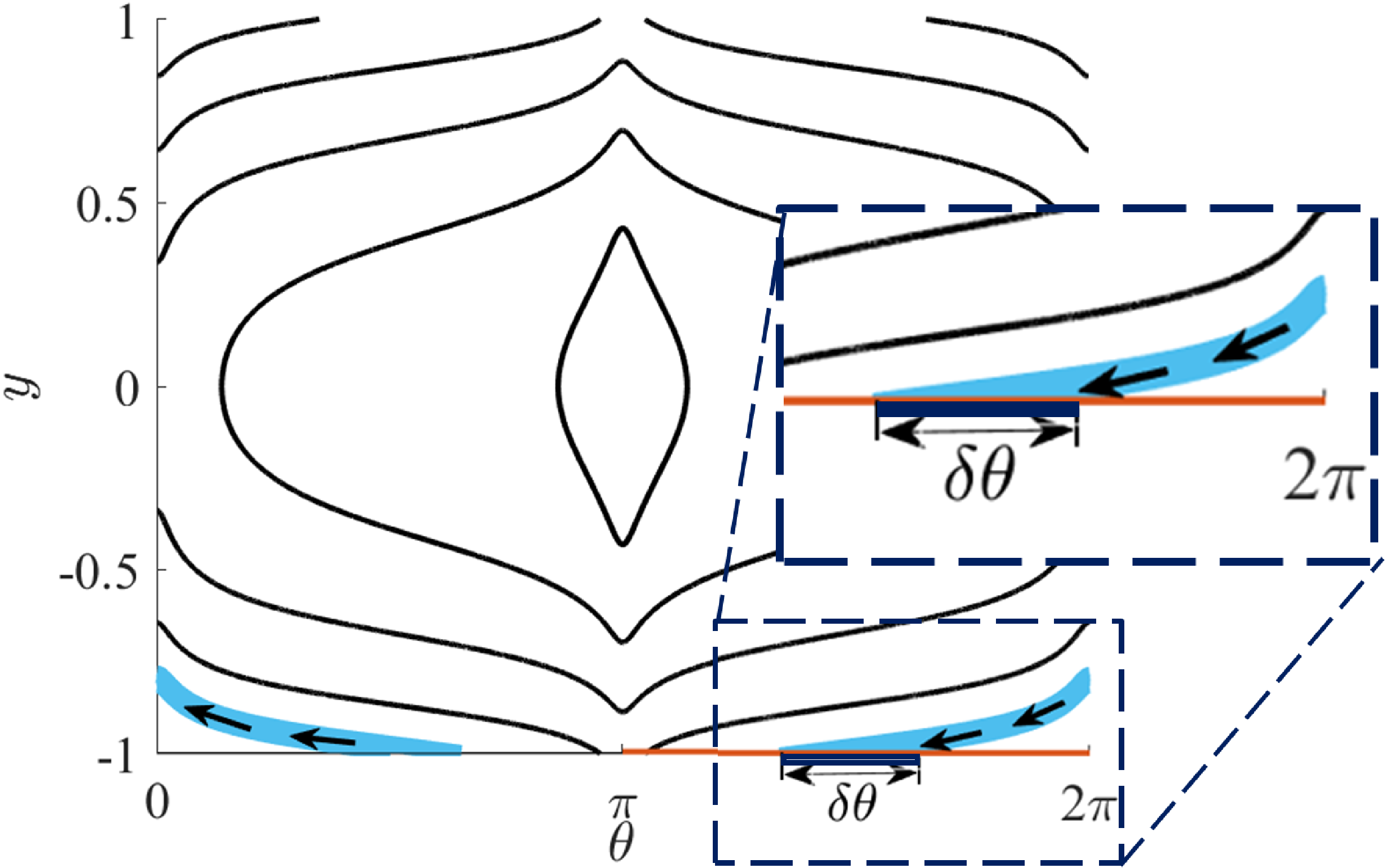}
                        \caption{     
              \label{AccumSchematic}
}
    \end{subfigure}
        \begin{subfigure}[H!]{0.32\textwidth}
         \centering        \includegraphics[width=\textwidth]{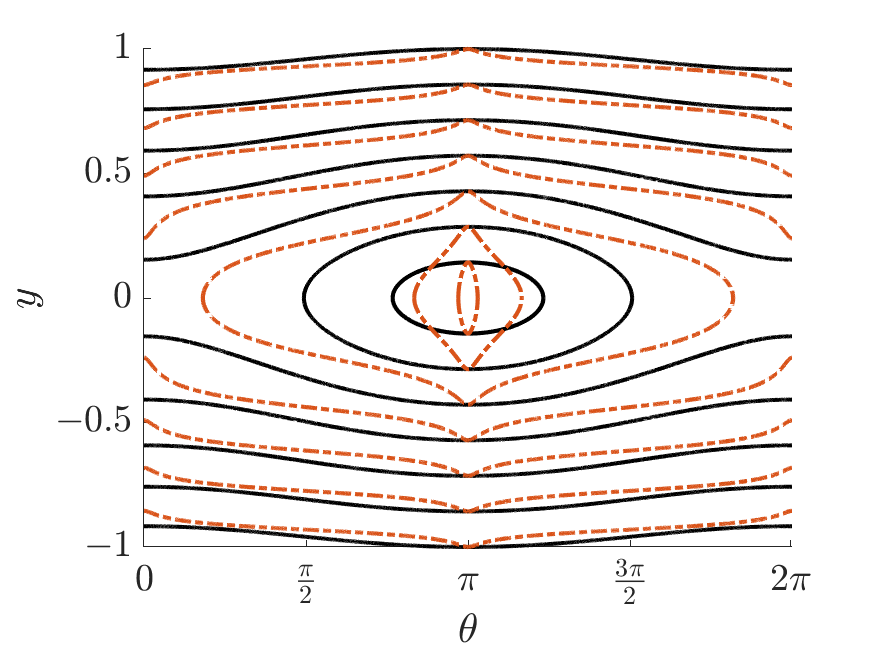}
                \caption{
              \label{HamiltonianNu0pt04}
}
    \end{subfigure}
            \begin{subfigure}[H!]{0.32\textwidth}
         \centering        \includegraphics[width=\textwidth]{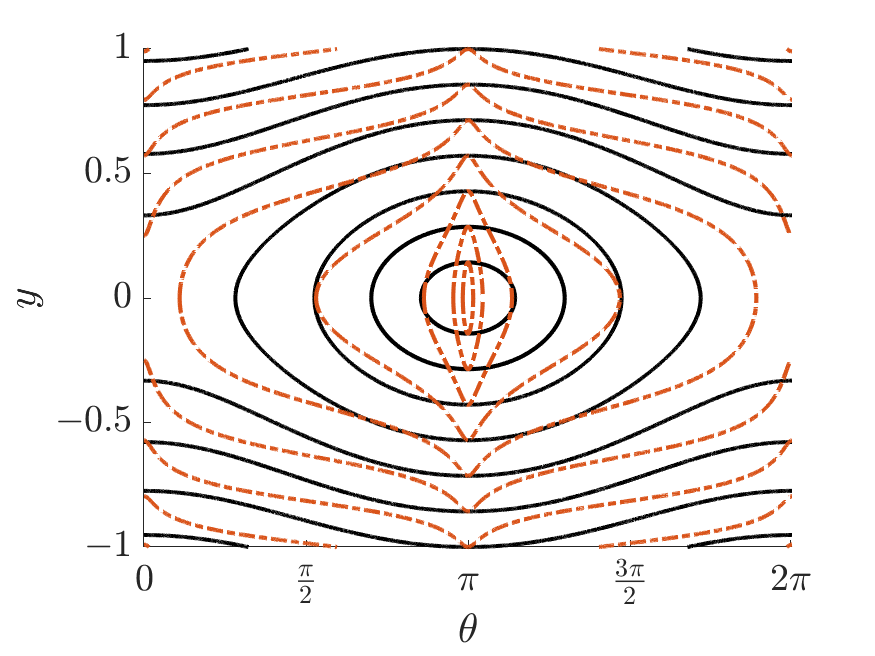}
                \caption{
              \label{HamiltonianNu0pt1}
}
    \end{subfigure}
                \begin{subfigure}[H!]{0.32\textwidth}
         \centering        \includegraphics[width=.99\textwidth]{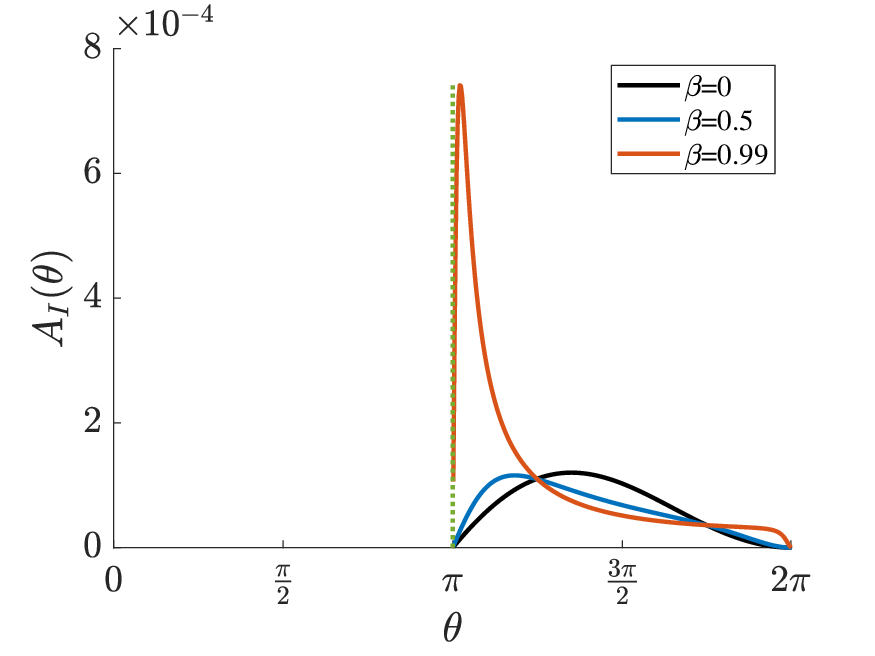}
                \caption{
              \label{AccumPdfNu0pt04}
}
    \end{subfigure}
\begin{subfigure}[H!]{0.32\textwidth}
         \centering        \includegraphics[width=.99\textwidth]{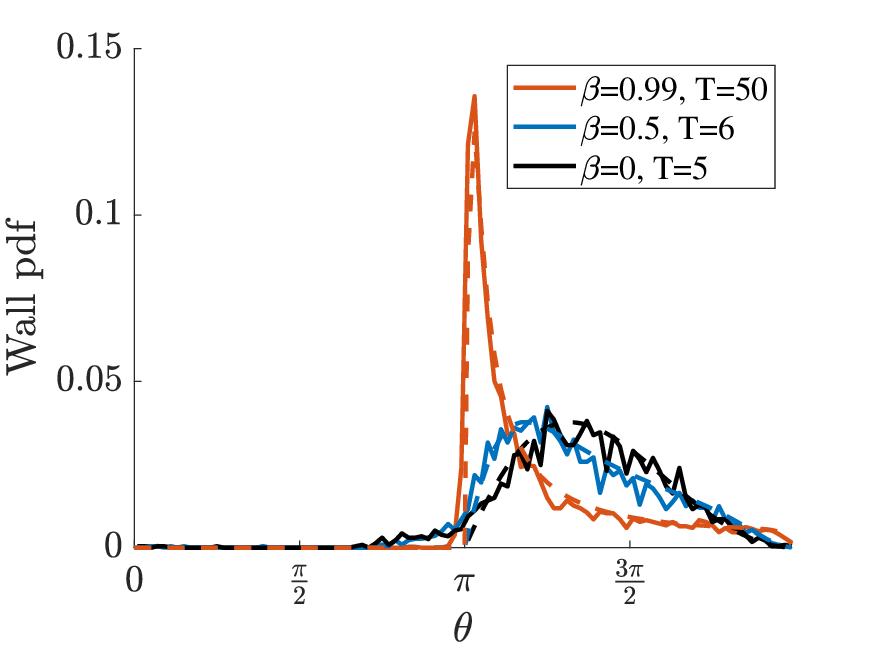}
                \caption{ \label{AccIndex_Matching_PerfAbsorbing}
}
    \end{subfigure}
                \begin{subfigure}[H!]{0.32\textwidth}
         \centering        \includegraphics[width=.99\textwidth]{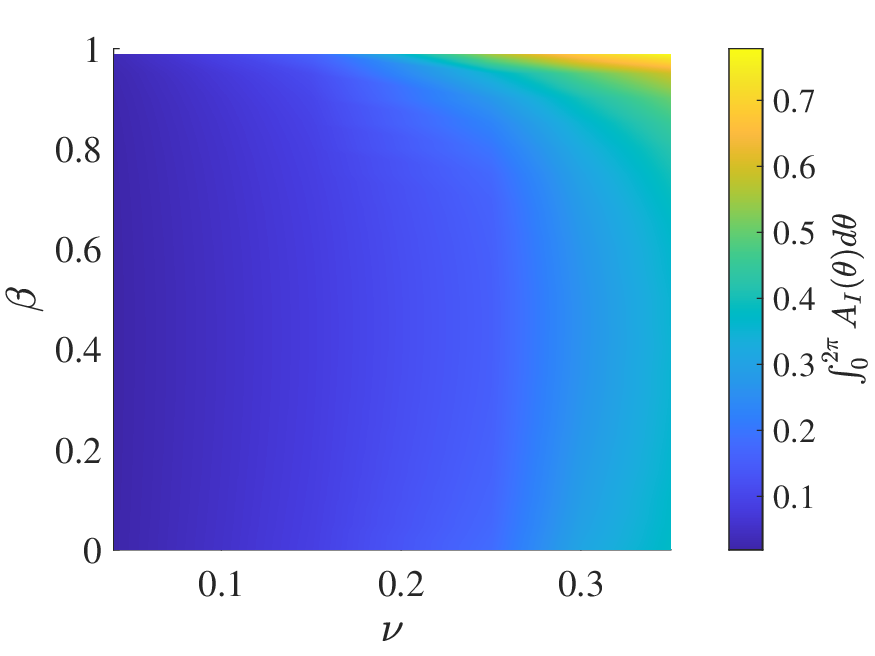}
                \caption{
              \label{AccIndTotalWallProportion}
}
    \end{subfigure}
\caption{ 
\label{AccInd}
{Deterministic dynamics and the accumulation index.}
(a) Schematic of the accumulation index highlighting the area of phase space (blue) from which cells will interact with the bottom wall over orientation space $\delta\theta$. The cell trajectories are shown via arrows; (b) Streamlines at constants of motion for $\nu=0.04$, $\beta=0,0.99$ (solid black and dash-dotted red, respectively); (c) Streamlines at constants of motion for $\nu=0.1$, $\beta=0, 0.99$ (solid black and dash-dotted red, respectively); (d) 
Accumulation index (proportion of 
{initially uniformly distributed cells in the phase space that are incident upon the bottom wall at angles $\theta$}) for $\beta=0, 0.5, 0.99$, for $\nu=0.04$; 
(e) Distribution of wall interactions with absorbing boundary conditions (solid lines) for $Pe=10^4$ for $\beta=0,0.5,0.99$ with $T_\mathrm{sim}=5,6,$ and 50, respectively, and the corresponding accumulation index distributions (dashed lines); and
(f) Proportion of total area of phase space incident on the bottom wall 
{$\int_0^{2\pi}A_I(\theta;\beta,\nu)\,\mathrm{d}\theta$}
{as a function of shape, $\beta$, for various swimming speeds, $\nu$.} 
}
\end{figure} 
{
To analyse shape effects on wall interactions we consider the deterministic problem, in which we keep the purely deterministic drift term and remove diffusion dynamics, such that
\begin{align}
            \derivone{y}{t}&=\nu\sin\theta,\\
            \derivone{\theta}{t}&=y(1-\beta\cos2\theta).
\end{align}
From this, we derive constants of motion for the dynamics (similar to 
\cite{zottl2013periodic}), by eliminating time dependence and solving
    \begin{align}
        \derivone{y}{\theta}=\dfrac{\derivone{y}{t}}{\derivone{\theta}{t}}=\dfrac{\nu\sin\theta}{y(2-\beta\cos2\theta)}.
    \end{align}
Integrating, we find constants of motion, $H$, with $y,\theta, \beta$ and $\nu$ dependence, such that
    \begin{align}
        H=\dfrac{y^2}{2\nu}+
        \dfrac{1}{\sqrt{2\beta(1+\beta)}}\mathrm{arctanh}\left(\sqrt{\frac{2\beta}{1+\beta}}\cos\theta\right)+\mathrm{C},
    \end{align}
    where $C$ is a constant dependent on particle initial conditions $(\theta_0, y_0)$.
From this we derive the trajectories  $y(\theta;H,\beta,\nu)$ in phase space $\theta$--$y$, for any particle with
 initial condition $(\theta_0, y_0)$ and corresponding constant of motion $H$:
\begin{align}
y(\theta;H,\beta,\nu)=\sqrt{
{2\nu}
\left(H-\dfrac{1}{\sqrt{2\beta(1+\beta)}}\mathrm{arctanh}\left(\sqrt{\frac{2\beta}{1+\beta}}\cos\theta
\right)\right)
}.
\end{align}
    
From the trajectories we  extract information regarding the expected wall interactions, trajectory times, and develop a novel accumulation index determining the distribution of expected wall interactions in the case of a uniformly seeded domain. 
 Example trajectories are shown in figure \ref{HamiltonianNu0pt04} for 
$\nu=0.04$, where $\beta=0$,and $\beta=0.99$.}
{ 
The shape of the trajectories themselves are dependent upon the elongation of the swimmers as highlighted in figures \ref{HamiltonianNu0pt04} and \ref{HamiltonianNu0pt1}, where the black lines correspond to spherical swimmers ($\beta=0$),  and the red dash-dotted lines correspond to $\beta=0.99$. 
Elongated swimmers undergo increasing strain effects, 
such that swimmers spend extended times oriented with the flow direction ($\theta=0,\pi$). With increased elongation, the reorientation in phase-space $\left(\parone{y}{\theta}\right)$ steepens about $\theta=0$ and $\theta=\pi$, thus leading to a change in total area enclosed by trajectories through $\theta=\pi, y=\pm1$ and the cell trajectories upon wall approach.}

Supposing there is an initial, uniform distribution over the entire phase plane $\theta\times y\in [0,2\pi)\times [-1,1]$, the 
{
accumulation index, $A_I$, is  defined as
\begin{align}
    \int^{\theta+\delta\theta}_\theta A_I(\theta') \,\mathrm{d}\theta'=\frac{I_W(\theta,\theta+\delta\theta)}{N},
\end{align}
where $I_W(\theta,\theta+\delta\theta)$ is the total number of swimmers that interact with the bottom wall at $y=-1$ with orientations  ranging in $[\theta, \theta+\delta\theta]$ (see schematic in figure \ref{AccumSchematic}), and $N$ is the total number of swimmers.
}
The accumulation indices for orientations of incidence captured in figure 
\ref{AccumPdfNu0pt04} 
 correspond to the velocity ratio $\nu=0.04$. 
For a fixed centreline flow velocity, the increased accumulation index for $\nu=0.1$ results from the increased swimming velocity $V_s$ enabling swimmers to traverse larger vertical distances prior to shear-induced reorientation. This, in turn, allows larger proportions of swimmers in an initially uniformly distributed domain to interact with the walls. 

Further points of interest include the orientation $\theta_{peak}$ at which maximal wall interactions occur. In the accumulation index 
{(figure \ref{AccInd}d)} there is a shift in the peak interaction orientation  $\theta_{peak}$ from  $\theta_{peak}\approx 3\pi/2$ to  $\theta_{peak}\approx\pi$, with cell elongation. 
We find similar shape-based shifts in peak wall-interaction orientation with the absorbing boundary condition in figure \ref{AccIndex_Matching_PerfAbsorbing}. 

The absorbing boundary condition distributions are shown to be in agreement with the accumulation index in the case of  small rotational diffusion for
shape dependent simulation run times $T_\mathrm{sim}$. 
We consider the role of swimmer shape for wall interactions as elongation affects swimming trajectories in sheared flows, especially in linearly varying shear flows. 
Trajectories, in turn, affect the time it takes for cells with specific initial positions and orientations to to swim and rotate before cells encountering the walls. 
We find that the accumulation index captures the wall interactions in short run times only,  as the clear shape dependent shift in peak interaction orientations (figure \ref{AccIndex_Matching_PerfAbsorbing})  disappears for sufficiently long run times, highlighting the transience of the accumulation index distribution. 
{
In figure               \ref{AccIndTotalWallProportion}, the total proportion of swimmers which interact with the bottom wall $\int_0^{2\pi}A_I(\theta)\,\mathrm{d}\theta$ are shown for a range of shape factors and velocity ratios. 
For small swimming velocities, for swimmers of all considered shape factors, only a small proportion of swimmers are expected to interact with the lower wall. 
The proportion of wall interactions increases monotonically with swimming velocity, and increases fastest for $\beta>0.9$, with over 70\% of swimmers interacting with the bottom wall for $\nu>0.3$. 
}

\begin{figure}
    \centering
\begin{subfigure}[H!]{0.32\textwidth}
        \centering
        \includegraphics[width=\textwidth]{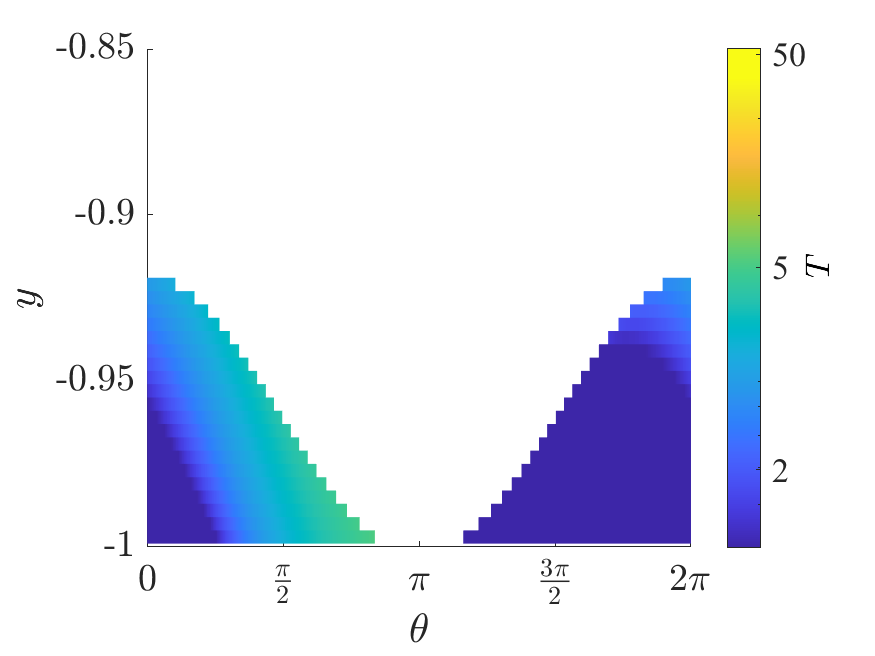}
                        \caption{     
              \label{TermTimebeta0}
}
    \end{subfigure}
\begin{subfigure}[H!]{0.32\textwidth}
        \centering
        \includegraphics[width=\textwidth]{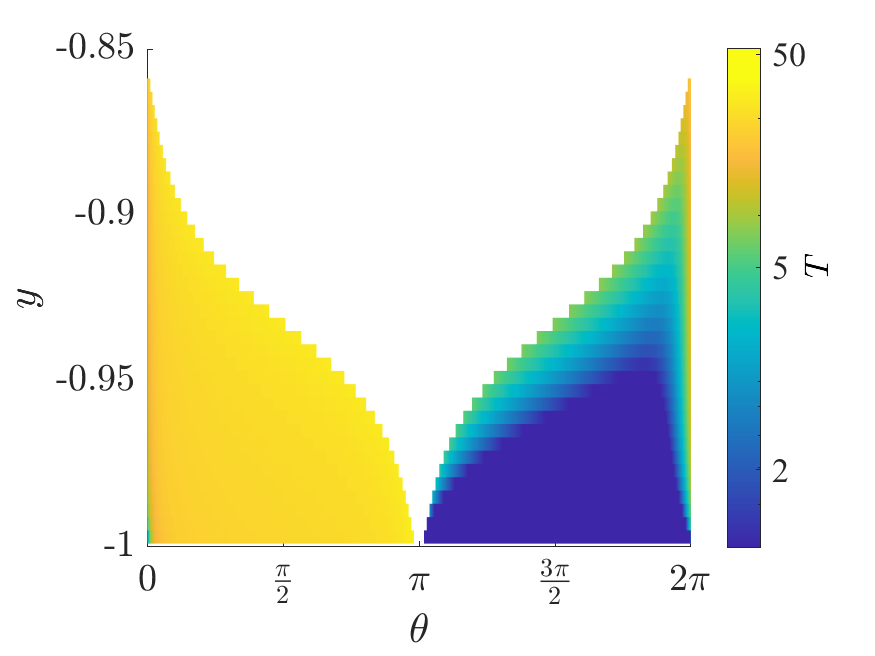}
                        \caption{     
              \label{TermTimebeta0pt99}
}
    \end{subfigure}
\caption{Shape dependence of time taken for trajectories beginning at $(\theta_0, y_0)$ to reach an absorbing wall condition at $y=-1$, $\theta\in[\pi,2\pi)$. From this we can extrapolate the total number of wall interactions by swimmers in the ‘trapped’ domain over a fixed total runtime.
For $\nu=0.04$, (a) $\beta=0$ and (b) $\beta=0.99$. 
\label{TermTimes}
}
\end{figure}

While swimmer shape affects the orientations at which swimmers are most likely to interact with the bottom wall 
 we find that  all particle trajectories are not of equal time duration. While separate identical particles on the same trajectories will have the same orbit duration in the deterministic problem, identical particles on different trajectories will have  different closed-loop orbits in phase space and with different orbit durations. In figure \ref{TermTimes}, the colour maps highlight the time taken for a trajectory starting at position $(\theta_0, y_0)$ to terminate at the bottom wall (i.e. at $y=-1, \theta\in[\pi, 2\pi]$) via an absorbing wall boundary condition. 
The longest trajectories have durations ranging from $T\approx 6$, for $\beta=0$, to $T\approx 45$, for $\beta=0.99$, indicating that slender, elongated cells can take over seven times longer to complete a single full orbit, compared to spherical cells. This indicates, that strongly elongated particles have over seven times fewer opportunities for wall interactions over the fixed time interval of the faster orbit. Although a larger proportion of cells are likely to come  into contact with walls at orientation $\theta_{peak}$ (as in figure               \ref{AccumPdfNu0pt04}), the particles
initially oriented about $\theta=\pi$ have the longest closed loop trajectories and consequent orbit durations, while cells about $\theta=0$ have the shortest closed loop trajectories and corresponding orbit times. When considering a Lagrangian perspective, this acts as a limiting factor for the number of wall interactions per interaction orientation. The number of orbits that a particle can undergo over a fixed simulation runtime $T_\mathrm{sim}$ is of biological interest, as it affects the probability of biofilm formation due to increased opportunities for cell attachment.
{Finally, we note that the accumulation index captures the deterministic limit of the perfectly absorbing boundary $\mathcal{A}$ (as shown in figure \ref{AccIndex_Matching_PerfAbsorbing}) for runtimes corresponding to the longest closed-loop orbit durations calculated for each cell shape in figure \ref{TermTimes}.}

\section{Conclusions \label{Sec:Conclusions}}
Using a finite element framework for continuum distributions of dilute suspensions of microswimmers we have studied the coupled relationship between the bulk flow cell dynamics and the boundary dynamics. 
We find that in order to capture the dynamics of different individual-based wall dynamics (specular reflection, uniform random reflection, absorbing boundaries), it is necessary to be cautious in the the choice of constraints for continuum model approximations. We find that a doubly periodic Poiseuille continuum approximation yields an equilibrium approximation for IBM microswimmer dynamics in a wall-bounded Poiseuille flow with specular reflection. We find that this continuum model effectively captures the macroscopic suspension distributions such as peaks of accumulation and cell depletion at the walls due to low-shear trapping in $y-\theta$ phase space. This  is especially noteworthy as this offers justification for the use of doubly periodic Poiseuille flow models like \cite{vennamneni2020shear} to capture simple bounded domains with a reflective wall condition. We also find that a constant boundary approximation in the equilibrium continuum model yields good agreement with an IBM with random reflections  capturing additional secondary peaks of cell accumulation near the walls. 
 Both results further justify the use of these continuum approximations in the study of wall interactions for the case of dilute microswimmer suspensions.  As the long-term limit of an absorbing boundary condition is all diffusing cells being attached to the wall, we developed a time evolving continuum model with a zero Dirichlet boundary condition which effectively captures the evolution of wall absorption and we have quantified the time and diffusion dependence of wall absorption for elongated particles.

The shape of the swimmers and rotational diffusion experienced by the swimmers is shown to significantly affect the orientation distributions. From a Eulerian perspective, there are no-preferred cell orientations for spherical cells, while more elongated swimmers exhibit a clear preference for orientation up and downstream. This preference 
has smallest orientational spread for $\beta=0.99$, for which the distributions are most peaked at angles just above $\theta=\{0,\pi\}$ (i.e. pointing downstream and out of the wall and pointing up stream and into the wall), with approximately 40\% of cells being shown to interact with the walls with incidence angles $\theta\in[\pi-0.25,\pi+0.5]$. 
From a Lagrangian perspective, this is due to  elongated swimmers spending over 60\% of their orbits aligned with the flow ($|\theta-\pi|<0.1$ and $|\theta-2\pi|<0.1$). On decreasing the rotational P\'eclet number, $Pe$, the spread of maximum wall incidence shifts from $\theta_{peak}=\pi$ to $\theta_{peak}=3\pi/2$ as diffusion dominates deterministic dynamics. For the case of an absorbing boundary condition,
 when decreasing the rotational diffusion, the wall-incidence distributions tend towards the distributions as captured by the novel accumulation index based on deterministic trajectories for runtimes corresponding to shape-dependent orbits, highlighting the importance of bulk flow swimmer trajectories on wall-interaction distributions.

The deterministic dynamics of individual trajectory dynamics in the phase plane $\theta$--$y$ capture 
{the} shift in peak orientation distribution 
from $\theta_{peak}=3\pi/2$ to $\theta_{peak}=\pi$ for spherical to highly elongated swimmers via the accumulation index. The perpendicular approach of spherical swimmers towards surfaces  and the parallel approach of elongated swimmers towards walls, have been observed  for both Chlamydomonas \citep{buchner2021hopping} and bacteria \citep{berke2008hydrodynamic}, respectively, in experiments and numerical studies which include hydrodynamic interactions. Our results suggest that the orientational preferences are influenced by the fundamental bulk behaviours of different shaped swimmers. 

We find that in the absence of diffusion, elongated particles take over seven times as long before interacting with the wall compared to a spherical swimmer. It is possible that elongated swimmers, therefore, must maximise each opportunity they have near the wall.  
Once near a wall, elongation leads to increased resistance to random Brownian rotation allowing swimmers to remain oriented parallel to flows for longer periods which improves their chemotactic sampling accuracy. Additionally, longer periods of  {alignment} with walls allow for longer periods of mechanosensing, which increases the chances of surface attachment being initiated.  

While we have considered multiple idealised wall interaction models, true biological wall interactions do not follow pin-ball dynamics,  uniformly random reflections, nor perfect absorption, especially in due to hydrodynamic interactions. In the literature there is ongoing study into how individual swimmers interact with surfaces in the absence of flows, and even then swimmers are shown to reorient due to hydrodynamic forces. One of nuances picked up by the accumulation index is that the bulk flow dynamics affects the likelihood of how cells approach the wall. For actual accumulation, this is also dependent on the attachment mechanisms of different swimmers, and attachment rates which are not accounted for here. Ideally, future models will combine the models with hydrodynamic interactions.
{For microswimmers in nature, there  exist further variables which affect the likelihood of attachment and reorientation like pili attachment location \citep{melville2013type,jain2012type,proft2009pili}, chemical signals \citep{wadhams2004making}, hydrodynamic stresses \citep{boyle2006quantification,conrad2018confined} and cell deformability \citep{yoshida2020soft}.}
Further experimental data regarding pili, and observed attachment rates at different cell orientations are required to refine the models to specific swimmer types and to draw further conclusions regarding the likelihood and speed of initial biofilm formation.

\section*{Acknowledgement}
 We thank the UKRI for support through the grant EP/S033211/1 Shape, shear, search \& strife; mathematical models of bacteria. We thank the oomph-lib group for various discussions on continuum modelling.
We are grateful to the referees of this paper for their constructive feedback on the first manuscript of this paper.

\section*{Declaration of Interests}
The authors report no conflict of interest.

\section*{Data access statement}
All data and codes supporting this study are provided as supplementary information accompanying this paper at \textcolor{red}{[insert DOI]}.
}
\appendix
\FloatBarrier

\section{Cell trajectories \label{Appendix:cellTrajectories}}
The effect of rotational diffusion in $x$-$y$ space is highlighted in figure \ref{ExampleTrajectories}, where we neglect translational diffusion for simplicity and the IBM is augmented with an $x$--direction advection  term such that $\boldsymbol{X}_t=(y(t), \theta(t), x(t))$:
\begin{subequations}
        \begin{align}
        \boldsymbol{\mu}(y,\theta,t)&=\boldsymbol{\dot{X}}_t 
=
\begin{pmatrix}
        \nu\sin\theta\\ y(1-\beta\cos2\theta) \\\hline1-y^2+\nu\cos\theta
\end{pmatrix},\\
        \boldsymbol{\sigma}(y,\theta,t)&=\left(\begin{array}{c c|c}
          \sqrt{\frac{2}{\mathrm{Pe}_T}}& 0&0\\
          0& \sqrt{\frac{2}{\mathrm{Pe}}}&0\\\hline
          0&0& 0
        \end{array}\right)\label{diffusionTermeq2}.
        \end{align}
\end{subequations}

\section{IBM with specular reflection: Origin of the wall depletion region \label{Appendix:Depletion}}

For the case of the stochastic individual based model regions of accumulation occur in the $\theta$--$y$ phase space, as observed for the doubly periodic Poiseuille flow and the wall-bounded formulation in section \ref{Sec:Specular}. However, in the case of the wall-bounded distribution with specular reflection at the walls, a drop in cell accumulation occurs around $\theta=0$ at  $y=-1$, and   $\theta=2\pi$ at $y=1$ which increases with increased run time.  
One factor which can contribute to this drop in local cell density is a time-stepping effect due to the discrete nature of the numerical method in the IBM. 

\begin{figure}
    \centering
\begin{subfigure}[H!]{0.4\textwidth}
        \centering
        \includegraphics[width=.95\textwidth]{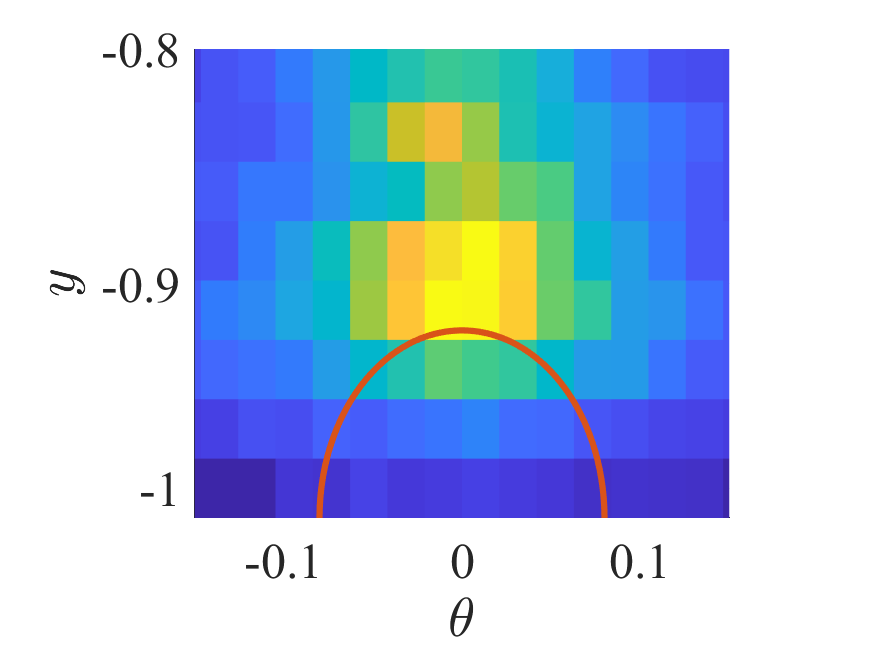}
        \caption{
        \label{IBMDetermGapdt0pt001}}
    \end{subfigure}
\begin{subfigure}[H!]{0.4\textwidth}
        \centering
        \includegraphics[width=.95\textwidth]{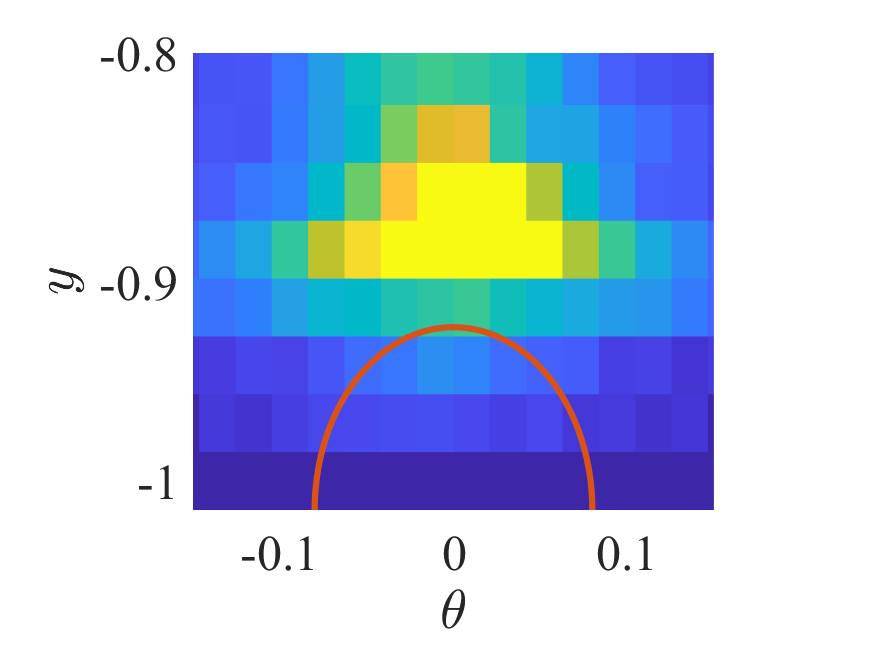}
        \caption{
        \label{IBMDetermGapdt0pt1}}
    \end{subfigure}
\begin{subfigure}[H!]{0.4\textwidth}
        \centering
        \includegraphics[width=.95\textwidth]{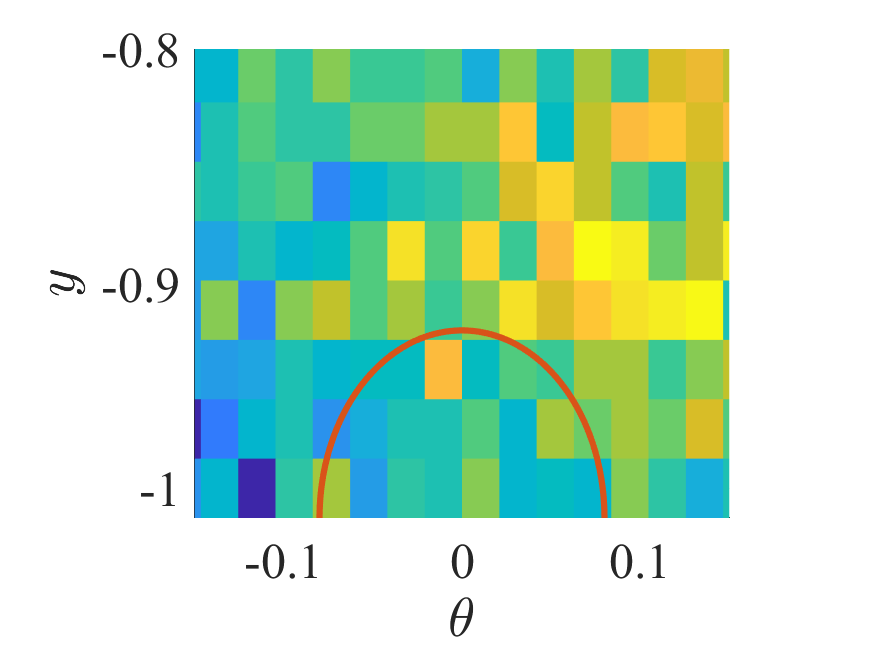}
        \caption{
        \label{IBMPe10Gapdt0pt01}}
    \end{subfigure}
\begin{subfigure}[H!]{0.4\textwidth}
        \centering
        \includegraphics[width=.95\textwidth]{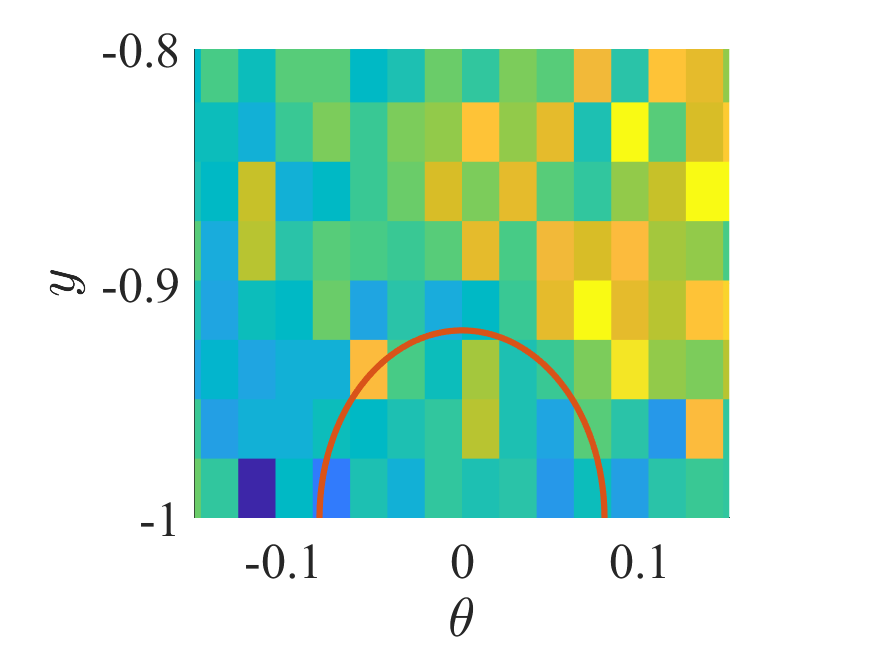}
        \caption{
        \label{IBMPe10Gapdt1}}
    \end{subfigure}
                    \begin{subfigure}[H!]{0.4\textwidth}
         \centering        \includegraphics[width=.95\textwidth]{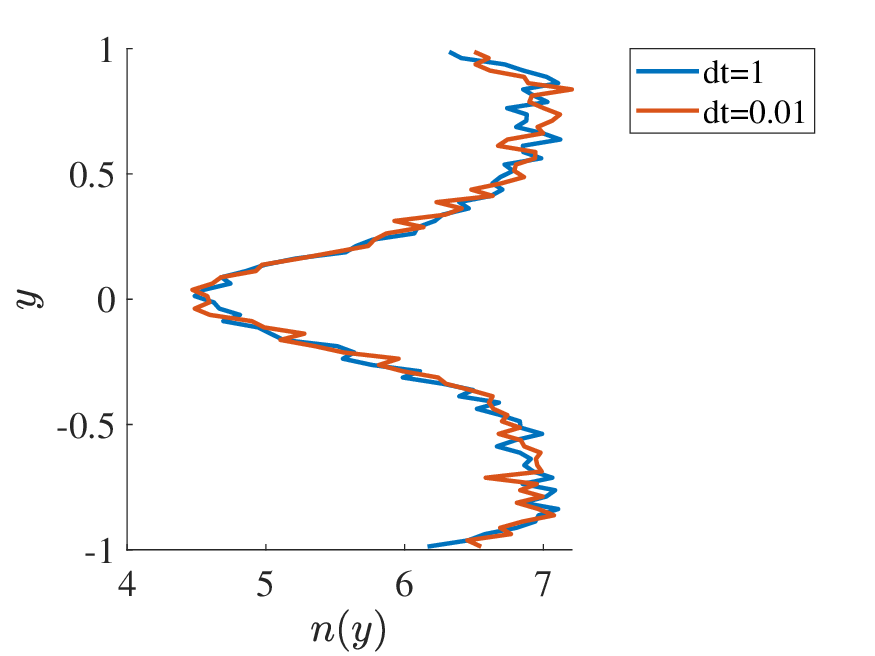}         
                \caption{
\label{NearWallBehaviourGapCellDistribution}
}
    \end{subfigure} 
            \begin{subfigure}[H!]{0.4\textwidth}
         \centering        \includegraphics[width=.95\textwidth]{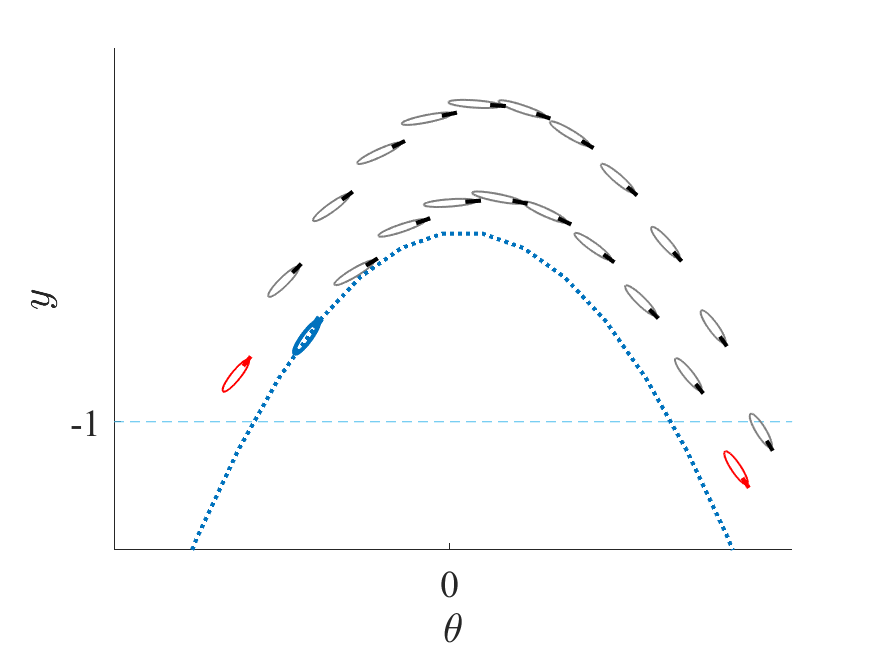}
                \caption{
              \label{GapSchematic}
}
    \end{subfigure}
\caption{Figures to highlight the sensitivity of the IBM to finite time steps, and how these affect the observed boundary interactions.
(a)-(e): IBM Poiseuille flow, for  $\beta=0.99$, $\nu=0.04$.  Purely deterministic IBM for $T_\mathrm{sim}=600$ near bottom wall in (a) \& (b), with (a): $dt=10^{-3}$ and (b): $dt=0.1$.
 (c)-(e): IBM Poiseuille flow, for  $\beta=0.99$, $\nu=0.04$, $Pe=10$, $Pe_T=10^6$, for $T_\mathrm{sim}=600$ near bottom wall in (c)\& (d), (c): $dt=0.01$ and (d): $dt=1$. 
(e): Cell density distribution $n(y)$  for diffusive case ($Pe=10$ and $Pe_T=10^6$) for dt=1 (blue line) and dt=0.01 (red line). 
 (f): Schematic of deterministic trajectory of a particle at bottom wall in continuous time (blue dotted line) highlighting the trajectory deviation for particles of finite time step. Red particle on the right overshoots the wall, and is reflected to the red particle on the left. 
}
\end{figure} 
To illustrate this, consider the case of a purely deterministic system such that the cells must all follow predetermined trajectories.  However, as time is discretised the time steps are of finite size. 
As shown in the schematic in figure \ref{GapSchematic}, if a swimmer (the blue particle) begins on a deterministic trajectory given by the dotted blue line, due to discrete step sizes, the swimmer will gradually drift further from the continuous trajectory with consecutive steps. This effect is compounded when the last step in the orbit(see the red particle on the right) undergoes specular reflection (the red particle on the left) to a position firmly outside its previous deterministic trajectory. With each cycle of reflection, the particle moves further from $\theta=0$, and contributes to local cell  depletion.
In a fully deterministic case for $\beta=0.99$, over a run time $T=600$, we have found a 50\% cell depletion in a radius $r_\epsilon(=0.08)$ about $(\theta,y)=(0,-1)$, when increasing step size from $\mathrm{d}t=10^{-3}$ to $\mathrm{d}t=0.1$ as seen in figures \ref{IBMDetermGapdt0pt001} and \ref{IBMDetermGapdt0pt1}.  


In the diffusive case for $\beta=0.99$ with $Pe=10$ and $Pe_T=10^6$,
over a run time $T=600$, we find enhanced cell depletion  about $(\theta,y)=(0,-1)$,
when increasing step size from $\mathrm{d}t=0.01$ to $\mathrm{d}t=1$ as seen in figures 
\ref{IBMPe10Gapdt0pt01} and 
\ref{IBMPe10Gapdt1}. 
In this case, at $T=600$, we have found a 20\% cell depletion in a radius of $r_\epsilon$ about $(\theta,y)=(0.2,-1)$. While decreasing time step size does not completely remove the cell depletion at the wall (see figure \ref{NearWallBehaviourGapCellDistribution}), it does decrease it.

\bibliographystyle{jfm}
\bibliography{Edits2JFM_BulkBoundaries}

\end{document}